\documentclass[twocolumn,twocolappendix]{aastex7}

\newcommand{\Lya}{Ly{$\alpha$}}
\newcommand{\Lyb}{Ly{$\beta$}}

\newcommand{\CII}{C\,{\sc ii}}

\newcommand{\OIII}{O\,{\sc iii}}
\newcommand{\Ha}{H{$\alpha$}}
\newcommand{\Hb}{H{$\beta$}}

\newcommand{\taueff}{\tau_\mathrm{eff}}
\newcommand{\xiion}{\xi_\mathrm{ion}}
\newcommand{\fesc}{f_\mathrm{esc}}

\usepackage{ulem}
\usepackage{comment}
\usepackage{float}
\usepackage{amsmath}
\usepackage{placeins}
\usepackage{hyperref}

\definecolor{mygreen}{rgb}{0.0, 0.4, 0.0}

\received{}
\revised{}
\accepted{}

\shorttitle{Galaxy--IGM connection in the EoR}
\shortauthors{Kashino et al.}

\graphicspath{{./}{./figs/}}

\begin{document}

\title{EIGER VII. The evolving relationship between galaxies and the intergalactic medium in the final stages of reionization}

\correspondingauthor{Daichi Kashino}

\author[0000-0001-9044-1747]{Daichi Kashino}
\affiliation{National Astronomical Observatory of Japan (NAOJ), 2-21-1, Osawa, Mitaka, Tokyo 181-8588, Japan}
\email[show]{kashinod.astro@gmail.com}

\author[0000-0002-6423-3597]{Simon J.~Lilly}
\affiliation{Department of Physics, ETH Z{\"u}rich, Wolfgang-Pauli-Strasse 27, Z{\"u}rich, 8093, Switzerland (emeritus)}
\email{}

\author[0000-0003-2871-127X]{Jorryt Matthee}
\affiliation{Institute of Science and Technology Austria (ISTA), Am Campus 1, 3400 Klosterneuburg, Austria}
\email{}

\author[0000-0003-0417-385X]{Ruari Mackenzie}
\affiliation{Laboratory of Astrophysics, {\'E}cole Polytechnique F{\'e}d{\'e}rale de Lausanne (EPFL), Observatoire de Sauverny, 1290 Versoix, Switzerland}
\email{}

\author[0000-0003-2895-6218]{Anna-Christina Eilers}
\affiliation{Department of Physics, Massachusetts Institute of Technology, Cambridge, MA 02139, USA}
\affiliation{MIT Kavli Institute for Astrophysics and Space Research, Massachusetts Institute of Technology, Cambridge, MA 02139, USA}
\email{}

\author[0000-0002-3120-7173]{Rongmon Bordoloi}
\affiliation{Department of Physics, North Carolina State University, Raleigh, 27695, North Carolina, USA}
\email{}

\author[0000-0003-3769-9559]{Robert A.~Simcoe}
\affiliation{Department of Physics, Massachusetts Institute of Technology, Cambridge, MA 02139, USA}
\affiliation{MIT Kavli Institute for Astrophysics and Space Research, Massachusetts Institute of Technology, Cambridge, MA 02139, USA}
\email{}

\author[0000-0003-2895-6218]{Rohan P.~Naidu}\thanks{NHFP Hubble Fellow}
\affiliation{MIT Kavli Institute for Astrophysics and Space Research, Massachusetts Institute of Technology, Cambridge, MA 02139, USA}
\email{}

\author[0000-0002-5367-8021]{Minghao Yue}
\affiliation{MIT Kavli Institute for Astrophysics and Space Research, Massachusetts Institute of Technology, Cambridge, MA 02139, USA}
\email{}

\author[0000-0003-4491-4122]{Bin Liu}
\affiliation{Department of Physics, North Carolina State University, Raleigh, 27695, North Carolina, USA}
\email{}

\begin{abstract}
We present a comprehensive analysis of the relationship between galaxies and the intergalactic medium (IGM) during the late stages of cosmic reionization, based on the complete \textit{JWST} EIGER dataset. Using deep NIRCam $3.5\,\mathrm{\mu m}$ slitless spectroscopy, we construct a sample of 948 [\OIII]$\lambda5008$-emitting galaxies with $-21.4\lesssim M_\mathrm{UV}\lesssim -17.2$ spanning $5.33<z<6.97$ along six quasar sightlines.  We correlate these galaxies with \Lya\ and \Lyb\ transmission measured from high-resolution quasar spectra across multiple redshift intervals.  We find clear redshift evolution in the correlation between galaxy density and transmission: it is suppressed in overdense regions at $z<5.50$, while enhanced at $5.70<z<6.15$.  The intermediate range exhibits a transitional behavior.  Cross-correlation measurements further reveal excess absorption within $\sim 8$\,cMpc of galaxies at low redshifts, and enhanced transmission at intermediate scales ($\sim$5--20\,cMpc) at $z>5.70$.  Statistical tests using mock catalogs with realistic galaxy clustering but no correlation with the transmission field confirm that the observed correlations are unlikely to arise by chance.  The evolving signals can be explained by stronger absorption in overdense regions, combined with the competing influences of local radiation fields and the rising background radiation.  While local radiation dominates ionization of the surrounding IGM at earlier times, the background becomes increasingly important, eventually surpassing the impact of nearby galaxies.  These results support an inside-out progression of reionization, with ionized regions originating around clustered, star-forming galaxies and gradually extending into underdense regions.
\end{abstract}



\section{Introduction}
\label{sec:intro}

The Epoch of Reionization (EoR) represents a pivotal phase in cosmic history, marking the transition of intergalactic hydrogen from a neutral to an ionized state \citep[e.g.,][]{2001PhR...349..125B,2006ARA&A..44..415F,2022LRCA....8....3G}. This epoch represents the last major phase transition of the universe, shaping the properties of the intergalactic medium (IGM) and influencing galaxy formation and evolution \citep[e.g.,][]{2012MNRAS.427..311W,2020MNRAS.494.2200K,2023MNRAS.525.5932B}.  A central goal of modern astrophysics is to uncover the physical processes driving this transition, with particular emphasis on identifying the sources of ionizing radiation and understanding the (related) evolution of the spatial ionization structure of the IGM \citep[e.g.,][]{2007MNRAS.377.1043M,2009MNRAS.394..960C,2020MNRAS.492..653H,2021MNRAS.503.3698H}.  

Among various potential sources, ultraviolet radiation from massive stars in star-forming galaxies has long been considered the primary ionizing source responsible for reionization \citep{1987ApJ...321L.107S,1999ApJ...514..648M,2000ApJ...535..530G,2004ApJ...613....1F,2006MNRAS.369.1625I,2009ApJ...703.1416F,2012ApJ...746..125H,2018PhR...780....1D}.  However, ongoing debates persist regarding the relative contributions of low-mass galaxies and active galactic nuclei (AGNs) in this process \citep{2014MNRAS.440..776Y,2015MNRAS.451.2544P,2018A&A...613A..44G,2018MNRAS.474.2904P,2019ApJ...879...36F,2015MNRAS.453.2943C,2017MNRAS.465.3429C,2017MNRAS.468.4691D,2024MNRAS.527.6139S,2024arXiv240111242D,2025arXiv250203683J,2024arXiv240915453A}.  
Furthermore, several more exotic mechanisms, such as dark matter annihilation or decay \citep{2006MNRAS.369.1719M}, primordial globular clusters \citep{2002MNRAS.336L..33R,2009MNRAS.395.1146P,2018MNRAS.479..332B,2021MNRAS.504.4062M}, and mini/micro-quasars \citep{2004ApJ...604..484M,2011A&A...528A.149M}, have been discussed as potential contributors.

While these more exotic mechanisms provide intriguing possibilities, the ``galaxy-driven'' hypothesis remains the prevailing framework, and has underpinned state-of-the-art cosmological simulations of the EoR \citep[e.g.,][]{2018MNRAS.479..994R,2020MNRAS.496.4087O,2022MNRAS.511.4005K,2024MNRAS.531.3406B,2025MNRAS.539.2790C}.  These models broadly reproduce the global reionization history within existing constraints \citep[e.g.,][]{2017MNRAS.465.4838G}.  However, many key uncertainties remain.  In particular, simulations often assume simplified prescriptions for the production and escape of ionizing photons, making it difficult to assess whether their success reflects correct physics or the tuning of the parameters of the prescriptions.

Observationally, the galaxy-driven scenario remains inadequately tested. An indirect approach has estimated the ionizing photon budget of galaxies based on insights into the UV luminosity function, ionizing photon production efficiency ($\xiion$), and the escape fraction ($\fesc$) of ionizing photons, and compared these estimates against the requirements for sustaining reionization.  A consensus has emerged that this photon budget may well be sufficient--or at least it is not in significant contradiction--to explain the ionization of hydrogen across the universe \citep{2015ApJ...802L..19R, 2015MNRAS.451.2030D,2015ApJ...811..140B,2020ApJ...892..109N}.  
However, it is not well understood how $\xiion$ and $\fesc$ vary with galaxy properties such as $M_\mathrm{UV}$, or where the faint-end cutoff of the UV luminosity function lies.  Accounting for galaxies that are too faint to be detected, or for which standard calibrations of $\xiion$ and $\fesc$ may not be applicable, is crucial for accurately assessing the photon budget.  Addressing these issues requires additional observational constraints, such as a joint study of galaxies and the IGM.

A crucial step toward testing the reionization scenarios more directly is the precise characterization of the spatial correlation between galaxies and the ionized structures of the IGM.  This can be achieved through combining \Lya\ (and also \Lyb) forest spectra of luminous quasars in the EoR with galaxy surveys along these sightlines.  Before the advent of the \textit{JWST}, observations with the Subaru Hyper Suprime-Cam (HSC) had suggested a large-scale ($\sim 40\,\mathrm{cMpc}$) positive correlation between galaxy density and mean \Lya\ transmitted flux at $z \sim 5.7$ \citep{2018ApJ...863...92B,2020ApJ...888....6K,2022MNRAS.515.5914I,2021ApJ...923...87C,2023ApJ...955..138C}.  These findings suggest that the ionizing radiation field is enhanced around galaxies at relatively higher redshifts during the tail end of the EoR.
However, these measurements used samples of galaxies selected either as narrow-band Lyman-$\alpha$ emitters (LAEs) or as Lyman-break galaxies (LBGs), and the photometric redshift uncertainties and relatively sparse sampling of bright sources limited the ability to investigate small-scale correlations.  Furthermore, the visibility of LAEs itself is sensitive to the local opacity of the IGM, which may bias these conclusions.
Recently, a novel approach called ``photometric IGM tomography'' has been proposed \citep{2023MNRAS.523.1772K}.  This uses wide-field narrow-band imaging to trace the \Lya\ transmission from background galaxies (instead of quasars) and detect LAEs at the same redshift, thereby enabling two-dimensional mapping of IGM--galaxy correlations.  This overcomes the fundamental limitation of quasars in probing transverse structure.  While promising, the method currently lacks the sensitivity for robust results and awaits further development.

Spectroscopic surveys enable measuring the cross-correlation between the position of galaxies and the transmitted \Lya\ flux down to a few cMpc scales \citep{2003ApJ...584...45A,2005ApJ...629..636A}.  This metric has been particularly valuable for probing the interplay between galaxies and the surrounding IGM, providing a direct observational constraint that can be readily compared with cosmological simulations \citep{2022MNRAS.512.4909G,2025MNRAS.539.2790C}. 
At $z\lesssim4$, well after the completion of reionization, the mean \Lya\ transmission is known to monotonically increase as a function of distance from galaxies, reaching the background level at $\sim10\,\mathrm{cMpc}$ \citep{2021ApJ...909..117M,2024MNRAS.529.2794M}.  In contrast, at $z\sim5\textrm{--}6$, only tentative evidence of excess transmission around galaxies had been obtained in pre-JWST pioneering studies based on ground-based facilities \citep{2018MNRAS.479...43K,2020MNRAS.494.1560M}. 

The launch of \textit{JWST} has revolutionized the study of reionization.  Its unprecedented sensitivity and near-infrared capabilities have enabled the detection of a vast population of high-redshift galaxies. In the Emission-line galaxies and Intergalactic Gas in the Epoch of Reionization (EIGER) project (PID 1243; PI S.~J.~Lilly), \citet{2023ApJ...950...66K} revealed strong evidence that star-forming galaxies ionize the surrounding gas at $z\sim 6$, by measuring the cross-correlation between [\OIII]-emitting galaxies and the \Lya\ transmission.  However, these results could still be largely affected by cosmic variance due to the reliance on a single sightline, limiting their ability to represent the average view of the Universe \citep{2024arXiv241002850G}. 
Similarly, using the ASPIRE survey (PID 2078; \citealt{2023ApJ...951L...4W}), \citet{2024ApJ...976...93J} analyzed 14 quasar sightlines and reached consistent conclusions.  However, their study was limited by the quality of the quasar spectra and did not measure the galaxy-\Lya\ cross-correlation.  \citet{2025arXiv250307074K} used ASPIRE data in five fields with high-quality quasar spectra to measure the galaxy-\Lya\ cross-correlation over $5.4<z<6.5$, detecting excess transmission on scales of $\sim20$--40\,cMpc.  While this supports the presence of large ionized regions around galaxies, the relatively small galaxy sample size (49 across five fields) and the use of a single, broad redshift bin limit its ability to investigate redshift evolution.  As a result, the connection between galaxies and the IGM, and how it evolves, remains poorly constrained.

In this work, we address these limitations by using the full dataset across six quasar sightlines from the complete EIGER survey, increasing the galaxy sample in \citet{2023ApJ...950...66K} sixfold.  Compared to ASPIRE, our observations reach twice the sensitivity to emission lines, and around each quasar, provide a contiguous survey area more than twice as large.  EIGER thus provides the most complete galaxy sample along a set of quasar sightlines to date.  Coupled with over 100 hours of quasar spectroscopy, we conduct a high-precision analysis of the correlation between galaxies and IGM transmission.  This approach addresses the challenges posed by sparse sampling and cosmic variance, allowing us to probe the multi-scale transmission structures around galaxies and their redshift evolution during the late stages of reionization.  

This paper is organized as follows.  
\textsection\ref{sec:EIGER} provides an overview of our \textit{JWST}/NIRCam observations, the identification of emission-line galaxies, and the basic characteristics of the complete galaxy sample.
\textsection\ref{sec:QSO_spectra} describes the ground-based spectroscopy of the target quasars and measurement of \Lya\ and \Lyb\ transmission fluxes.
\textsection\ref{sec:results} presents the results from galaxy--transmission correlation analyses.
In \textsection\ref{sec:discussion}, we interpret our results and provide a coherent picture of the evolving connection between galaxies and the IGM during the final stages of cosmic reionization.
\textsection\ref{sec:summary} summarizes our results and their interpretation.  

Throughout this work, we adopt a flat $\Lambda$CDM cosmology with $\Omega_\Lambda = 0.69$, $\Omega_\mathrm{M}=0.31$, and $H_0 = 67.7\,\mathrm{km\,s^{-1}\,Mpc^{-1}}$ \citep{2020A&A...641A...6P}.  All magnitudes are quoted in the AB system.  Unless otherwise specified, distances are expressed in comoving units, hereafter denoted as cMpc.

\section{The EIGER survey}
\label{sec:EIGER}

The EIGER Collaboration observed a total of six $z \gtrsim 6$ luminous quasar fields using NIRCam in the WFSS mode.  These six quasars were selected to address a range of science, based on the properties of the \Lya\ forest and Gunn-Peterson troughs, the presence and ionization state of known metal absorption systems, the characteristics of the quasars themselves, and the availability of deep high-resolution spectra from ground-based observations.  Detailed descriptions of the survey design, observing program, data reduction, and the identification of [\OIII]-emitting galaxies are provided in \citet{2023ApJ...950...66K} and \citet{2023ApJ...950...67M}, which presented results from the first quasar field (J0100$+$2802).  We briefly summarize the program setup and the main steps of galaxy identification, and then present the newly extended sample of [\OIII]-emitters.  Table \ref{tb:target_quasars} summarizes the six target quasars, their observing periods, and the corresponding abbreviations used in this paper.  We refer the reader to the companion papers \citep{2023ApJ...950...68E,2024ApJ...966..176Y,2024ApJ...974..275E} for intrinsic quasar properties, \citet{2024ApJ...963...28B} for lower-redshift galaxy populations identified by different emission lines (\Ha\, He\,{\sc i}, S\,{\sc iii}, and Pa\,$\gamma$), and \citet{2024ApJ...963..129M} for broad-line \Ha-emitters at $z\sim4$--5. 

\begin{deluxetable*}{ccccc}
\tablecaption{Target quasars and observations \label{tb:target_quasars}}
\tablehead{
    \colhead{Identifier}&
    \colhead{$z_\mathrm{QSO}$}&
    \colhead{Observing periods}&
    \colhead{P.A.}&
    \colhead{Abbreviations}\\
    \colhead{}&
    \colhead{}&
    \colhead{(yyyy/mm/dd)}&
    \colhead{degree}&
    \colhead{}}
\startdata
 SDSS J010013.02+280225.8 &  6.3270$^\mathrm{a}$ & 2022/08/22--24  & 236 & J0100$+$2801, J0100\\
 SDSS J114816.64+525150.3 &  6.4189$^\mathrm{a}$ & 2022/12/22--28 & 280 & J1148$+$5251, J1148 \\
 ULAS J014837.63+060020.0 &  5.99$^\mathrm{a}$  & 2022/12/13, 2023/01/14--15 & 68 & J0148$+$0600, J0148 \\
 SDSS J103027.09+052455.0 &  6.308$^\mathrm{b}$ & 2023/05/03, 05/30--06/02, 2024/05/25 & 110 & J1030$+$0524, J1030 \\
 PSO J159.2257-02.5438    &  6.381$^\mathrm{b}$ & 2023/05/03, 06/02--03, 2024/05/31--06/01 & 108 & J159$-$02, J159 \\
 ULAS J112001.48+064124.3 &  7.0848$^\mathrm{a}$ & 2022/12/16--17, 2023/01/16--17 & 294 & J1120$+$0641, J1120
\enddata
\tablecomments{The columns denote the quasar identifiers, their redshifts (determined by means of a: [\CII]158\,$\mu$m or b: Mg\,{\sc ii} $\lambda\lambda2796,2803$ emission lines), the periods of the EIGER observations, the approximate position angle of the survey defined with respect to the telescope V3 axis, and the abbreviations used in this paper.}
\end{deluxetable*}

\subsection{\textit{JWST} NIRCam observations}
\label{sec:observations}

\begin{figure*}[t]
\centering
\includegraphics[width=3.5 in]{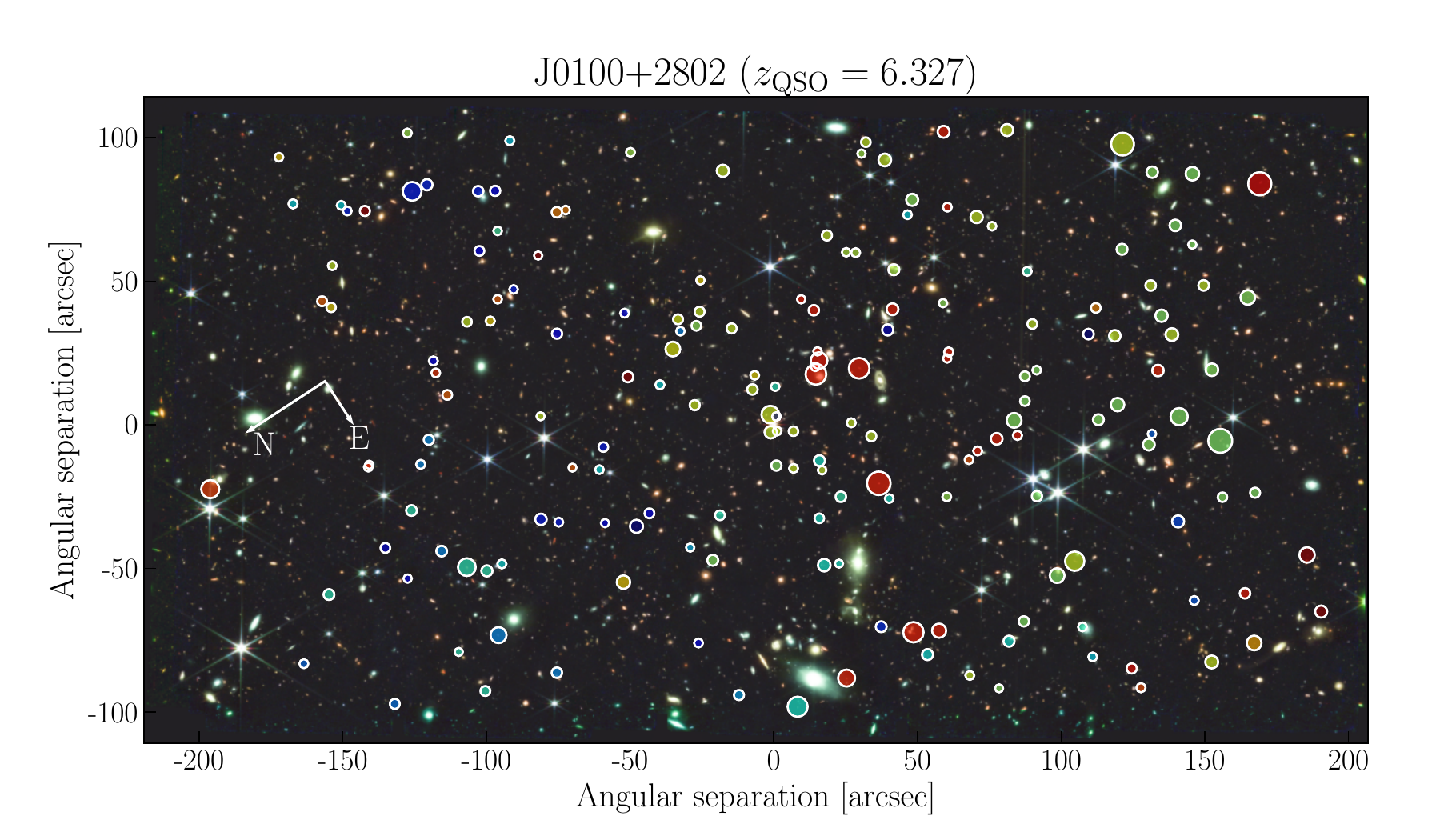}
\includegraphics[width=3.5 in]{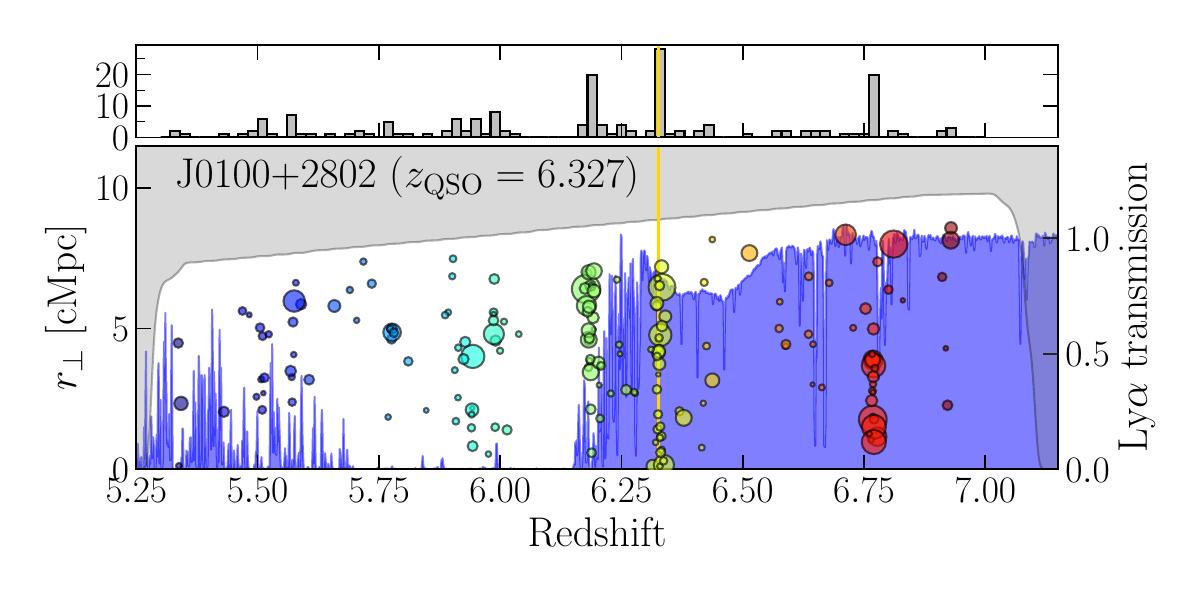}
\includegraphics[width=3.5 in]{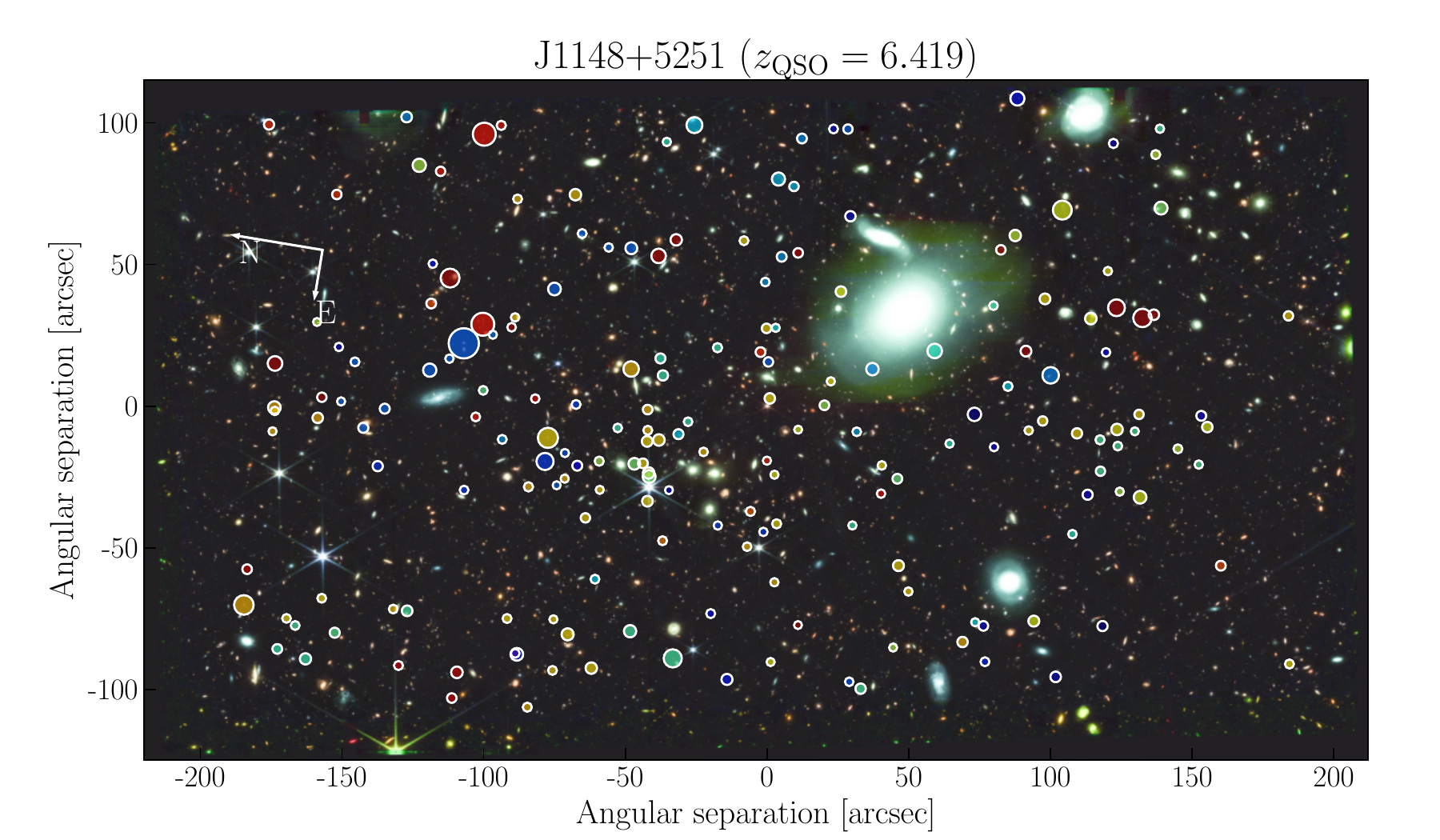}
\includegraphics[width=3.5 in]{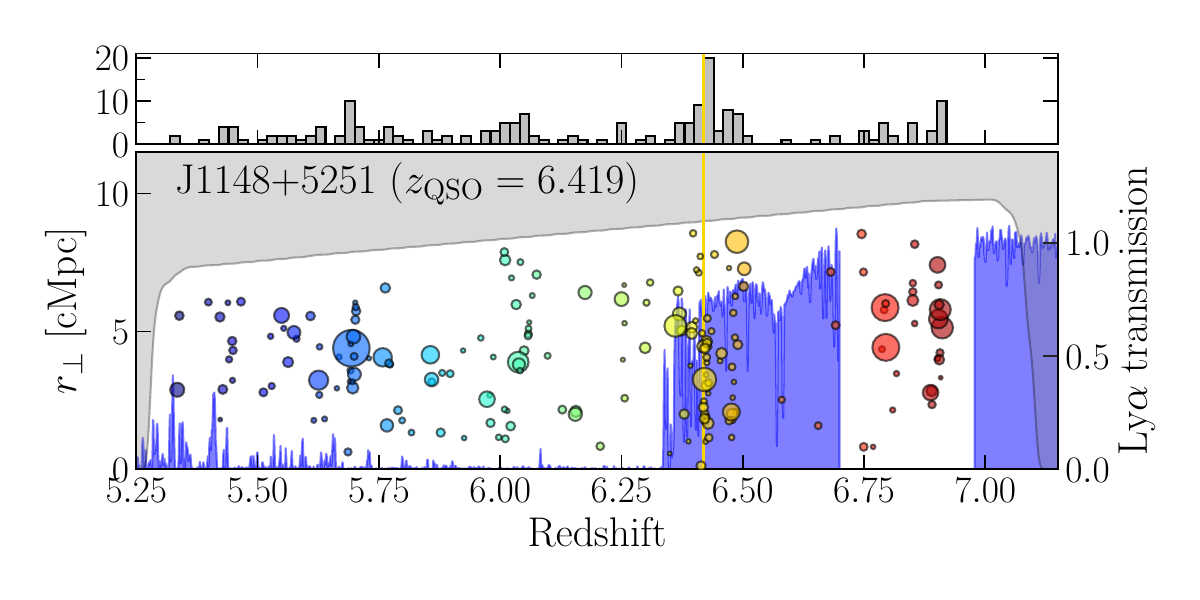}
\includegraphics[width=3.5 in]{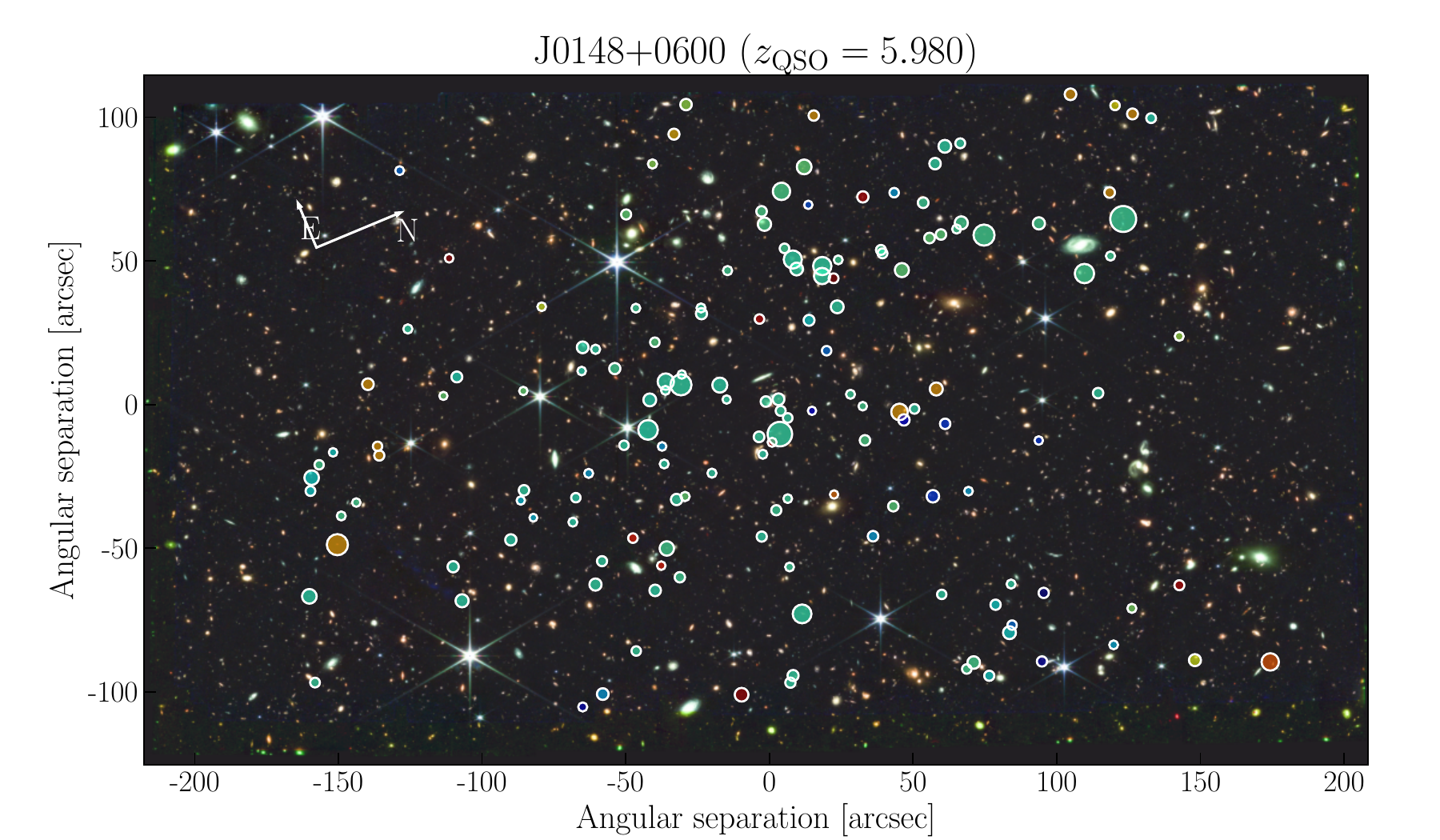}
\includegraphics[width=3.5 in]{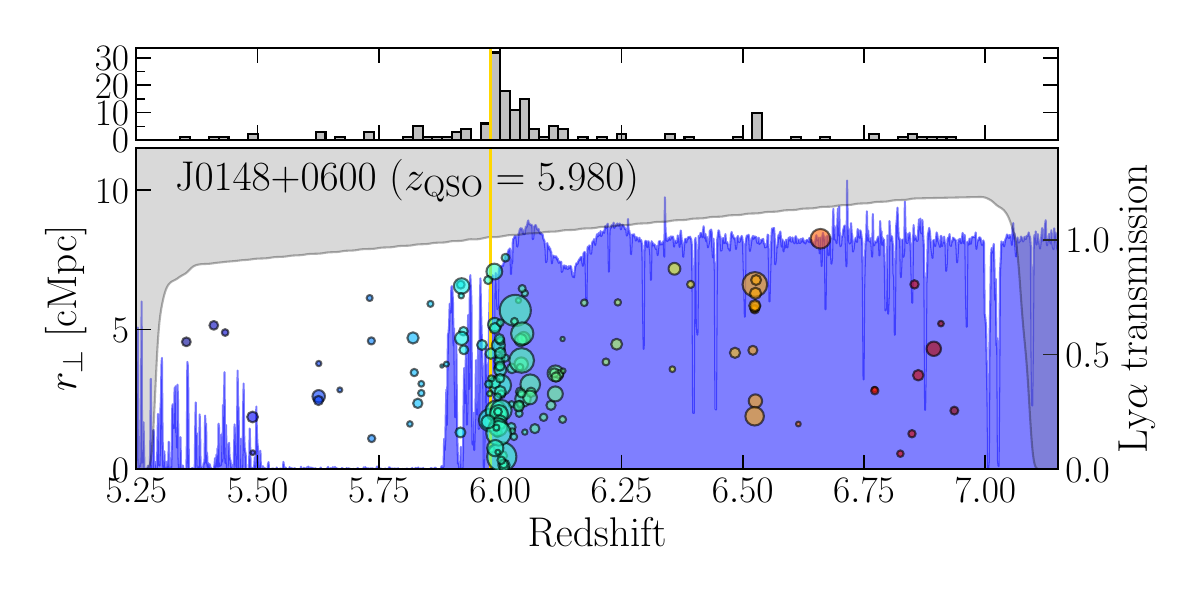}
\includegraphics[width=6.5in]{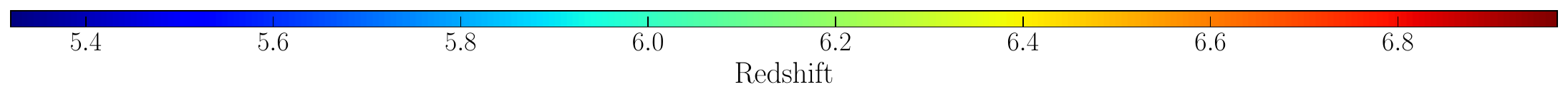}
\caption{
Observations of three quasar fields (J0100, J1148 and J0148) out of the six targets.  Left panels show the RGB images constructed from the NIRCam imaging data (F115W, F200W, and F356W).  The axes indicate angular separation from the quasar position.  Circles indicate the on-sky positions of the detected [\OIII]-emitters, with color-coded redshift and size-coded relative [\OIII] luminosity.
Right panels show the transverse separation of the [\OIII]-emitters from the quasar sightline as a function of redshift, with the redshift histogram ($\Delta z =0.02$) shown in the upper subpanels.  The symbols follow the same color and size coding as in the left panels.  The gray shaded regions define the forbidden region, which accounts for the redshift-dependent maximum field-of-view and the filter transmission curve ($5.3\le z \le 7.0$).  In the background, the transmission spectrum of each quasar is shown in filled blue (refer to the right-hand $y$-axis), where the wavelength is translated into the redshift for \Lya.  The gold vertical lines mark the redshift of each quasar.  
\label{fig:observations1}
}
\end{figure*}

\begin{figure*}[t]
\centering
\includegraphics[width=3.5 in]{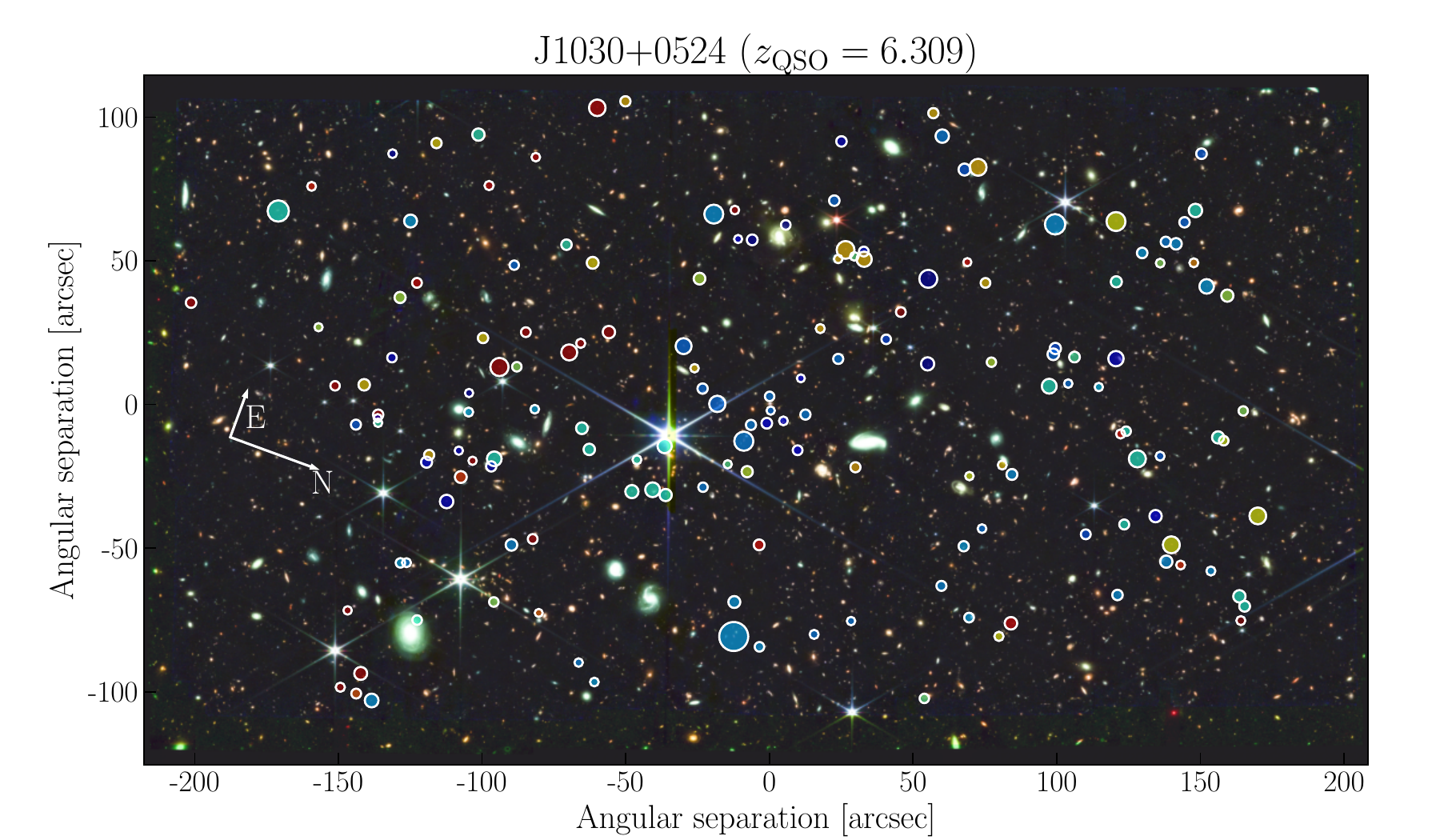}
\includegraphics[width=3.5 in]{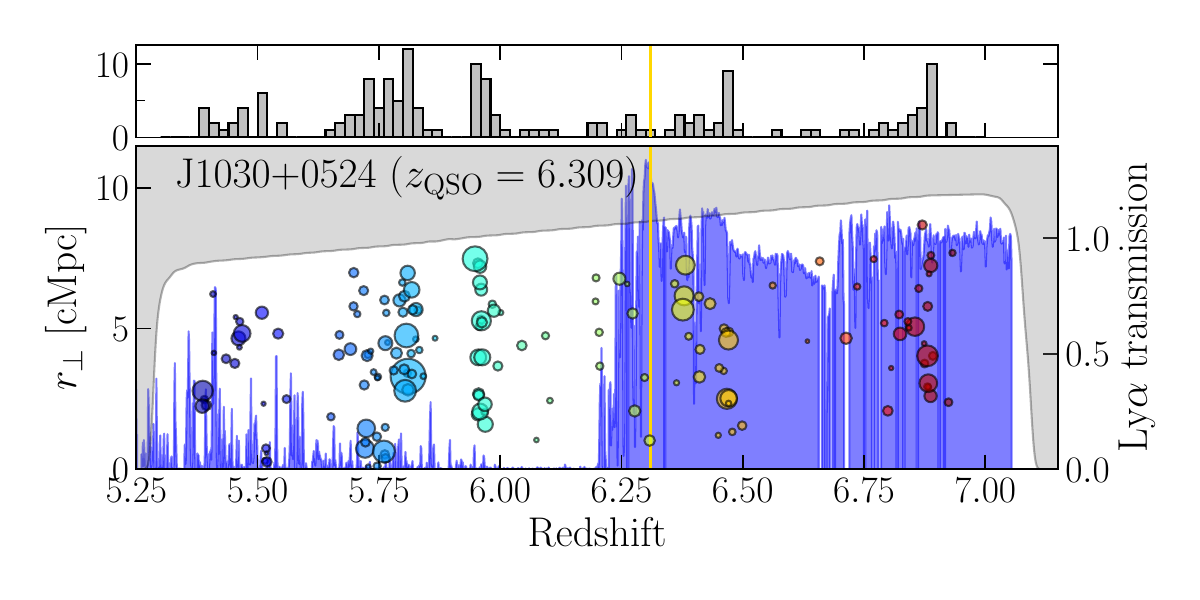}
\includegraphics[width=3.5 in]{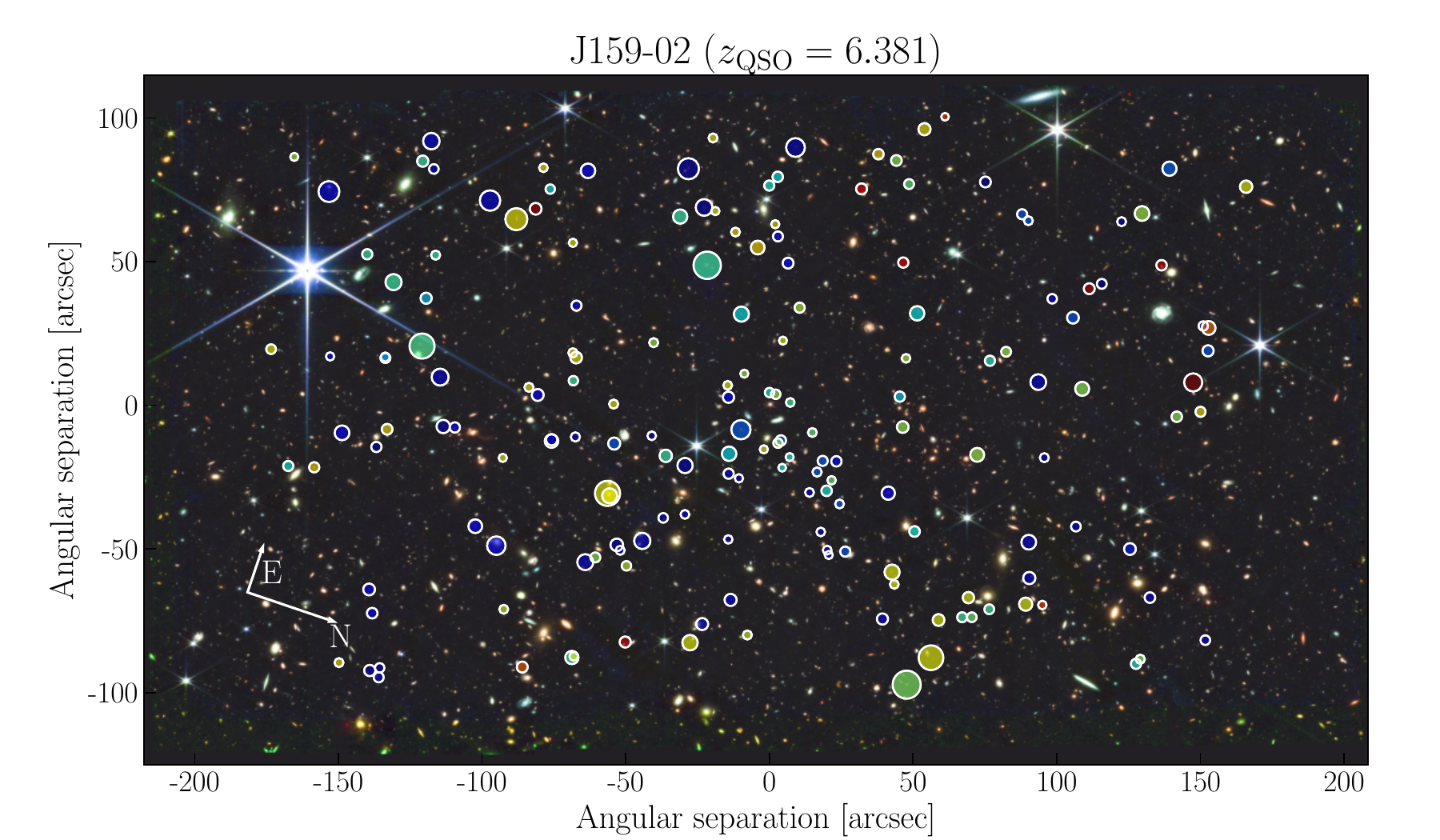}
\includegraphics[width=3.5 in]{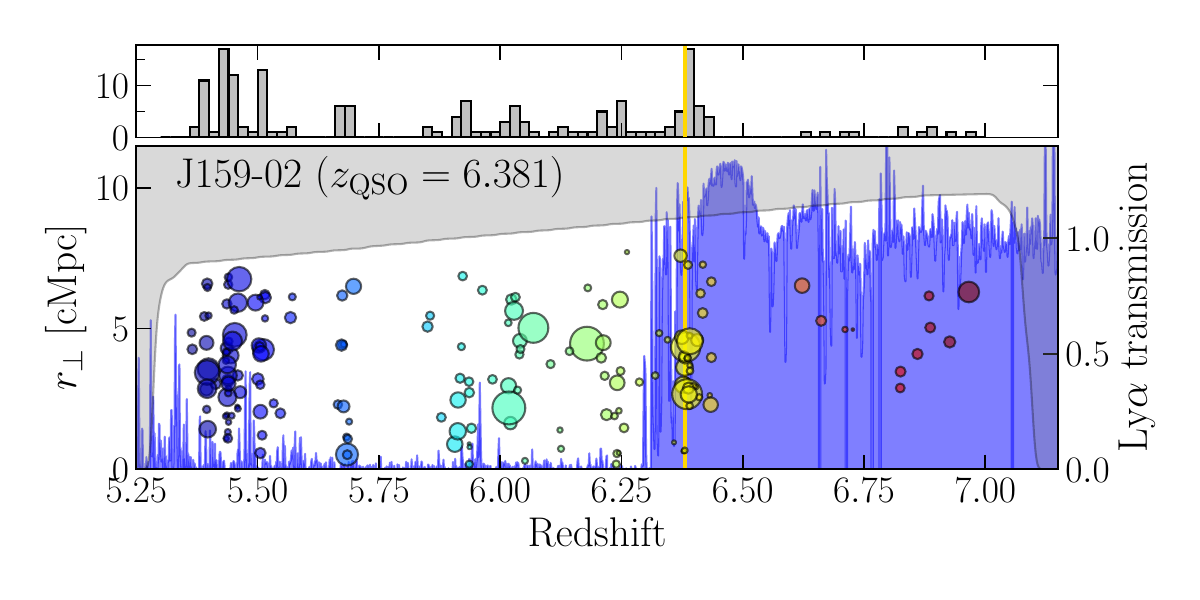}
\includegraphics[width=3.5 in]{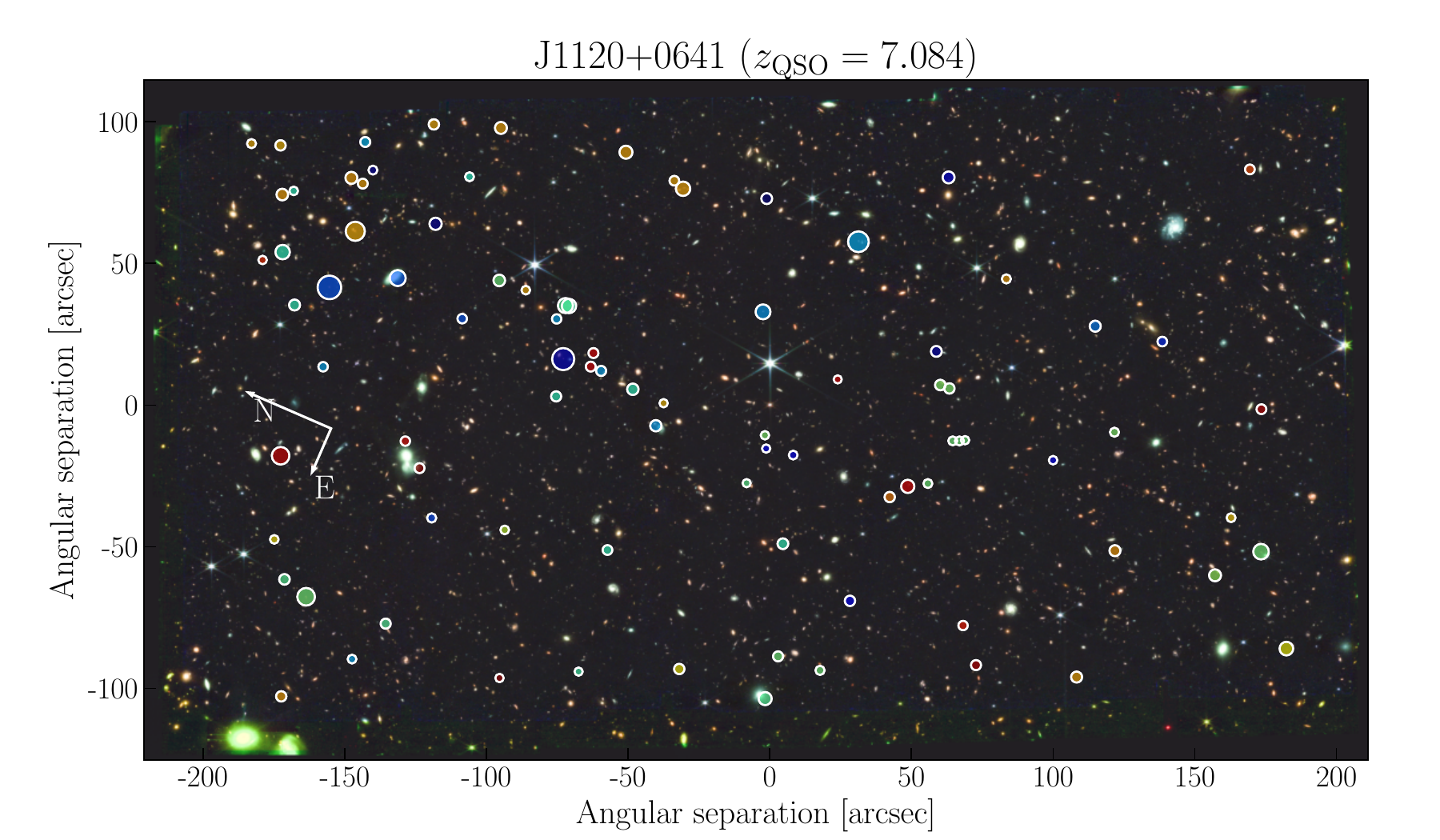}
\includegraphics[width=3.5 in]{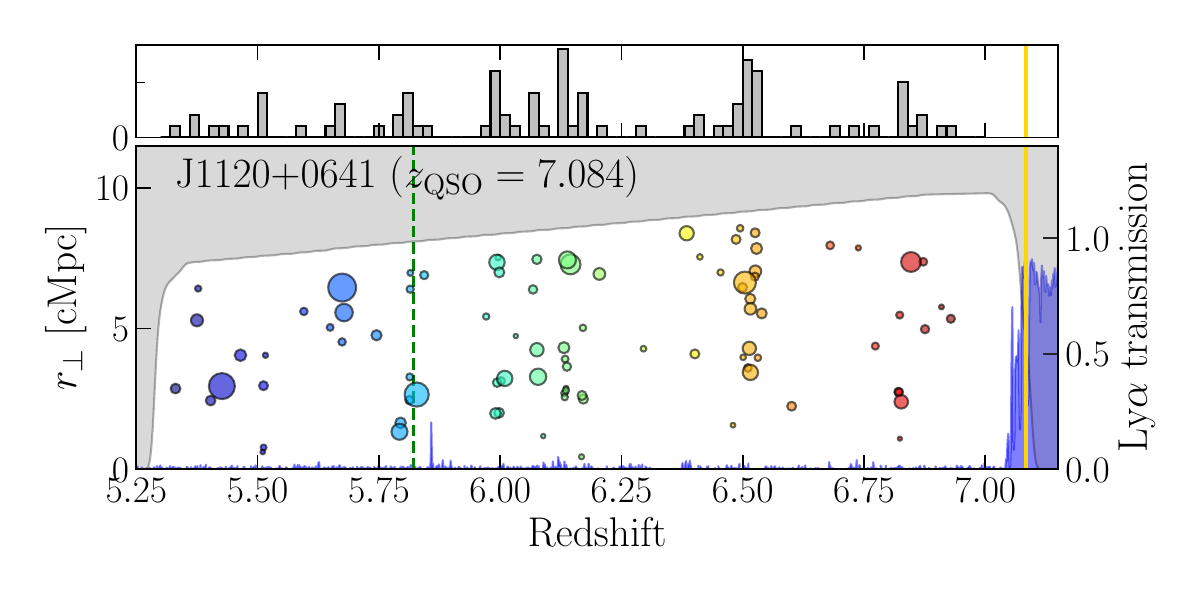}
\includegraphics[width=6.5 in]{figs/colorbar.pdf}
\caption{Same as Figure~\ref{fig:observations1}, but for the remaining three fields (J1030, J159, and J1120). The green dashed vertical line in the bottom right panel (J1120) indicates the position of the Ly$\beta$ line, below which the transmitted flux is affected by \Lyb\ absorption at higher redshifts.
\label{fig:observations2}}
\end{figure*}

The NIRCam imaging and WFSS observations of the six quasar fields were conducted between August 22, 2022 and June 2, 2024 (see Table \ref{tb:target_quasars}).  The Visit 2 observation of J1030 was truncated due to insufficient storage space on the telescope data recorder, but was re-conducted in May 2024.  Likewise, Visits 3 and 4 for J159 were initially skipped due to a failed guide star acquisition, but were later compensated through observations conducted in May--June 2024.

Each quasar field was observed over four visits, forming a $2 \times 2$ overlapping mosaic.  This setup provides a rectangular coverage of approximately $3\arcmin \times 6 \arcmin$ per field, centered on the quasar, as shown in Figures~\ref{fig:observations1} and \ref{fig:observations2}.  
Approximately $40\arcsec \times 40\arcsec$ around the quasar is covered by all four mosaic tiles, yielding the longest total exposure time.  

The WFSS observations utilized the Grism R in the long wavelength (LW) channel combined with the F356W filter.  This setup yields spectra covering wavelengths from 3.1 to 4.0\,$\mathrm{\mu m}$, dispersed along the detector rows with a spectral resolution of $R\approx 1500$.  This relatively high spectral resolution significantly facilitates the identification of emission-line signals in the WFSS data and their source objects in the direct images.  The two NIRCam modules, A and B, disperse the spectra in opposite directions.  In our $2\times2$ mosaic configuration, the central vertical strip of width $\approx 60\arcsec$ is covered by both modules, yielding oppositely dispersed spectra for each object within this region (see also Figure 1 of \citealt{2023ApJ...950...66K}).   This dual coverage helps associate emission lines, particularly singlets such as H$\alpha$, with their corresponding galaxies without ambiguity (see \citealt{2024ApJ...963...28B}).

Simultaneous imaging observations were performed during the WFSS exposures using the short wavelength (SW) channels, first with the F115W filter and then with F200W.  Each visit employed a three-point INTRAMODULEX primary dither and a four-point subpixel dither, yielding a total of 12 exposures per visit and module for the SW images (F115W and F200W), and 24 WFSS images.  By co-adding these exposures, the nominal total exposure time per visit is 4380 seconds per SW filter and 8760 seconds for WFSS.  The combination of small-scale dithering and the large-scale mosaic pattern results in spatial variations in the total exposure time across the field, ranging from 1.5\,ks in edge regions sampled by a single primary dither (with four-point subpixel dither) to 13\,ks for SW imaging, and from 2.9\,ks to 35\,ks for WFSS observations.

Following the WFSS observations in each visit, direct and two displaced ``out-of-field'' images were taken in the F356W filter, with F200W in the SW channel.  These three images have an exposure time of 526 seconds each, yielding a total exposure time of 1578 seconds across most of the coverage per visit and a maximum of 6.3\,ks in the overlap regions of the four visits.

The procedure described above is common to all the quasar fields.  However, after completing the observations in the first quasar field (J0100$+$2802), we noticed that a strip running along the bottom of the combined WFSS coverage lacks coverage in the direct images, because the combination of direct and out-of-field images do not fully cover the survey field defined by the primary dithers.  To address this issue, we requested additional exposures to ensure complete direct image coverage for all sources with dispersed NIRCam WFSS grism spectra.  This was approved and implemented for all the remaining five quasar fields.  

The additional imaging involved two visits, each consisting of a single exposure in F356W (simultaneously with F200W in the SW channel) with an exposure time of 526 seconds (using the same detector setup as in the direct and out-of-field imaging).  The central positions of these frames were slightly offset vertically from the originally planned visits in the lower row of the mosaic pattern.  As a result, in addition to the required extended coverage (vertically $+13$ arcseconds beyond the original bottom edge), approximately 50\% of the entire survey field received additional exposure time due to these new images.  These supplementary images also proved valuable for identifying the sources of so-called ``tadpole'' scattered light artifacts that severely contaminated the Module B Grism R data, as these artifacts often mimic emission-line signals albeit with unphysically high equivalent widths.

The reduction of the NIRCam imaging and WFSS data is entirely described in \citet{2023ApJ...950...66K}.  In brief, the WFSS data were processed using the \texttt{jwst} pipeline (version 1.8.2--1.9.4) and additional custom steps.  We first applied the \texttt{Detector1} stage of the pipeline to the raw exposure files, and then used \texttt{Spec2} solely to assign the WCS solution to each frame.  The grism images were subsequently processed using \texttt{Image2}.\footnote{This requires modifying the FITS header to trick the pipeline into treating the data as images.}  We then constructed emission-line images (referred to as the EMLINE images in \citet{2023ApJ...950...66K}) by applying a running median filter along each detector row to subtract the continuum emission.

\subsection{Identification of [\OIII]-emitting galaxies}
\label{sec:identification}

The spectral range covered by the F356W filter allows us to identify [\OIII]-emitters within the redshift range of $5.33  \lesssim z \lesssim 6.97$.  To construct robust galaxy samples, we employed two complementary methods, referred to as the ``backward'' and ``forward'' approaches, as described in detail in \citet{2023ApJ...950...66K}.  We used \texttt{grismconf}\footnote{\url{https://github.com/npirzkal/GRISMCONF}} with the NIRCam trace models (V4)\footnote{\url{https://github.com/npirzkal/GRISM_NIRCAM}}, to which we applied pixel-level corrections based on our independent analysis of commissioning data (PID 1076; PI N.~Pirzkal).

The backward approach begins with the co-added emission-line images of each visit.  Using SExtractor, we searched for combinations of emission features that could plausibly arise from [\OIII] doublets ($\lambda \lambda 4960, 5008$) or triplets including H$\beta$.  We then attempted to associate these detected lines with a source galaxy in the broadband image.  The separation of the detected lines, assuming an [\OIII] doublet and H$\beta$, provides a preliminary estimate of the redshift and, consequently, their observed wavelengths.  Using the grism trace model, we can obtain a rough estimate of the location of the source galaxy in the sky (i.e., in the imaging data).  Additionally, the appearance of the source galaxy--such as its shape and characteristic red color (high $m_\mathrm{F200W}-m_\mathrm{F356W}$) due to flux excess in F356W--further aids in source identification.  This process therefore allows us to reliably associate detected emission lines to their corresponding broadband sources in almost every case.

In contrast, the forward approach starts with a catalog of broadband sources detected in the imaging observations. From this catalog, we selected $\sim$20k sources brighter than $m_\mathrm{F356W} = 29\,\mathrm{mag}$.  For each of these broadband objects, we extracted 2D spectra from the EMLINE (i.e., continuum-subtracted) grism images of the individual exposures and coadded them.  SExtractor was run on each of these coadded images to detect emission-line signals.  We selected those in which a candidate [\OIII] doublet was detected at $>3\sigma$, spatially close to the spectral trace of the object and with a self-consistent separation in wavelength.

The two samples of [\OIII]-emitter candidates derived from these complementary approaches were then carefully inspected and reconciled by DK and JM.  Notably, our identification algorithm has been fine-tuned and improved since the publication of the early results from the EIGER survey \citep{2023ApJ...950...66K,2023ApJ...950...67M}.  These refinements have led to an increased number of [\OIII]-emitting galaxies identified in the first quasar field, J0100.  

The reconciled catalog was independently vetted by RB by extracting and measuring redshifts for each galaxy.  We estimate the redshift uncertainty to be about $70\,\mathrm{km\,s^{-1}}$ for the full sample ($\sim 20\,\mathrm{km\,s^{-1}}$ for high confidence detections), using spectra from the oppositely dispersed grism data (see also \citealt{2025arXiv250315587S}).

We have identified a total of 948 [\OIII]-emitting galaxies within the redshift range $5.33 < z < 6.97$ across the six targeted quasar fields: J0100 (180 galaxies), J1148 (191), J0148 (153), J1030 (161), J159 (174), and J1120 (89).  It should be noted that, as detailed in \citet{2023ApJ...950...67M}, [\OIII]-emitting clumps located within 2\arcsec\ were counted as a single ``system'' using a friends-of-friends algorithm.  For each system, we combined the fluxes of all components and assign its position to that of the brightest component.  Figures~\ref{fig:observations1} and \ref{fig:observations2} present the detected [\OIII]-emitters in each quasar field.  The full catalog of the [\OIII]-emitter sample is provided in Table \ref{tb:O3_sample}.

The relatively lower number of detections in the field of J1120 reflects cosmic variance, since the background noise level in the J1120 data is comparable to the average seen in the other fields.  In addition, the quasar redshift is not included in the redshift coverage for [\OIII]: we have found that four out of the five quasars whose redshifts fall within coverage reside in strong galaxy overdensities (see \citealt{2024ApJ...974..275E} and Mackenzie et al. in preparation for studies of quasar environments).  

\begin{deluxetable}{ccccc}
\tablecaption{[\OIII]-emitter sample\label{tb:O3_sample}\tablenotemark{a}}
\tablehead{
    \colhead{FIELD}&
    \colhead{ID}&
    \colhead{R.A.}&
    \colhead{Decl.}&
    \colhead{Redshift}}
\startdata
J0100 &    30 &  15.081560 & 28.053464 & 5.94 \\
J0100 &    70 &  15.057173 & 28.085925 & 5.73 \\
J0100 &   257 &  15.076980 & 28.057597 & 5.81 \\
J0100 &   391 &  15.061417 & 28.077917 & 5.98 \\
J0100 &   421 &  15.091936 & 28.036181 & 6.19 \\
J1148 &   884 & 177.123991 & 52.882304 & 6.48 \\
J1148 &  1160 & 177.124547 & 52.889755 & 6.85 \\
J1148 &  1443 & 177.112105 & 52.850369 & 6.04 \\
J1148 &  1660 & 177.111304 & 52.851583 & 5.66 \\
J1148 &  1808 & 177.114197 & 52.863445 & 5.56 \\
J0148 &   844 &  27.136666 &  5.977662 & 5.43 \\
J0148 &  1207 &  27.137072 &  5.979931 & 5.82 \\
J0148 &  1223 &  27.131826 &  5.992295 & 6.89 \\
J0148 &  1538 &  27.148854 &  5.954675 & 6.01 \\
J0148 &  1551 &  27.131086 &  5.997100 & 6.00 \\
J1030 &  1184 & 157.572617 &  5.389005 & 5.80 \\
J1030 &  1269 & 157.591242 &  5.439117 & 6.09 \\
J1030 &  1323 & 157.572733 &  5.387374 & 6.65 \\
J1030 &  1527 & 157.572781 &  5.385720 & 6.88 \\
J1030 &  1634 & 157.581731 &  5.408614 & 5.82 \\
J159  &  1740 & 159.204343 & -2.522699 & 6.17 \\
J159  &  1838 & 159.188635 & -2.571279 & 5.43 \\
J159  &  1982 & 159.188981 & -2.572305 & 5.39 \\
J159  &  2014 & 159.189532 & -2.571479 & 5.39 \\
J159  &  2066 & 159.194031 & -2.558448 & 6.66 \\
J1120 &  1054 & 170.032857 &  6.678901 & 6.07 \\
J1120 &  1060 & 170.051666 &  6.722512 & 6.52 \\
J1120 &  1496 & 170.041429 &  6.703598 & 6.90 \\
J1120 &  1526 & 170.018668 &  6.651731 & 6.51 \\
J1120 &  1635 & 170.037739 &  6.696733 & 6.03
\enddata
\tablenotetext{a}{The first five sources in each field are listed here.  The full catalog will be made publicly available upon publication.}
\end{deluxetable}

\subsection{Flux measurement}
\label{sec:flux_measurement}

We measure total emission-line fluxes of the galaxies based on an optimal 1D spectral extraction, as detailed in \citet{2023ApJ...950...67M}.  The extracted 1D spectral line profile of [\OIII]$\lambda 5008$ is fit with 1--3 Gaussian profiles to account for subcomponents seen in the collapsed profile, arising from multiple clumps and/or spatially extended galaxy morphologies.  The fitting procedure uses a least-squares algorithm with the Python package \texttt{lmfit}.  The same spectral profile (with different amplitudes) is then fit to the \Hb\ and [\OIII]$\lambda4960$ lines.

\begin{figure}[t]
\centering
\includegraphics[width=3.5 in]{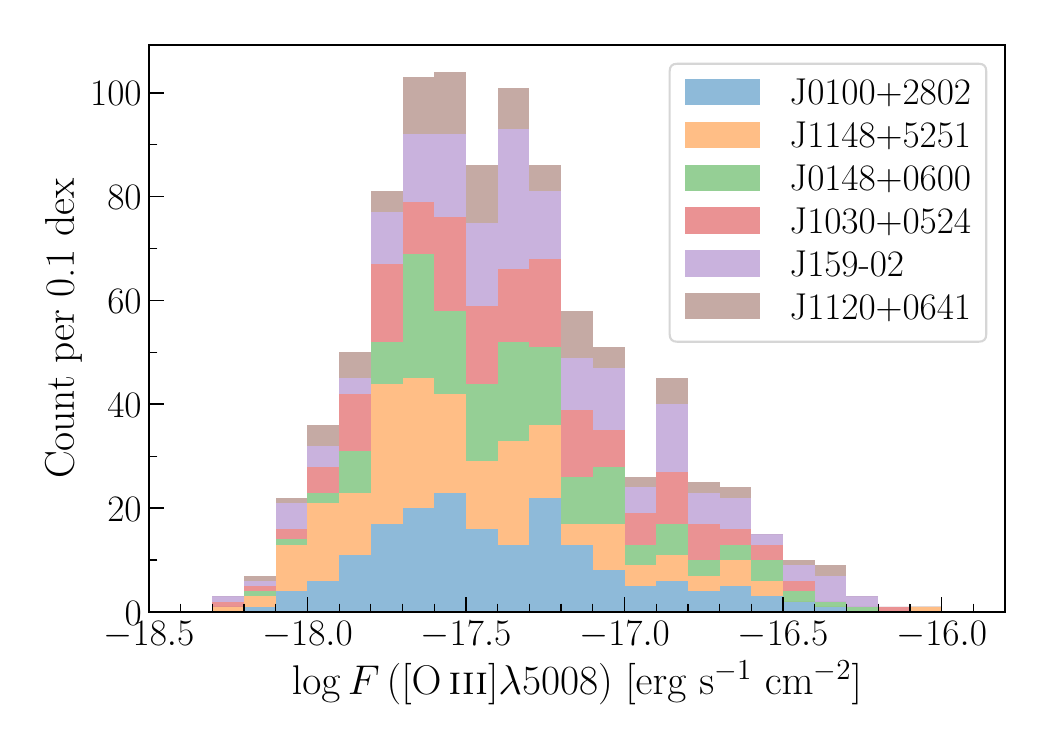}
\includegraphics[width=3.5 in]{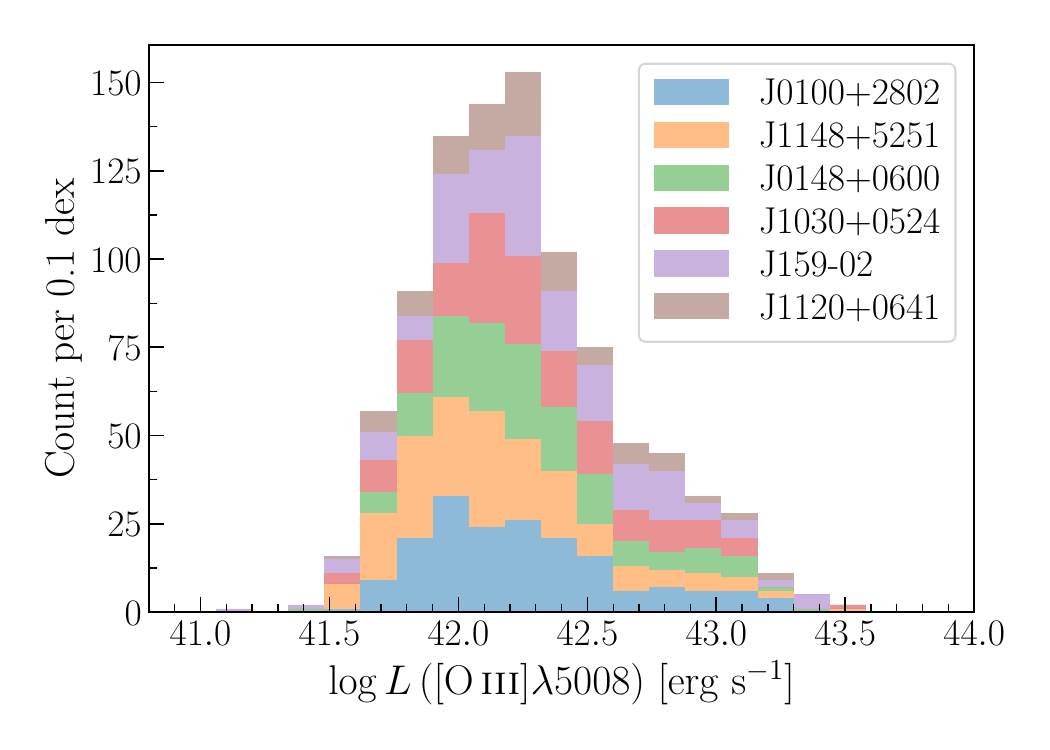}
\caption{Distributions of observed [\OIII]$\lambda 5008$ fluxes (upper panel) and the corresponding luminosities (lower panel).  The different colors represent the different quasar fields as shown in the legend.
\label{fig:hist_O3_fluxes}}
\end{figure}

Figure~\ref{fig:hist_O3_fluxes} shows the distribution of the observed [\OIII]$\lambda 5008$ fluxes ($F_{5008}$), which range from $\log (F_{5008}/\mathrm{erg\,s^{-1}\,cm^{-2}}) \approx -18.0$ to $-16.5$ (2.5--97.5th percentiles), with a median of $-17.42$.  There are no significant differences in the observed flux distribution between the survey fields: the median fluxes lie within $\log F_{5008} (\mathrm{erg\,s^{-1}\,cm^{-2}}) = -17.57$ to $-17.34$.  The corresponding [\OIII]$\lambda5008$ luminosities ($L_{5008}$) range from  $\log L_{5008} (\mathrm{erg\,s^{-1}}) \approx 41.6$ to $43.1$, with a median of 42.21.  

The UV luminosities of our galaxies, estimated from the observed F115W magnitudes, range from $M_\mathrm{UV}\approx -21.5$ to $-17.2$, with a median $-19.16$.

\subsection{Detection completeness}
\label{sec:completeness}

The selection function of [\OIII]-emitters is nontrivial due to the combination of the four-point mosaic, the dithering pattern, the different sensitivities of NIRCam modules A and B, and the wavelength-dependent throughput.  Additionally, the effective field of view per pointing, or the area over which an emission line can be captured by the detector, varies with the wavelength of the line.  Accurate modeling of completeness throughout the survey volume is crucial for deriving the luminosity function and for constructing random and mock catalogs used in correlation and clustering analyses.

The evaluation of completeness proceeds as follows (see R.~Mackenzie et al. in prep. for more details).  For each field, we first construct a noise cube that maps the root-mean-square (RMS) pixel noise as a function of sky position and wavelength.  To achieve this, we create coadded 2D spectra for numerous random points distributed across the survey field using the forward spectral extraction method, and measure the noise levels in the coadded 2D spectral images over the wavelength window.  These noise estimates are then combined and averaged on a 3D linear grid in steps of $1.2\arcsec$ along the spatial axis and 9.75\AA\ along the wavelength axis (identical to the wavelength grid of the science spectra).

Next, artificial emission-line signals mimicking the [\OIII] doublet, with various intrinsic fluxes, are injected into the coadded 2D spectral images, and SExtractor is used to evaluate the recovery rate.  The detection threshold is set at $\mathrm{SNR}=3$ for the [\OIII]$\lambda 4960$ line, consistent with the real data.  Here, SNR is defined as the ratio of \texttt{FLUX\_APER} to \texttt{FLUXERR\_APER} within a 5-pixel ($0.3\arcsec$) diameter aperture.

As might be expected, we find that the recovery rate, or completeness, is tightly correlated with the ratio of the injected flux to the pixel noise level ($F/\mathrm{RMS}$), and this correlation is largely independent of the pixel noise level itself, wavelength, and sky position.  Thus, a single conversion function from $F/\mathrm{RMS}$ to completeness is derived for each field.  Using this conversion, the completeness cube can be constructed from the noise cube for a given line flux or luminosity.

\begin{figure}[t]
\centering
\includegraphics[width=3.5 in]{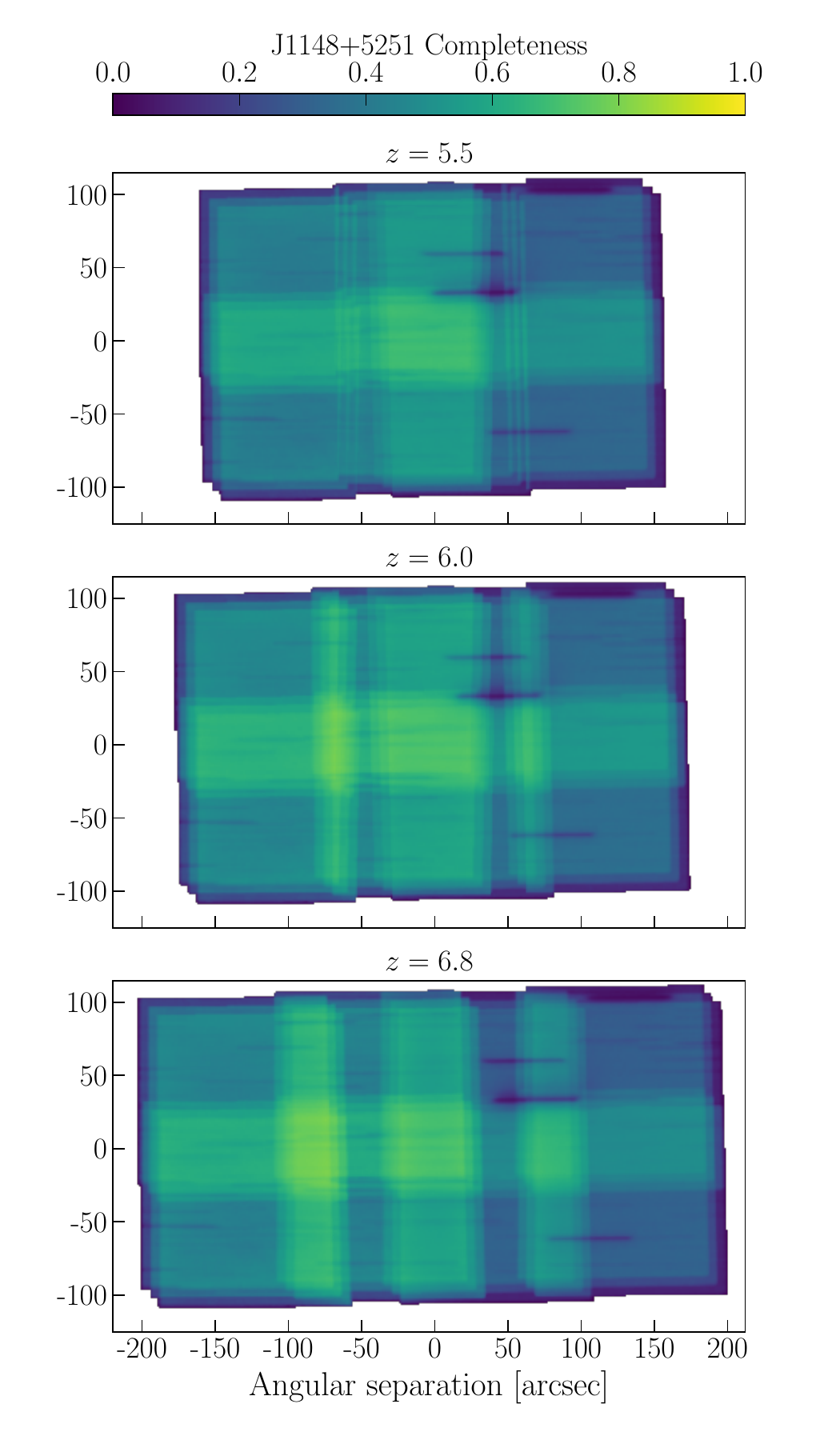}
\caption{Completeness maps for the field of J1148$+$5251.  The three panels show the completeness for [\OIII]-emitters with $L_{5008} = 10^{42}\,\mathrm{erg\,s^{-1}}$ across the survey area at different redshifts ($z = 5.5$, 6.0, and 6.8, from top to bottom).  Horizontal trails of lower completeness are caused by the presence of bright galaxies and stars within the field.  
\label{fig:completeness_maps}}
\end{figure}

Figure~\ref{fig:completeness_maps} shows the completeness cube for the J1148$+$5251 field.  The three panels display the completeness for [\OIII]-emitters with $L_{5008} = 10^{42}\,\mathrm{erg\,s^{-1}}$ across the survey area at different redshifts ($z = 5.5$, 6.0, and 6.8 from top to bottom).  The maximum width of the survey area increases with redshift because a larger portion of the dispersed spectra, particularly on the shorter-wavelength side, falls outside the detector for sources located near the field edges.  The large-scale completeness pattern reflects the $2\times2$ mosaic configuration (compare with Figure~\ref{fig:observations1}) and the lower throughput of Module B, which covers roughly the right half of the survey area.  The overall structure of the completeness cube is consistent across the other quasar fields.  Some specific horizontal trails of lower completeness are caused by the presence of very bright galaxies and stars within this particular field.  

\begin{figure}[t]
\centering
\includegraphics[width=3.5 in]{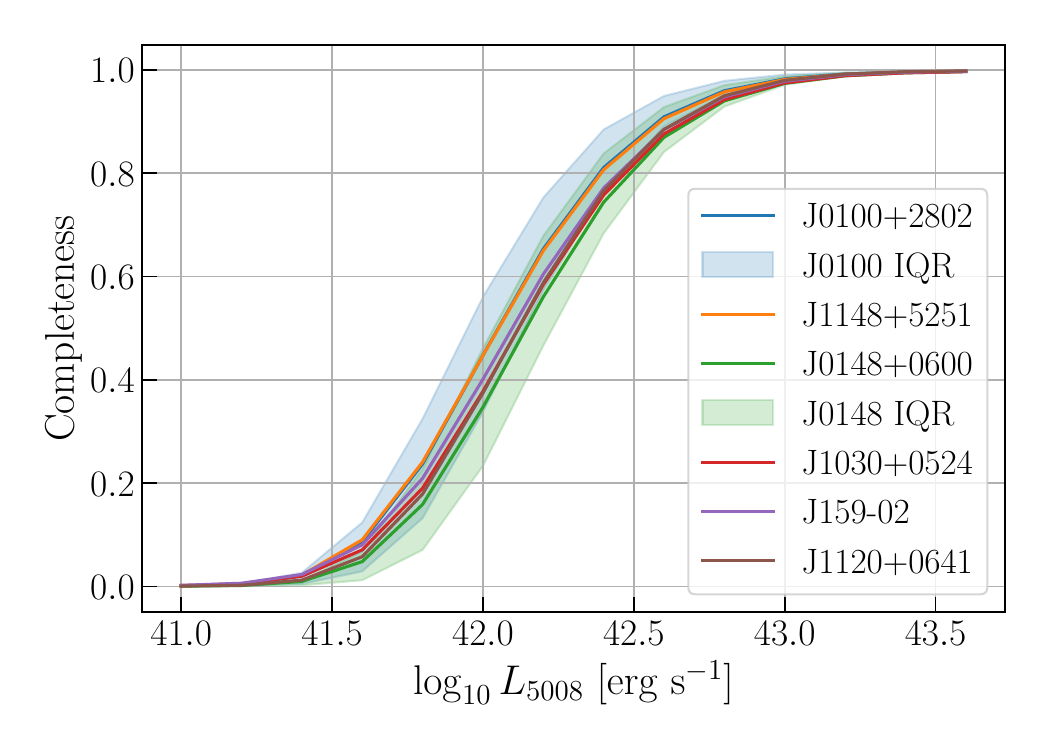}
\caption{Completeness over the entire survey volume ($5.32 < z < 6.97$) as a function of [\OIII]$\lambda 5008$ luminosity for each field.  Solid lines indicate the average completeness functions.  The blue and green shaded regions show the interquartile range (IQR) of the spatial variations in the J0100 and J0148 fields, respectively.  \label{fig:completeness}}
\end{figure}

Figure~\ref{fig:completeness} shows the completeness averaged over the entire survey volume ($5.32<z<6.97$) as a function of [\OIII]$\lambda 5008$ luminosity for each field.  The average completeness is 80\% at $2.5 \times 10^{42}\,\mathrm{erg\,s^{-1}}$ but 
drops to $\sim 40\%$ at $L_{5008}=10^{42}\,\mathrm{erg\,s^{-1}}$, with no substantial differences between the quasar fields.  The figure also illustrates the interquartile range (IQR; 25th-75th percentiles) of the spatial variations in completeness across each field (shown only for J0100 and J0148 for clarity; results are consistent across the other fields).  At 50\% completeness, the IQR is $\sim 20\%$.  These completeness models are used in the following analyses.

\section{IGM transmission measurements along the sightlines to the quasars}
\label{sec:QSO_spectra}

We measure the transmission as a function of redshift in the \Lya\ and \Lyb\ forests using ground-based spectroscopic observations of our target quasars.  The spectra were obtained with Folded-point InfraRed Echellette (FIRE; \citealt{2013PASP..125..270S}) on the Magellan Baade Telescope, X-Shooter \citep{2011A&A...536A.105V} on Unit Telescope 2 of the Very Large Telescope (VLT), and the Multi-Object Spectrometer For Infra-Red Exploration (MOSFIRE;  \citealt{2012SPIE.8446E..0JM}) and the Echellette Spectrograph and Imager (ESI; \citealt{2002PASP..114..851S}) on the Keck I Telescope.  The wavelength range used in the transmission analysis is covered by the X-Shooter VIS arm (for all quasars except J1148) or by ESI (for J1148).  Details on the processing of the ground-based quasar spectra can be found in \citet{2023ApJ...950...68E} and \citet{2024ApJ...969..162D}.  The resulting spectral resolution at the wavelengths used for transmission analysis is $R \approx 8000$ for X-Shooter and $R \approx 5000$ for ESI.  Assuming pure Hubble flow, these resolutions correspond to a comoving distance interval of $\Delta d_c \approx 0.4$--0.6\,cMpc at $z \sim 5.5$.  These resolutions are sufficiently high to resolve transmission structures along the sightlines on scales of interest, $\gtrsim 1$\,cMpc.

\begin{figure}[t]
\centering
\includegraphics[width=3.5in]{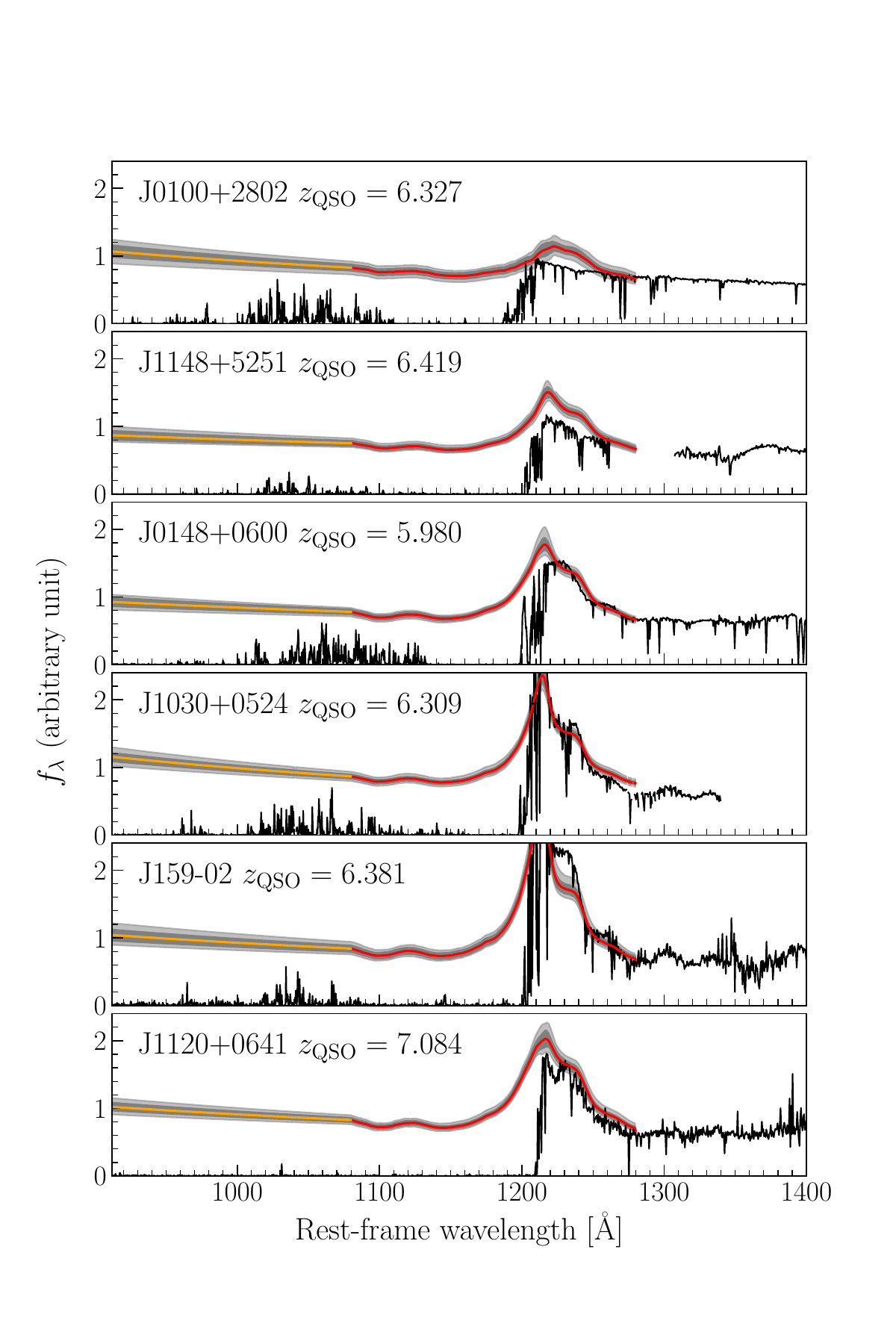}
\vspace{-12mm}
\caption{Quasar spectrum (black) and the intrinsic continuum model (red) for each quasar.  The extrapolated portions of the continua are shown in orange.  Light and dark gray shaded regions represent the 1$\sigma$ and 2$\sigma$ confidence intervals of the model, respectively.
\label{fig:QSO_spec_cont}}
\end{figure}

To measure the \Lya\ and \Lyb\ transmission, the intrinsic continuum spectrum blueward of 1280\,{\AA} was estimated for each quasar based on the shape of its redward spectrum.  This was achieved using a deep neural network trained on a sample of low-redshift ($0.1<z<3$) quasar spectra \citep{2021MNRAS.502.3510L}.  The resulting intrinsic spectrum, valid down to a rest-frame wavelength of 1080\,{\AA}, was extended to cover \Lyb\ forest region using a power-law fit to the continuum redward of \Lya.  The power-law slope $\gamma$ (defined as $f_\lambda \propto \lambda^{-\gamma}$) ranges from approximately 0.9 to 1.7, broadly consistent with typical values of $z \sim 4$ luminous quasars \citep{2016MNRAS.462.2478C}.  Figure~\ref{fig:QSO_spec_cont} shows the continuum model for each target quasar.  The typical uncertainties in the continuum models range from 4 to 8\%.  In the main analysis below, we neglect these uncertainties (see Section \ref{sec:results} for justification).

Figures~\ref{fig:observations1} and \ref{fig:observations2} show the transmission spectra for each quasar, defined as the flux spectrum normalized by the predicted intrinsic continuum.  The wavelength is translated into redshift for the \Lya\ line using its rest-frame wavelength ($\lambda_\alpha=1215.67\text{\AA}$, i.e., $z_\alpha = \lambda_\mathrm{obs}/\lambda_\alpha -1$).  

For all quasars except J1120$+$0641, the redshift of \Lya\ at the observed wavelength of the \Lyb\ line ($\lambda_\beta = 1025.72\text{\AA}$) of the quasar is below the lower boundary of the redshift window for detecting [\OIII]-emitters in the grism survey.  Therefore, the transmission within the window is purely due to \Lya\ opacity.  For J1120, with $z_\mathrm{QSO}=7.085$, however, the portion of the spectrum below $z_\alpha \lesssim 5.82$ is affected by the \Lyb\ absorption at higher redshifts.  Indeed, the spectrum of J1120 shows significantly reduced transmission at these redshifts compared to the other quasars (see Figure~\ref{fig:observations2}).  We exclude this portion of the J1120 spectrum from our analysis.

\begin{figure}[t]
\centering
\includegraphics[width=3.4in]{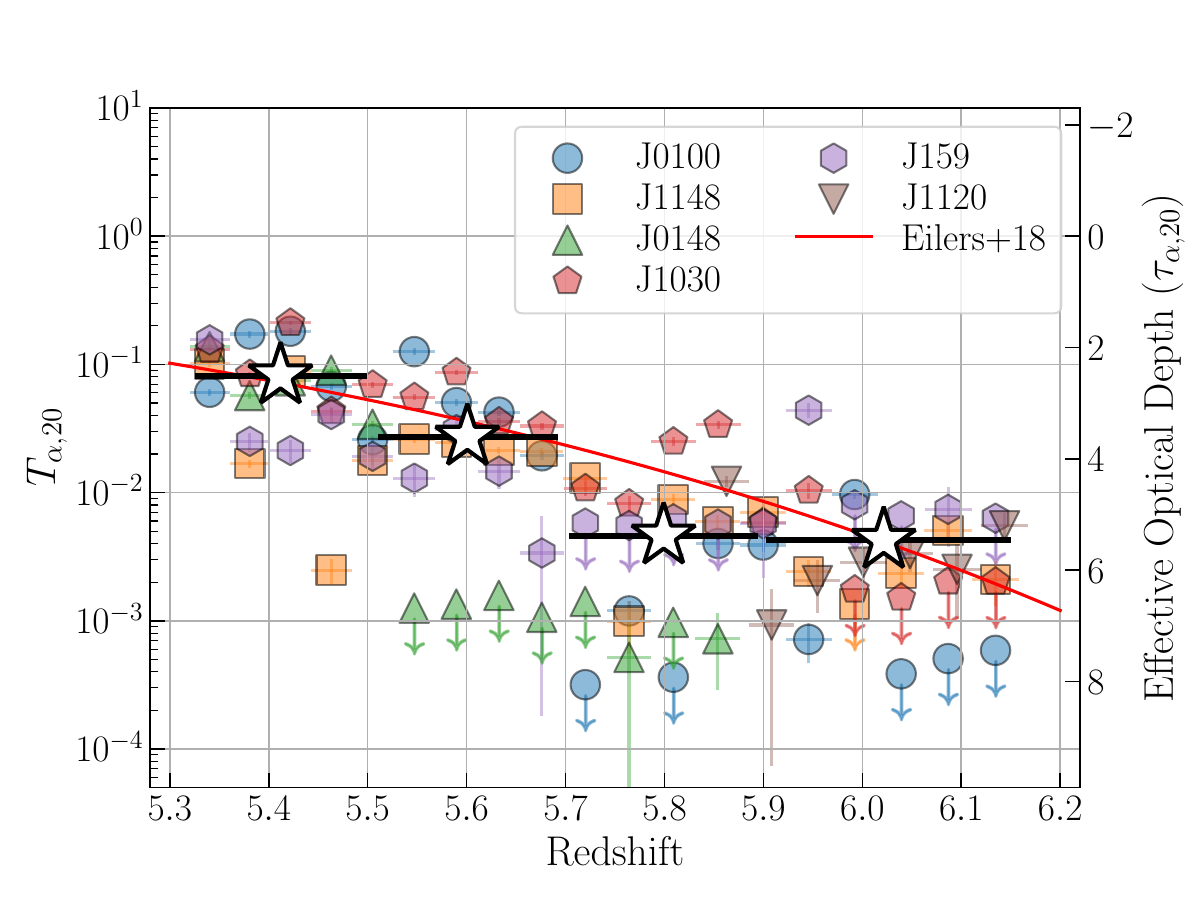}
\includegraphics[width=3.4in]{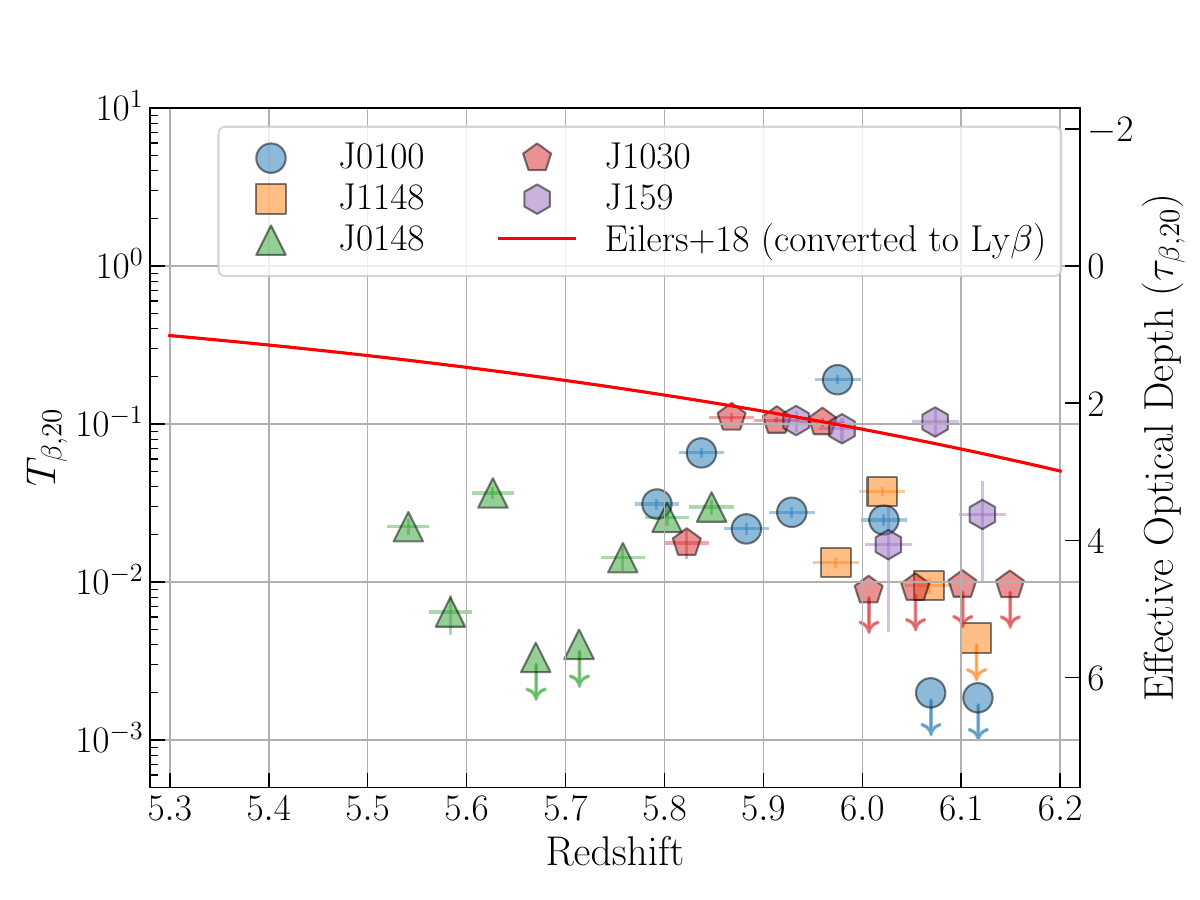}
\caption{Upper panel: \Lya\ transmission measured in 20\,cMpc bins along the sightlines toward our six target quasars.  Different symbols represent different quasars.  Down arrows indicate 2$\sigma$ upper limits for non detections ($<1\sigma$).  The star symbols indicate the mean transmission in four bins of redshifts.  The right-hand $y$-axis shows the corresponding effective optical depth.  The red curve represents the cosmic volume-averaged trend from \citet{2018ApJ...864...53E}.  Lower panel: similar to the upper panel, but showing \Lyb\ transmission measured in the same binsize, statistically corrected for foreground \Lya\ absorption.  The red curve is converted to \Lyb\ from the same result shown above (see text).      
\label{fig:z_vs_T}}
\end{figure}

The upper panel of Figure~\ref{fig:z_vs_T} shows the \Lya\ transmission measured in 20\,cMpc bins along the sightlines ($T_{\alpha,20}$\footnote{This is defined as $T_{\alpha,20} = \int_\mathrm{20\,\mathrm{cMpc}} T_\alpha(
D_\mathrm{c})\,dD_\mathrm{c}/(20\,\mathrm{cMpc})$ where $D_\mathrm{c}$ is comoving distance along the sightline, and the integration is performed over each 20\,cMpc interval.}).  The uncertainties in continuum modeling (4--8\%) are negligible for the purpose of this figure.  For reference, the corresponding effective optical depth ($\tau = -\ln T$) is shown on the right-hand $y$-axis.  However, our analyses are performed using transmission, rather than optical depth, to rigorously account for flux errors in the quasar spectra.  The figure also includes the cosmic mean transmission as a function of redshift derived by \citet{2018ApJ...864...53E} for comparison.  The overall trend, declining with increasing redshift, is consistent with the cosmic mean.  Additionally, the scatter in transmission (in log-scale) increases significantly from $\sigma(\tau)\sim0.4$ at $z<5.5$ to $\sigma(\tau)\sim 0.7$ at $z>5.5$, as reported by early studies compiling many quasar sightlines \citep{2018ApJ...864...53E,2018MNRAS.479.1055B,2022MNRAS.514...55B}.  The deep Gunn-Peterson trough of the J0148 sightline \citep{2015MNRAS.447.3402B} is clearly visible as a series of persistently low transmission values (green triangles) over the redshift range $5.5 \lesssim z \lesssim 5.9$.  

To observe the trend, the mean transmission levels within four approximately equal redshift bins ($z = [5.32,5.50]$, $[5.50,5.70]$, $[5.70,5.90]$, and $[5.90,6.15]$) are shown in the figure.  These mean values are consistent with the cosmic mean in all but the third bin, where a sharp drop from the second bin results in a mean transmission that remains nearly constant across the redshift range $5.70 < z < 6.15$.  It might be worth noting that a constant mean transmission across different redshifts does not imply that the average neutral fraction remains unchanged with redshift, because the effective optical depth approximately scales as $\taueff \propto \langle x_\mathrm{HI}\rangle (1+z)^{3/2}$ \citep{2015PASA...32...45B}.  However, the difference in the $(1+z)^{3/2}$ factor between $z=5.70$ and 6.15 is only about 10\%, which is indeed much smaller than the nearly fourfold change in transmission between this bin and the adjacent intermediate redshift bin.  Therefore, this trend in the large-scale mean transmission motivates us to conduct a galaxy--transmission correlation analysis within three redshift regimes, divided at $ z = 5.5 $ and $ z = 5.7 $ (see Section \ref{sec:results}).

We also measure \Lyb\ transmission using the \Lyb\ forest region of the spectra.  This region is affected by foreground \Lya\ absorption at lower redshifts, $z_{\alpha}^\mathrm{fg} = (z_{\beta} + 1 )(\lambda_\alpha/\lambda_\beta)-1$.
To statistically correct for this foreground absorption, we assume the cosmic mean transmission at the corresponding absorption redshift.  However, it is important to note that spatial variations in the foreground \Lya\ absorption are not negligible.  According to \citet{2022MNRAS.514...55B} (see their supplementary data), the optical depth measured over a scale of $30\,h^{-1}\,\mathrm{cMpc}$ has a scatter of $\approx 0.32$, corresponding to a factor of $\approx 1.4$ in transmission, around $z_\alpha^\mathrm{fg} \approx 4.9$ (corresponding to $z_\beta \approx 6$).  Consequently, \Lyb\ transmission measurements may be significantly distorted by \Lya\ absorption in the lower redshift foreground IGM.  It should be noted that while the \textit{presence} of a transmission spike in this wavelength range requires that the region is ionized (transmitting) at both $z_\beta$ and $z_\alpha$, since photons must have successfully passed through both redshifts, the \textit{absence} of a transmission spike does not necessarily imply \Lyb\ absorption at $z_\beta$, as the photons could instead have been subsequently absorbed by \Lya\ at $z_{\alpha}$.

The lower panel of Figure~\ref{fig:z_vs_T} shows the \Lyb\ transmission, corrected for foreground \Lya\ absorption, measured in 20\,cMpc bins along five quasar sightlines (excluding J1120$+$0640).  For a given neutral hydrogen density, the optical depth of \Lyb\ is smaller than that of \Lya\ by a factor of 6.2.\footnote{This factor 6.2 is given as $\tau_\alpha/\tau_\beta = (f_\alpha \lambda_\alpha)/(f_\beta\lambda_\beta)$, where $f_\alpha=0.4164$ and $f_\beta=0.0791$ are the oscillator strengths.}  However, in reality, the neutral hydrogen density may not be homogeneous, and the ratio of effective optical depths can deviate from this nominal value ($\sim 2$--3; \citealt{2005ApJ...620L...9O}).  We adopt $\tau_\alpha/\tau_\beta = 2.25$, which was empirically derived by \citet{2006AJ....132..117F}, and convert the cosmic \Lya\ transmission trend to its corresponding \Lyb\ trend ($T_\beta = T_\alpha e^{2.25}$).  The evolution of \Lyb\ transition within our survey fields is less clear compared to \Lya, due to the limited redshift coverage of the \Lyb\ forests and possibly the contamination from foreground \Lya\ absorption.  The deep trough in the J0148 sightline is again clearly visible, with the green triangles falling well below the cosmic mean.

\section{Results}
\label{sec:results}

\begin{figure}[t]
\centering
\includegraphics[width=3.4in]{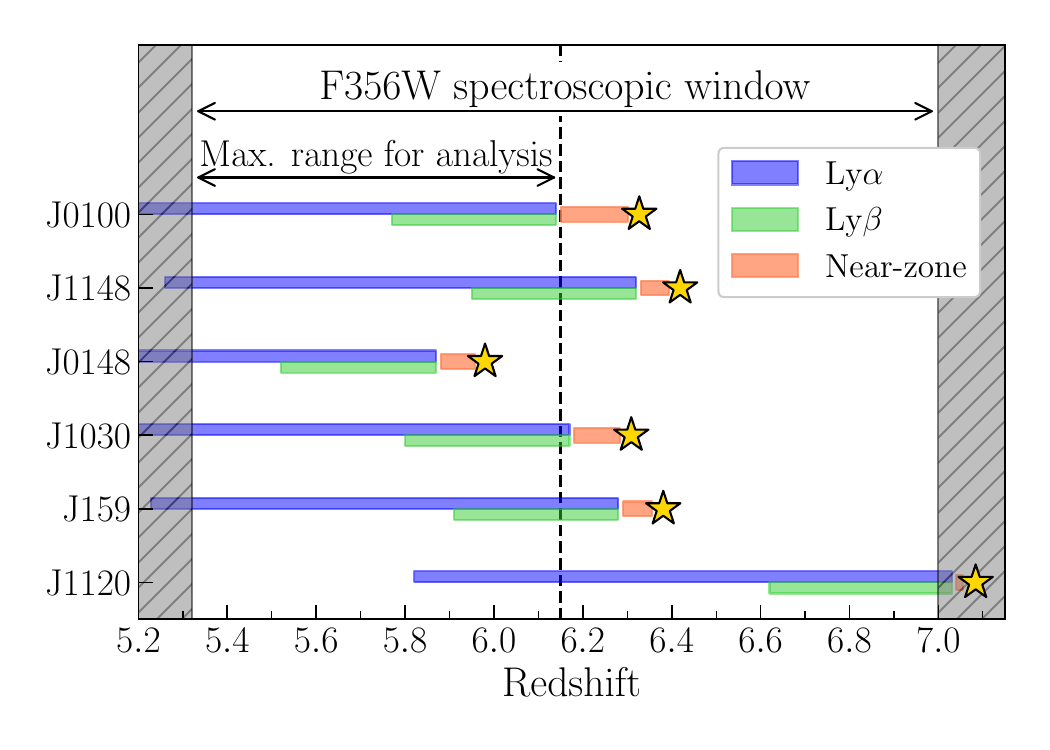}
\caption{Redshift intervals of the \Lya\ (blue bars) and \Lyb\ (green bars) forests for the target quasars.  Star symbols indicate the redshifts of the quasars, while red bars correspond to their near-zones.  The vertical dashed line marks the upper limit of the redshift range used for the galaxy--transmission analysis in this paper.  The gray shaded regions at the left and right edges denote ranges outside the spectral coverage for detecting [\OIII]-emitters.  
\label{fig:redshift_ranges}}
\end{figure}

\begin{figure*}[t]
\centering
\makebox[\textwidth]{\includegraphics[width=7.in]{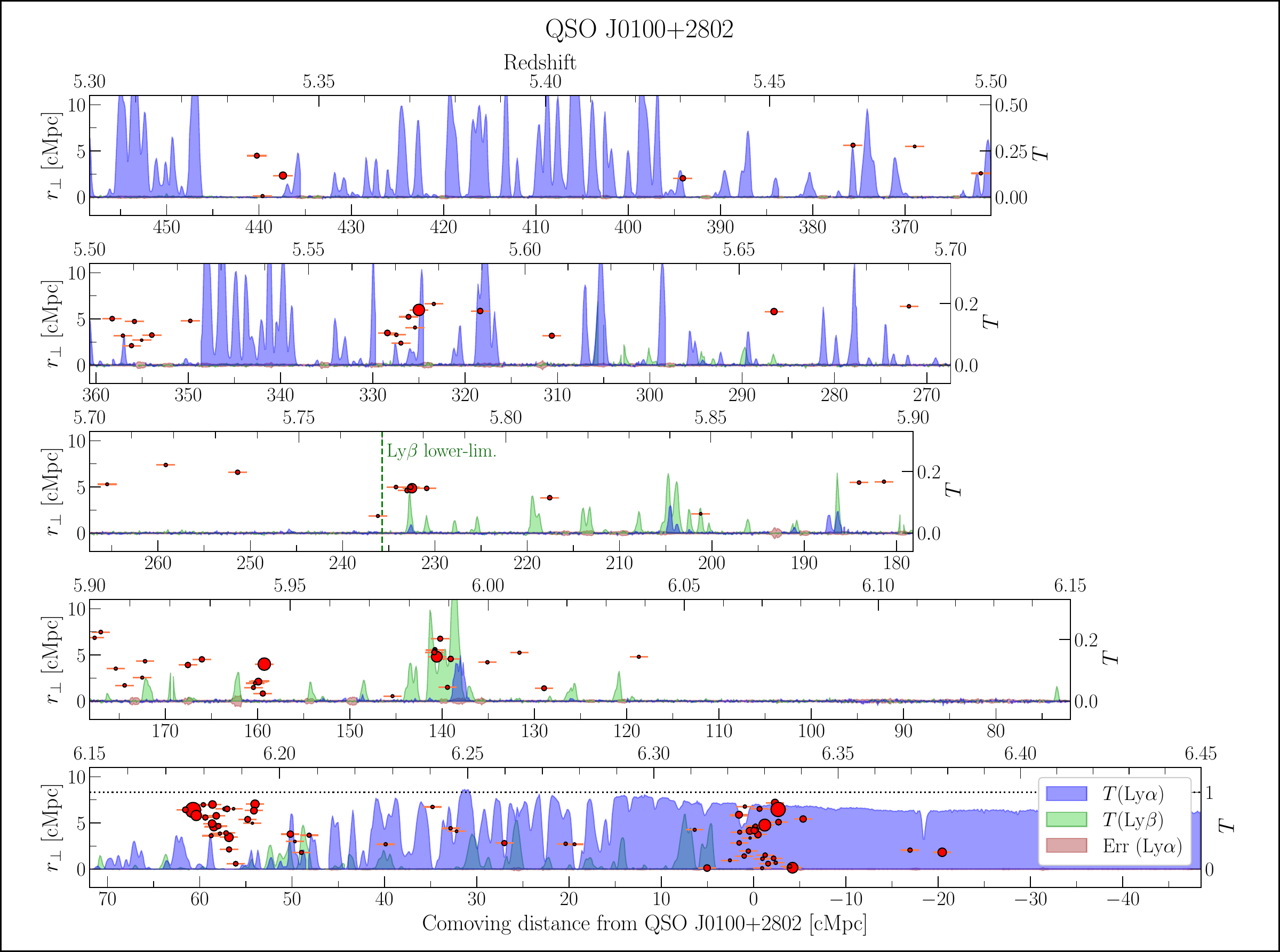}}
\caption{The distribution of [\OIII]-emitting galaxies, and the \Lya (blue) and \Lyb (green) transmission along the line of sight to QSO J0100$+$2802.  The error spectrum, corresponding to \Lya\ transmission), is shown in brown. 
 The red circles show the comoving transverse distances, $r_\perp$, to the detected [\OIII]-emitters from the quasar sightline, with different sizes coded according to the [\OIII] luminosity.  The vertical dashed line indicates the lower redshift limit of the \Lyb\ forest used in our analysis.
\label{fig:LoS_zoom_J0100}}
\end{figure*}

\begin{figure*}[t]
\centering
\makebox[\textwidth]{\includegraphics[width=6.8in]{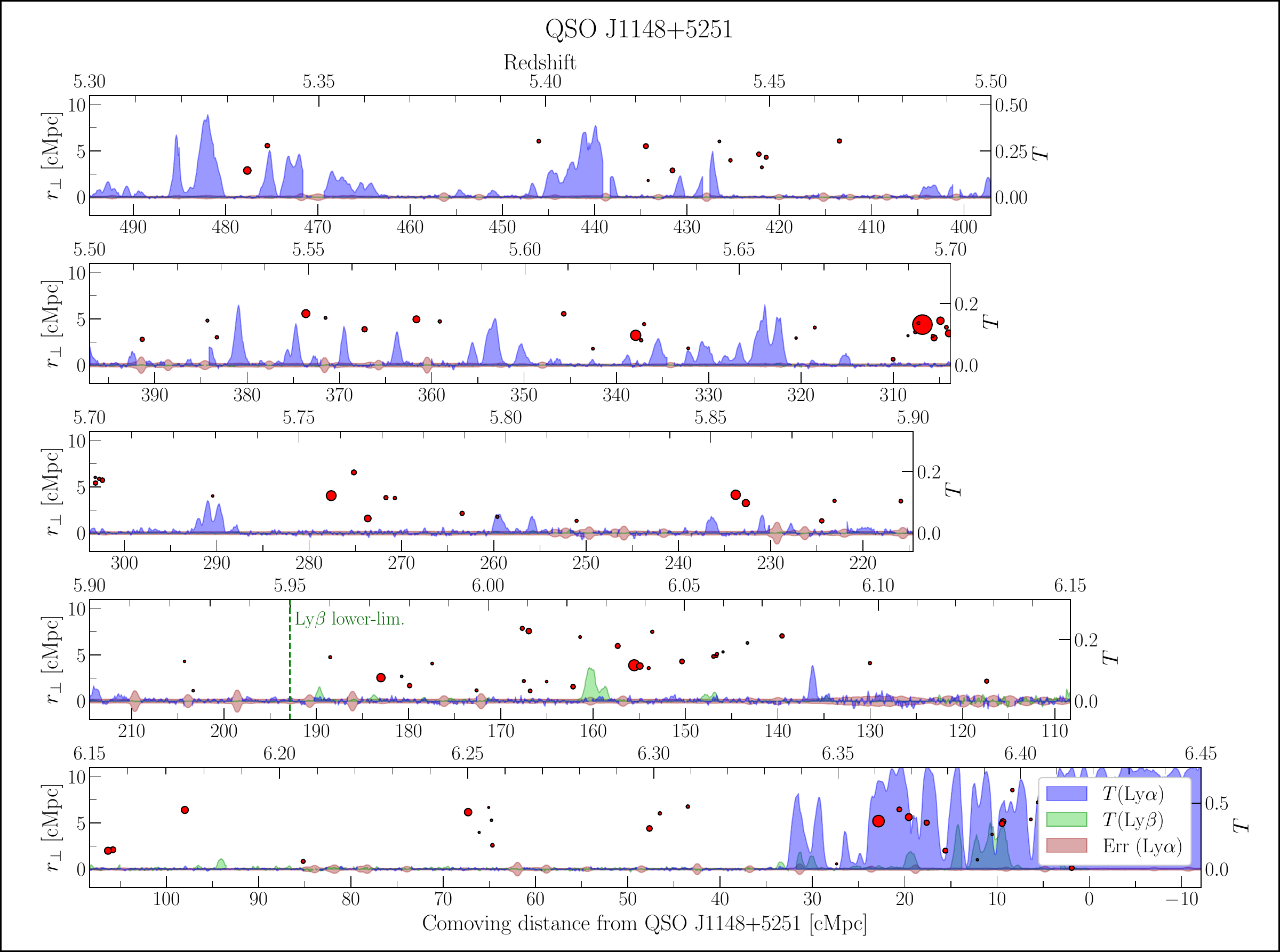}}
\caption{Same as Figure~\ref{fig:LoS_zoom_J0100}, but for QSO J1148$+$5251.
\label{fig:LoS_zoom_J1148}}
\end{figure*}

\begin{figure*}[t]
\centering
\makebox[\textwidth]{\includegraphics[width=6.8in]{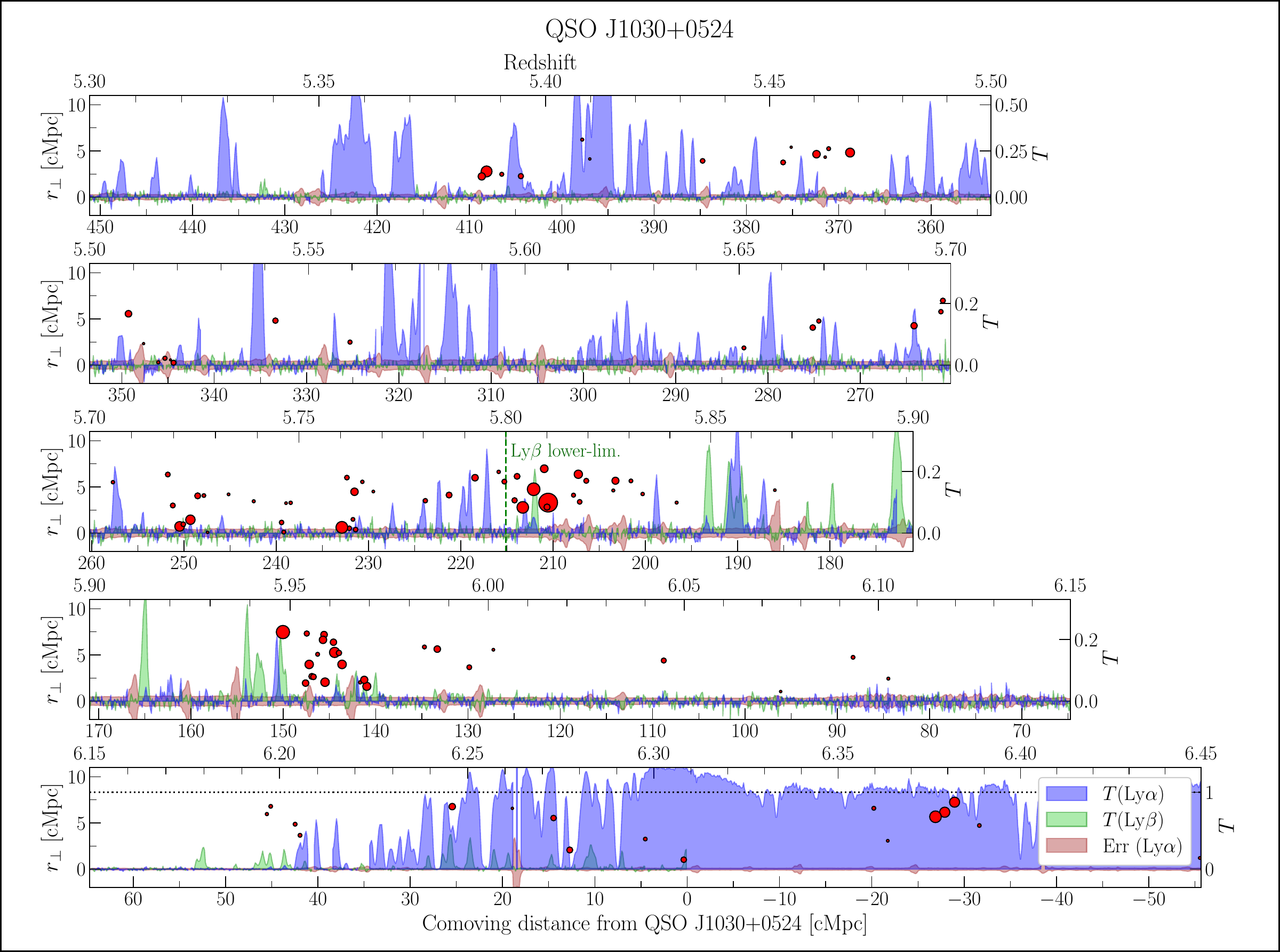}}
\caption{Same as Figure~\ref{fig:LoS_zoom_J0100}, but for QSO J1030$+$0524.
\label{fig:LoS_zoom_J1030}}
\end{figure*}

\begin{figure*}[t]
\centering
\makebox[\textwidth]{\includegraphics[width=6.8in]{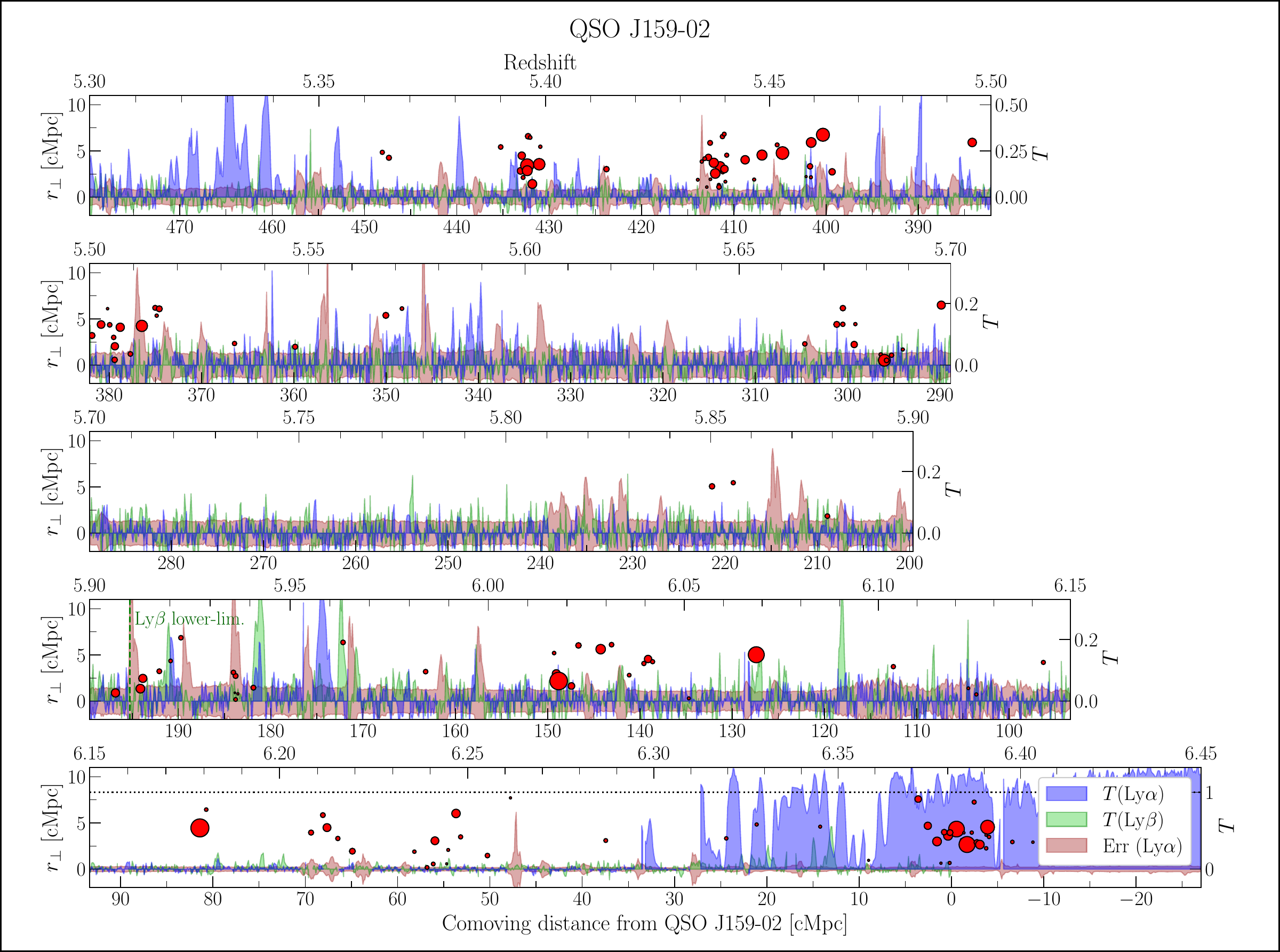}}
\caption{Same as Figure~\ref{fig:LoS_zoom_J0100}, but for QSO J159$-$02.
\label{fig:LoS_zoom_J159}}
\end{figure*}

\begin{figure*}[htbp]
  \centering
  \begin{minipage}{\textwidth}
    \centering
    \makebox[\textwidth]{\includegraphics[width=6.8in]{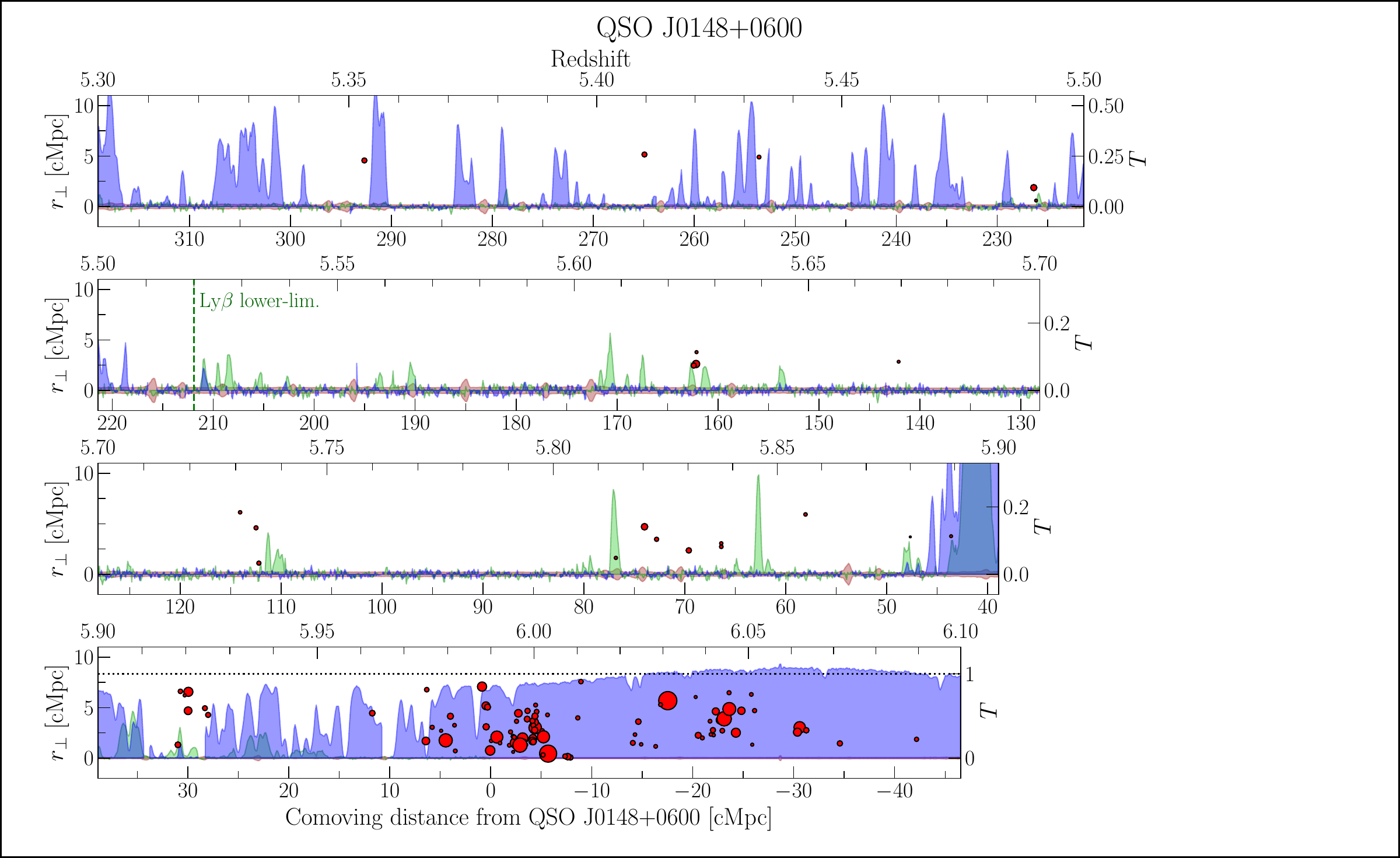}}
    \caption{Same as Figure~\ref{fig:LoS_zoom_J0100}, but for QSO J0148$+$0600.}
    \label{fig:LoS_zoom_J0148}
  \end{minipage}

  \vspace{0.5cm} 

  \begin{minipage}{\textwidth}
    \centering
    \makebox[\textwidth]{\includegraphics[width=6.8in]{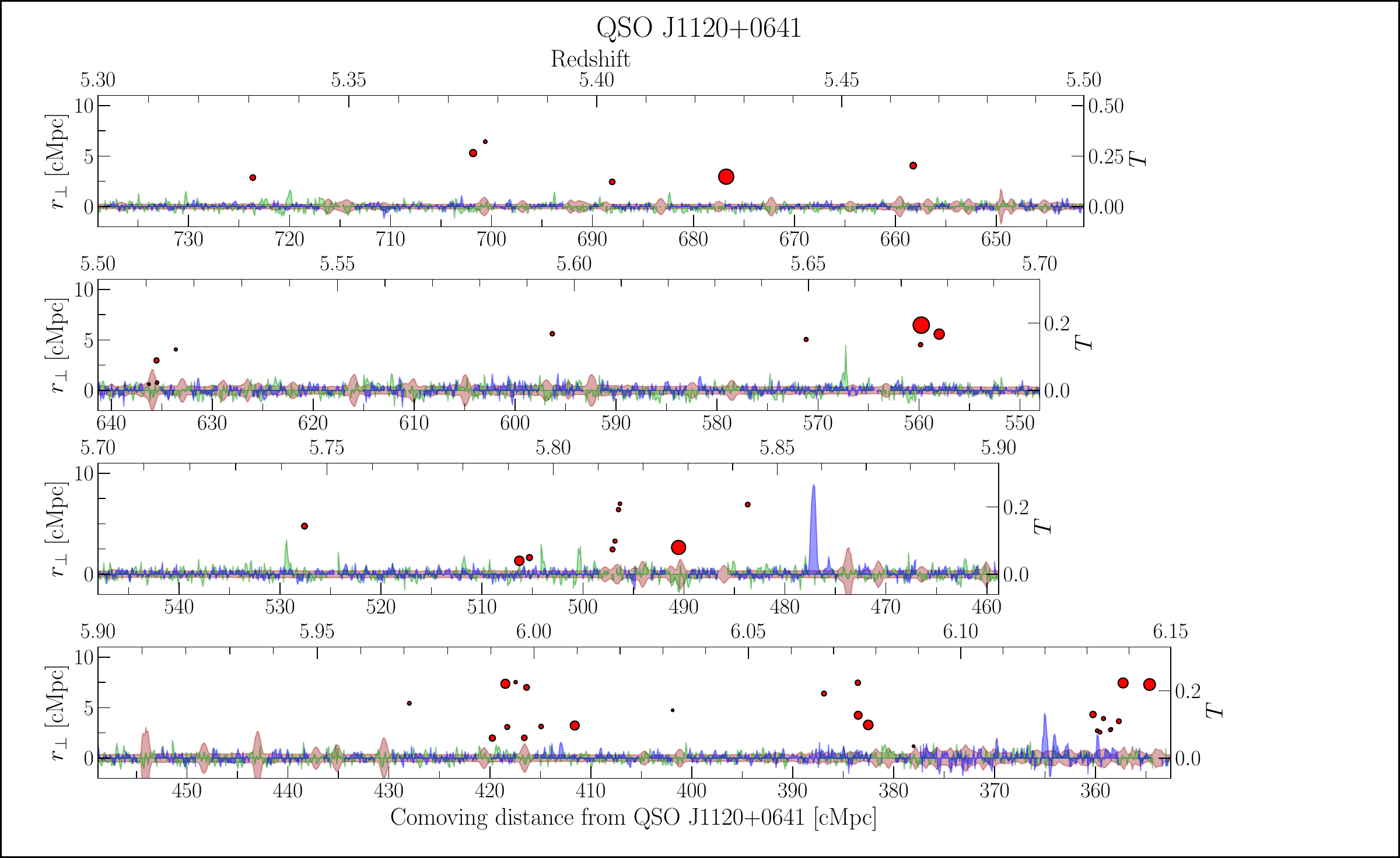}}
    \caption{Same as Figure~\ref{fig:LoS_zoom_J0100}, but for QSO J1120$+$0641. The quasar redshift lies well beyond the upper limit of the redshift range shown here.  The pure \Lya\ forest region used in the analysis is limited to $5.84 < z < 6.15$ (see Figure~\ref{fig:redshift_ranges}).}
    \label{fig:LoS_zoom_J1120}
  \end{minipage}
\end{figure*}

Here we present our statistical measurements regarding the relationship between the detected [\OIII]-emitting galaxies and the IGM transmission measured along the sightlines to the quasars.  Figure~\ref{fig:redshift_ranges} shows the redshift intervals of the \Lya\ and \Lyb\ forests that are used in our analysis.  We set the maximum redshift for our analysis to be $z_\mathrm{max} = 6.15$, as \Lya\ absorption becomes almost completely saturated above this redshift.  Quasar spectra generally do not exhibit significant transmission spikes beyond this redshift.  The minimum redshift is set to $z_\mathrm{min}=5.32$, which corresponds to the lower boundary of our [\OIII]-emitter sample, limited by the spectral window of the F356W filter.

With the sightlines of six quasars, our dataset encompasses a total \Lya\ forest spectral length of 1891\,cMpc.  In the following analyses, we divide the entire redshift range into three subranges, $5.32<z<5.50$, $5.50<z<5.70$, and $5.70<z<6.15$, in order to investigate the redshift evolution of the galaxy--IGM connection.  The path-length-weighted mean redshifts of these subranges are $\langle z\rangle = 5.41$, 5.60, and 5.92, respectively.  The corresponding lengths of the \Lya\ forest spectra are 438, 466, and 986\,cMpc, respectively.  As noted in Section \ref{sec:QSO_spectra}, this binning is motivated by the trend of the mean transmission measured along our six sightlines (see Figure~\ref{fig:z_vs_T}).  In \citep{2023ApJ...950...66K}, we previously divided the \Lya\ forest region into two redshift bins at $z=5.7$, based on the observation that the incidence frequency of transmission spikes changes dramatically across this redshift.  With the increased number of sightlines now available, we are able to further subdivide the lower range ($z < 5.7$) by introducing an additional boundary at $z=5.50$, near the midpoint of that range.  The \Lyb\ forest analysis is conducted only for the highest redshift range ($5.70 < z < 6.15$), yielding a total path length of 573\,cMpc.  

Recently, the level of scatter caused by cosmic variance in Galaxy-\Lya\ cross-correlation measurements based on limited \Lya\ forest path lengths was assessed using the THESAN simulation \citep{2024arXiv241002850G}.  The study found that although measurements derived from these limited path lengths cannot fully eliminate the effects of cosmic variance, they can suppress it to a degree sufficient for statistically detecting meaningful signals, as shown below.

Figures~\ref{fig:LoS_zoom_J0100}--\ref{fig:LoS_zoom_J1120} provide a detailed view of the distribution of detected [\OIII]-emitting galaxies along the line of sight to each quasar.  The $x$-axis represents the comoving distance from the quasar, with the corresponding redshift indicated on the upper axis, while the $y$-axis shows the transverse comoving distance.  To ensure an undistorted representation of the galaxy distribution, equal scaling is applied to both the axes.  The distribution of galaxies is compared against the \Lya\ and \Lyb\ transmission, depicted in blue and green, respectively, in the background.

\clearpage

\subsection{Sightline-to-sightline large-scale correlation between galaxy density and IGM transmission}
\label{sec:ngal_vs_T_LoS}

\begin{figure}[t]
\centering
\includegraphics[width=3.4in]{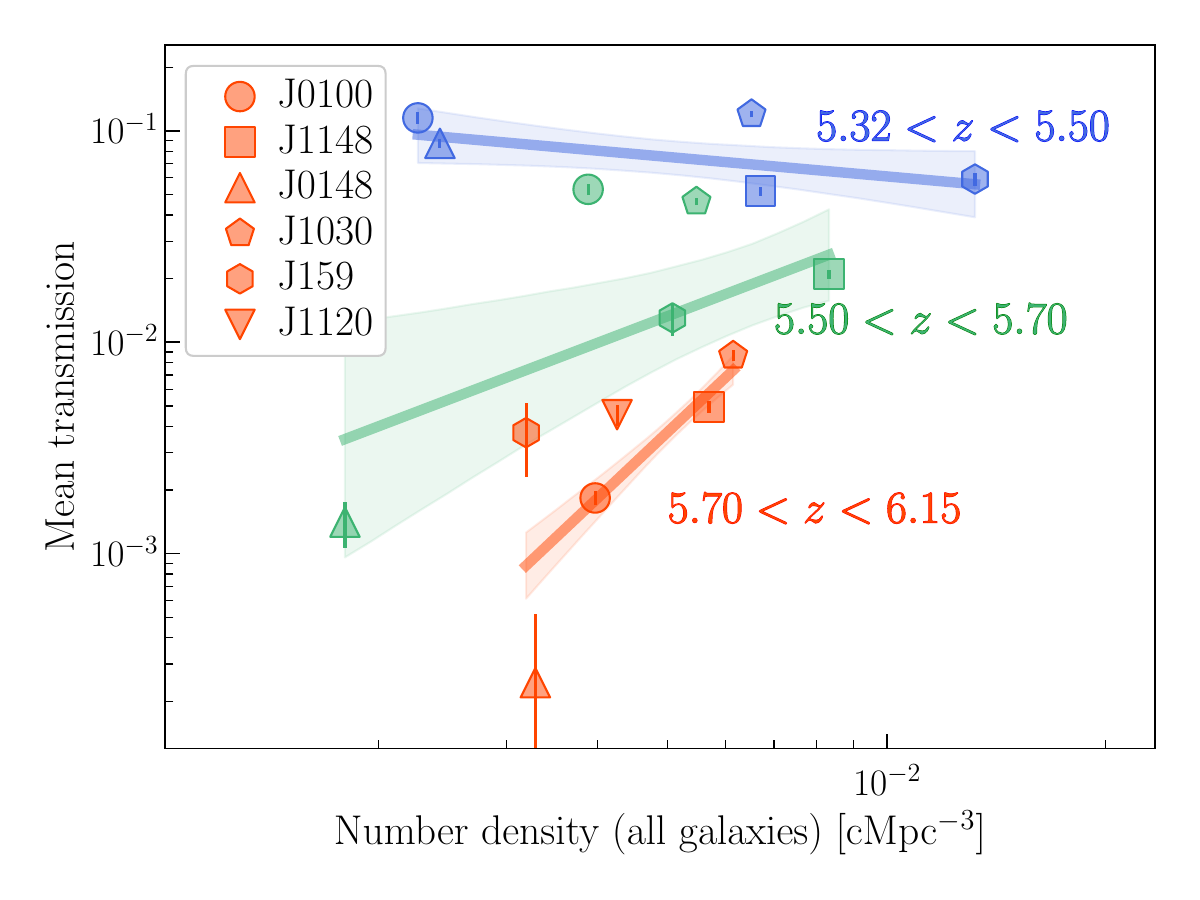}
\caption{Mean transmission versus galaxy number density measured along individual sightlines within three redshift ranges: $5.32<z<5.50$, $5.50<z<5.70$, and $5.70<z<6.15$, as labeled. Different symbols represent different sightlines, as indicated in the legend.  Vertical error bars reflect uncertainties in the quasar spectra and continuum modeling (4–8\%; see Section \ref{sec:QSO_spectra}).  Solid lines and shaded regions indicate the best-fit linear relations (in log-log space) and their $1\sigma$ uncertainties for each redshift range.
\label{fig:ngal_vs_T_LoS}}
\end{figure}

\begin{deluxetable}{cccc}
\tablecaption{Mean transmission and galaxy number densities over the lines of sight\label{tb:ngal_vs_T_los}}
\tablehead{
    \colhead{Field}&
    \colhead{$5.32<z<5.50$}&
    \colhead{$5.50<z<5.70$}&
    \colhead{$5.70<z<6.15$}}
\startdata
\multicolumn{4}{c}{Transmission (\%)} \\
J0100 & $11.529 \pm 0.013$ & $05.294 \pm 0.011$ & $0.183 \pm 0.008$\\
J1148 & $05.189 \pm 0.027$ & $02.100 \pm 0.021$ & $0.494 \pm 0.020$\\
J0148 & $08.739 \pm 0.038$ & $00.140 \pm 0.033$ & $0.024 \pm 0.027$\\
J1030 & $12.022 \pm 0.064$ & $04.637 \pm 0.062$ & $0.865 \pm 0.043$\\
J159  & $05.919 \pm 0.223$ & $01.303 \pm 0.218$ & $0.373 \pm 0.140$\\
J1120 & -                  & -                  & $0.454 \pm 0.049$\\
\hline
\multicolumn{4}{c}{Number densities ($\mathrm{cMpc}^{-3}$)} \\
J0100 & 0.0022 & 0.0038 & 0.0039 \\
J1148 & 0.0067 & 0.0083 & 0.0056 \\
J0148 & 0.0024 & 0.0018 & 0.0032 \\
J1030 & 0.0065 & 0.0054 & 0.0061 \\
J159  & 0.0132 & 0.0050 & 0.0031 \\
J1120 & -      & -      & 0.0042 \\
\hline
\multicolumn{4}{c}{Spearman rank correlation coefficients} \\
$\rho$    & -0.6  & 0.3   & 0.83 \\
$p$-value & 0.285 & 0.624 & 0.041 \\
\enddata
\tablenotetext{a}{This table summarizes the data plotted in Figure~\ref{fig:ngal_vs_T_LoS}, along with the results of the Spearman rank correlation analysis.}
\end{deluxetable}

As a starting point, we present in Figure~\ref{fig:ngal_vs_T_LoS} the mean \Lya\ transmission as a function of the total number density of galaxies across different fields and different redshift bins.  These values represent averages of both transmission and galaxy number density over very large (albeit essentially one-dimensional) scales, up to 100-200 cMpc.  Here, the number density is calculated as follows:
\begin{equation}
n = \sum_i \frac{1}{V_\mathrm{eff}(L_{5008,i})}
\label{eq:galdensity_Veff}
\end{equation}
where $V_\mathrm{eff}(L_{5008,i})$ is the effective volume for the $i$-the object, computed by integrating the completeness for its $L_{5008}$ over the relevant field and redshift range.  Even when the number density is computed without considering the $L_{5008}$-dependence of completeness, the qualitative trend remains unchanged.  Table~\ref{tb:ngal_vs_T_los} summarizes the plotted data and the results of the Spearman rank correlation analysis.  Log-log space linear fits to the data are shown in the figure.

Interestingly, at $z<5.5$, the number density and transmission appear to be anti-correlated, although the statistical significance is marginal.  In contrast, an opposite, positive correlation is seen at $z>5.7$.

In the intermediate redshift bin, the situation is complicated by a notable sightline (green triangle) exhibiting both very low galaxy density and very low mean transmission.  This data point corresponds to the long Gunn-Petersen trough along the J0148 sightline \citep{2018ApJ...863...92B}, and deviates significantly from the trend defined by the other sightlines.  Rather than being a mere outlier, it likely reflects a physically meaningful coincidence of low galaxy density and low transmission \citep{2018ApJ...863...92B,2020ApJ...888....6K}.  This may suggest that in such extremely underdense regimes, a positive correlation--similar to that seen in the highest-$z$ bin--could emerge.  We revisit this possibility in Section \ref{sec:interpretation}, in the context of additional results presented later.

Overall, these observed trends suggest a correlation between the galaxy distribution and the ionization state of the IGM on very large scales ($\gtrsim 100\,\mathrm{cMpc}$).  In particular, the strong correlation between the overall number density of galaxies and the mean IGM transmission at $ z > 5.7$ already provides compelling evidence that star-forming galaxies, or something with a similar spatial distribution, played an important role in cosmic reionization.

In the following sections, we investigate the connection between galaxies and the IGM transmission in more detail using various statistical metrics.

\subsection{Galaxy-transmission correlation}
\label{sec:gal_vs_T}

\begin{figure*}[t]
\centering
\includegraphics[width=7.0in]{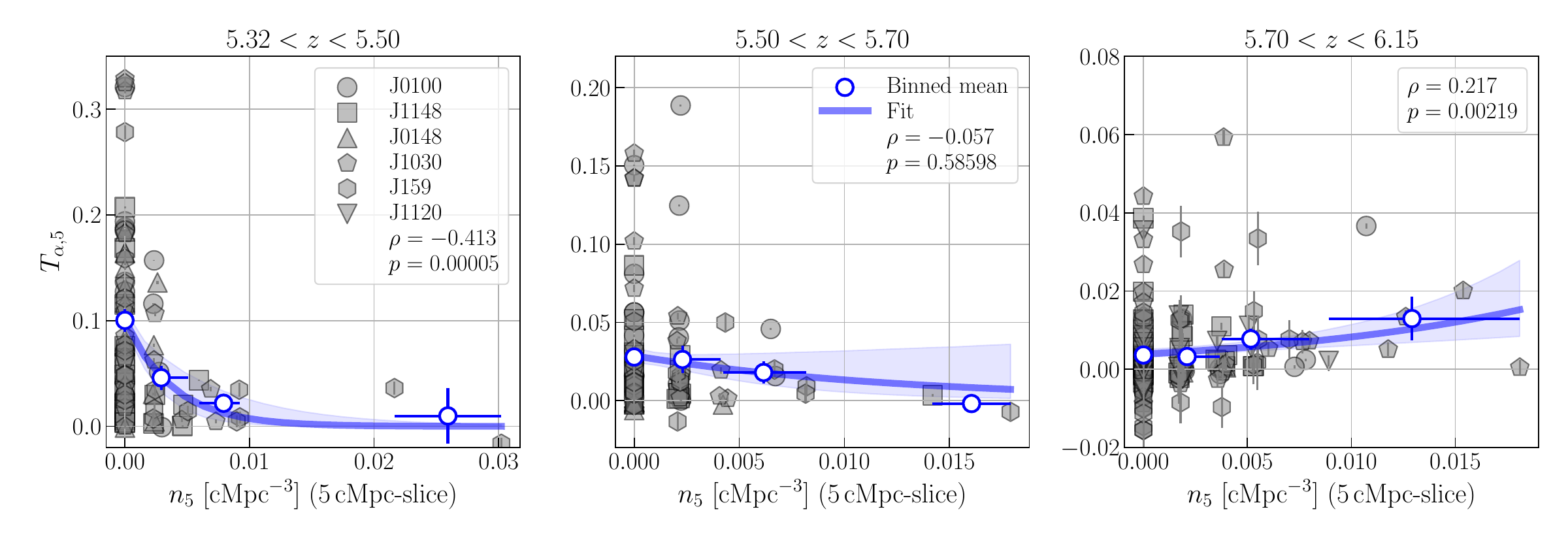}\vspace{-3mm}
\includegraphics[width=7.0in]{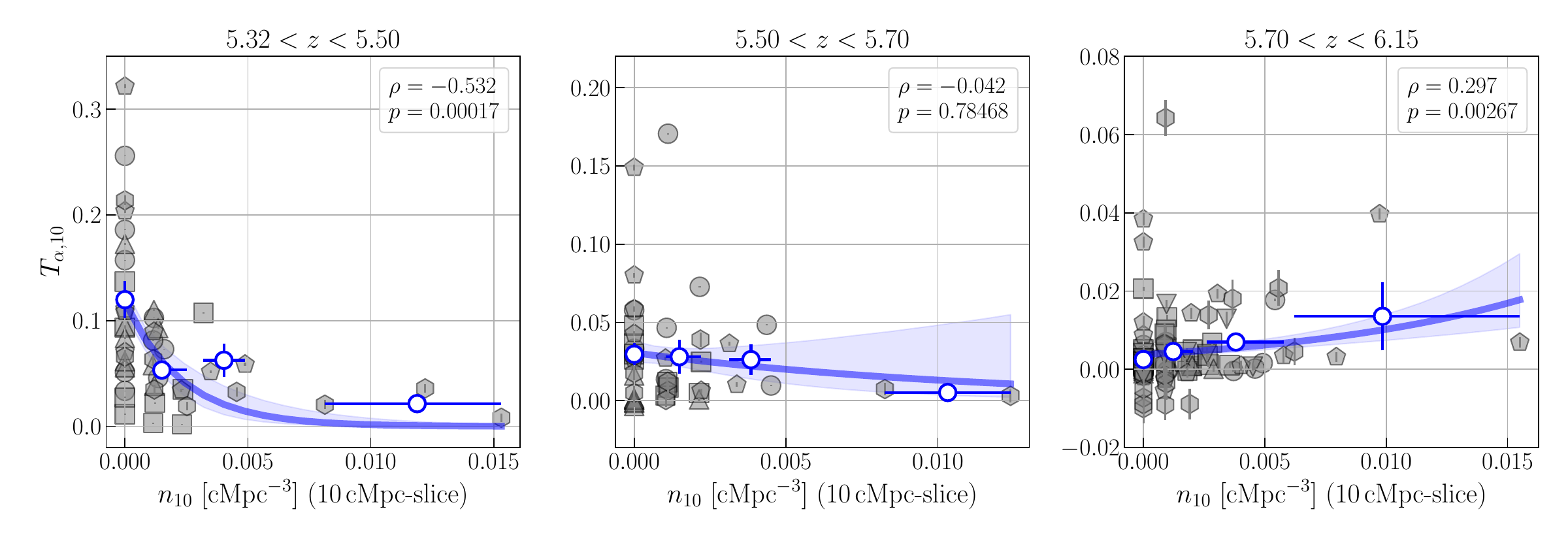}\vspace{-3mm}
\includegraphics[width=7.0in]{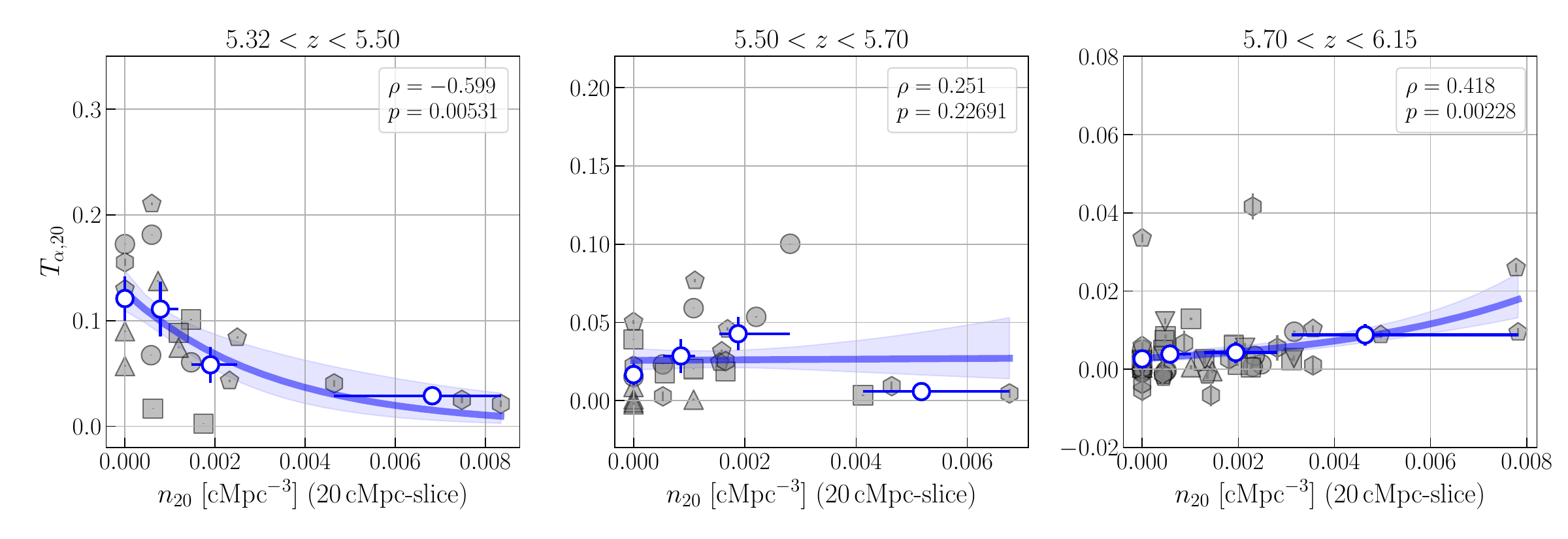}\vspace{-3mm}
\caption{Correlation between the number density of galaxies with [\OIII]$\lambda$5008 luminosity $L_{5008} \ge 10^{42}\,\mathrm{erg\,s^{-1}}$ and the \Lya\ transmission, for three redshift ranges.  The three rows correspond to three different binning scales along the sightlines (5, 10, and 20\,cMpc from top to bottom).  Different symbols are used for different quasars.  The blue-rimmed circles indicate the mean values in four bins of galaxy number density, with the vertical error bars indicating the errors on the means and the horizontal error bars indicating the binsizes.  The blue curves are the best fits (a linear relation between $N$ and $\ln T$) to the individual data points, with the shaded region indicating the uncertainties corresponding to the central 68 percentiles.  In each panel, the Spearman rank correlation coefficient $\rho$ and corresponding $p$-value are provided.
\label{fig:gal_vs_T_Lya}}
\end{figure*}

We investigate the correlation between the local galaxy abundance and \Lya\ transmission in more detail.  The former is represented by the galaxy number density, $n_X$, measured within narrow redshift slices of a length $X$\,cMpc across the entire survey area in each quasar sightline.  For fiducial analysis, we limit the sample to those above [\OIII]$\lambda$5008 luminosity $L_{5008} \ge 10^{42}\,\mathrm{erg\,s^{-1}}$, at which the average completeness is $\approx 40\%$, to reduce statistical uncertainties arising from the inclusion of fainter, low-completeness sources.  Due to the small number of galaxies in each bin, the number density is estimated by dividing the count by the nominal survey volume of each sightline slice, instead of applying Equation \ref{eq:galdensity_Veff}.  The corresponding mean \Lya\ transmission within the $X$-cMpc slices is denoted in $T_{\alpha,X}$.  

This correlation analysis treats all sightline slices equally and thus represents a ``volume-centric'' approach, in contrast to the ``galacto-centric'' cross-correlation analysis presented in the next subsection.

We perform this analysis in three different redshifts ranges, $z=[5.32, 5.5], [5.50, 5.70]$ and $[5.7, 6.15]$, using varying bin sizes from 5 to 20\,cMpc.  This division is motivated by the evolution of the average transmission across our six sightlines (see Figure~\ref{fig:z_vs_T}).  It was also found to strike a balance between achieving sufficient statistical precision and preserving variations between the redshift bins, thereby allowing us to clearly capture redshift evolution clearly.  

Figure~\ref{fig:gal_vs_T_Lya} shows the results.  The individual data points correspond to measurements in the individual (non-overlapping) sightline slices.  To clarify the trends, we calculate the mean \Lya\ transmission in four bins of number density.  We also fit the data points with a simple functional form that assumes a linear relation between $\ln T_{\alpha,X}$ and density.  The 1$\sigma$ confidence intervals are evaluated based on the nominal covariance matrix of the fitting parameters, assuming a Gaussian distribution.  In these procedures, we did not account for the uncertainties in the $T_\alpha$ measurements arising from noise in the spectral data of the quasars.  This choice avoids potential biases due to varying data quality across different quasars.  Notably, the conclusions remain unchanged even if the data points are weighted by their associated uncertainties. 

It is clear that the \Lya\ transmission significantly anti-correlates with galaxy density at $5.32<z<5.50$, similar to results from $z \sim 2$--4 studies \citep{2017ApJ...835..281M,2021ApJ...909..117M,2024MNRAS.529.2794M}.  The Spearman rank correlation coefficient is found to be $\rho\approx -0.4$ to $-0.6$, depending on the binning scale, but consistently associated with very small $p$-values ($\le 0.0053$).  The correlation becomes weaker or disappears in the intermediate redshift bin, and then turns positive in the highest redshift bin, with $\rho \approx 0.2$--$0.4$ with $p$-values $<0.003$.  This trend is robust across different binning scales.

\begin{figure}[t]
\centering
\includegraphics[width=2.8in]{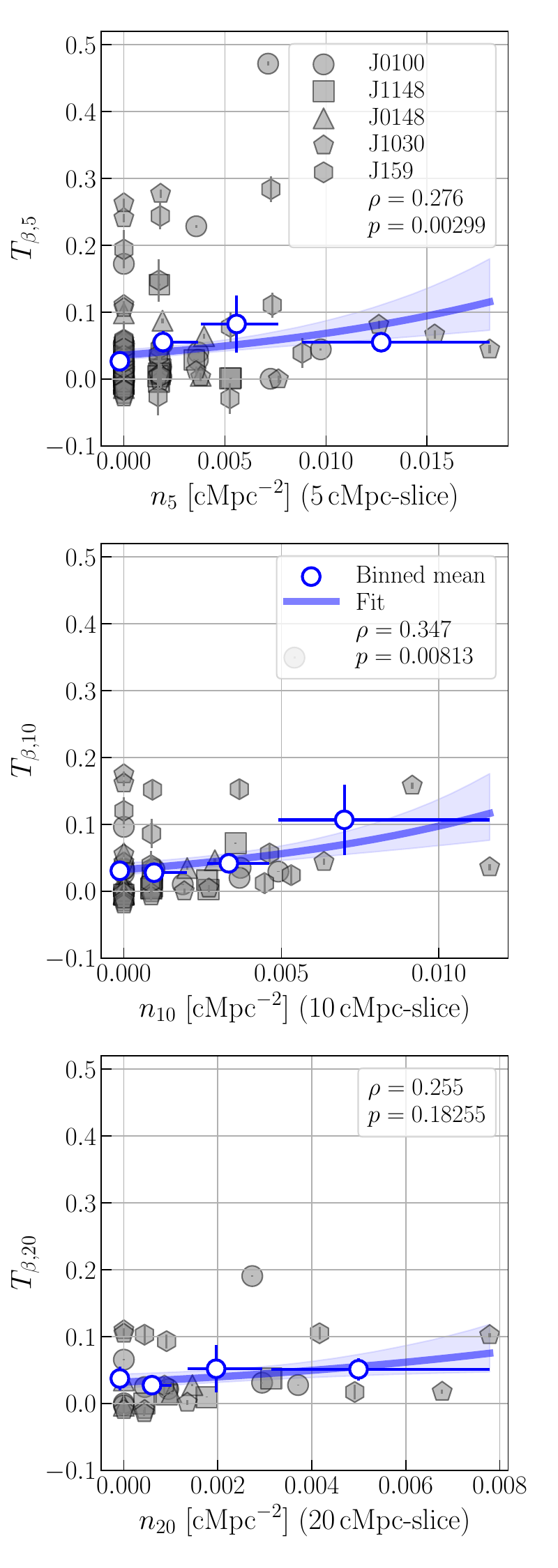}\vspace{-3mm}
\caption{Correlation between galaxy number density and \Lyb\ transmission within $5.70<z<6.15$, shown for different binning scales (5, 10, and 20\,cMpc, from top to bottom).  The symbols follow the same convention as in Figure~\ref{fig:gal_vs_T_Lya}.
\label{fig:gal_vs_T_Lyb}}
\end{figure}

We also analyze the same correlation using the \Lyb\ transmission within the highest redshift range.  Figure~\ref{fig:gal_vs_T_Lyb} shows the results for the three slice lengths.  A positive correlation is evident at high significance for binning scales $\le 10\,\mathrm{cMpc}$.  The 20\,cMpc binning yields a consistent trend, though with reduced statistical significant.  These results are fully consistent with the trend of \Lya\ transmission in the highest redshift bin and further strengthen the identification of a positive correlation between galaxy density and transmission.


\begin{figure*}[t]
\centering
\includegraphics[height=3in]{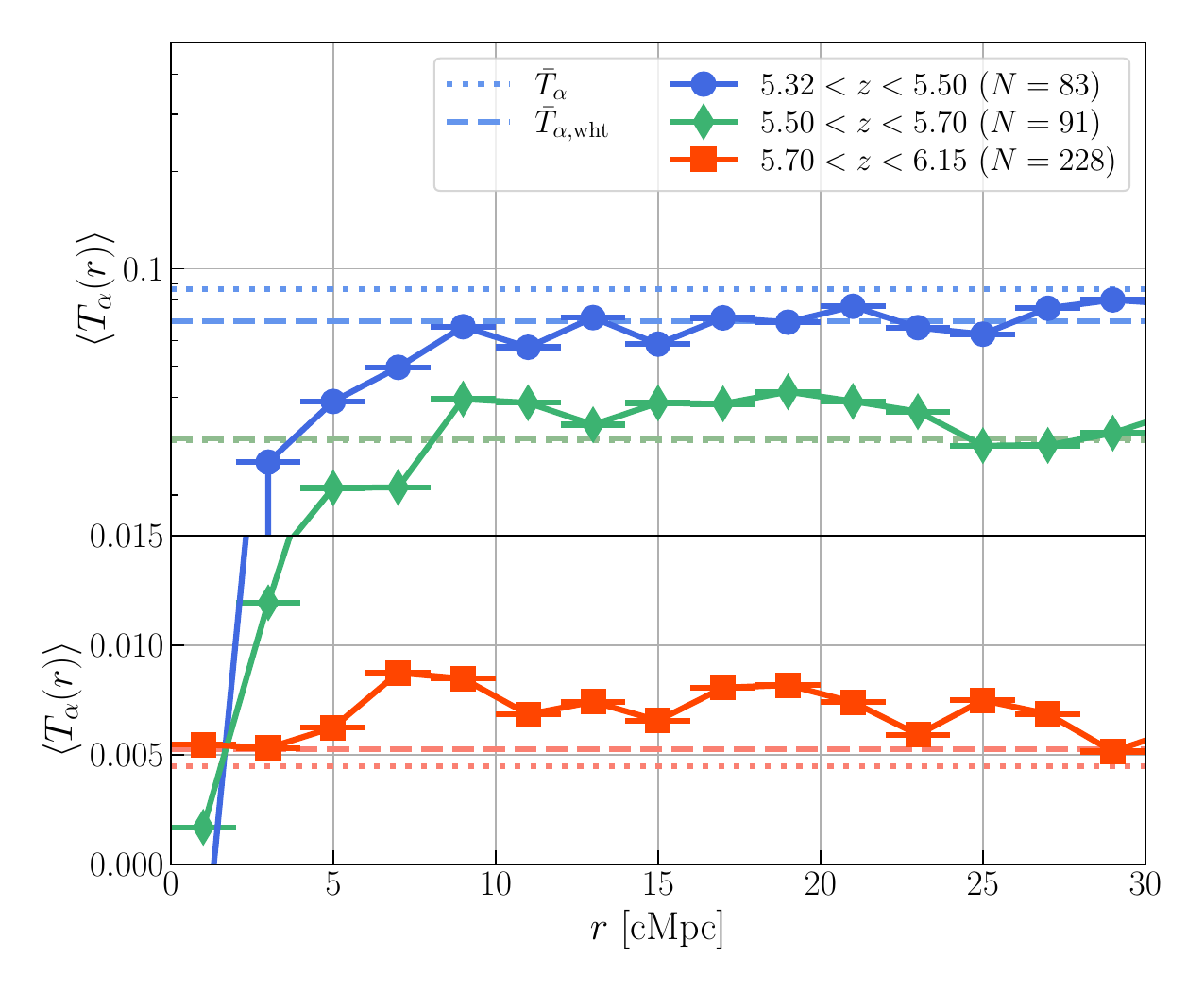}
\includegraphics[height=3in]{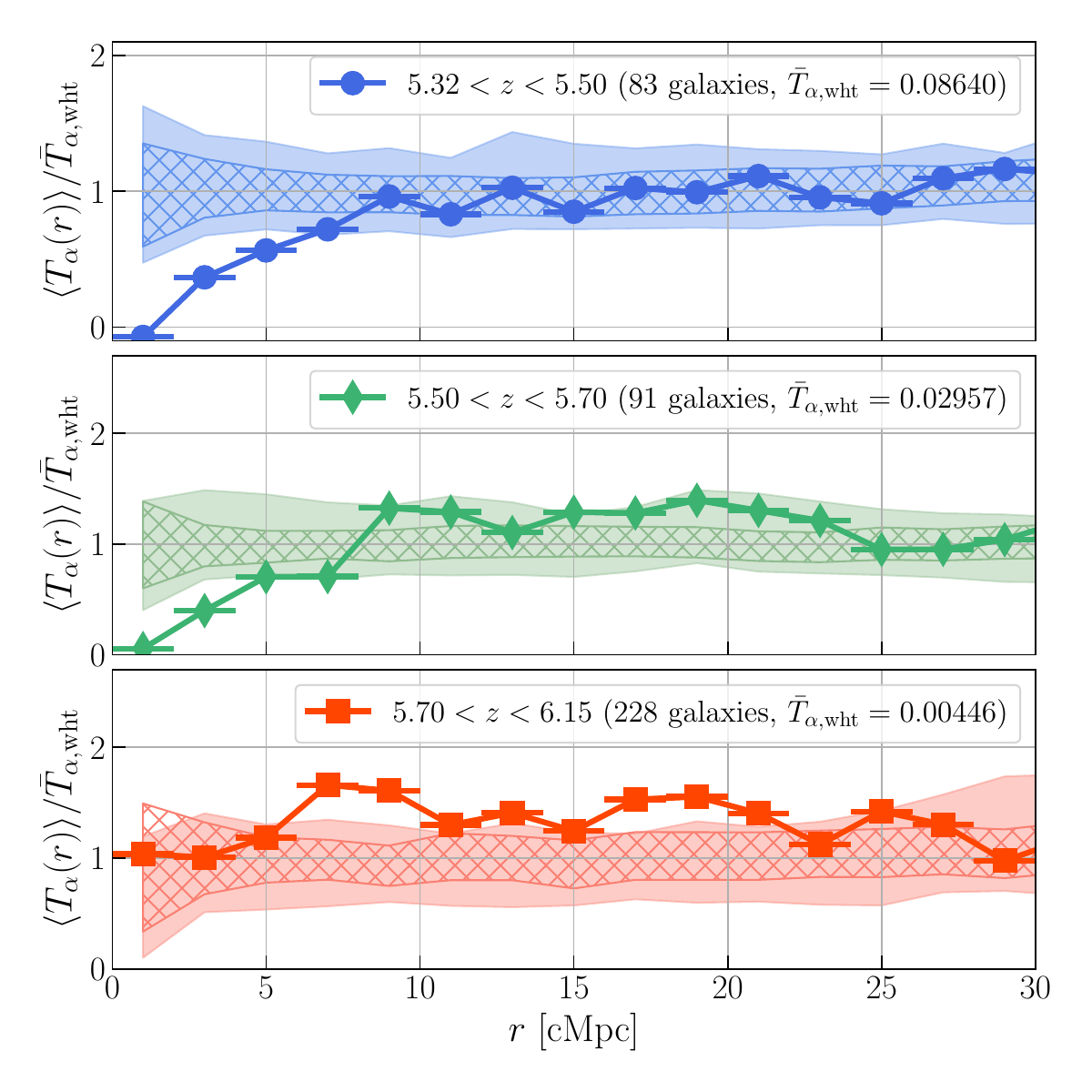}
\caption{
Left panel: \Lya\ transmission curves measured in three redshift bins along the sightlines of our six quasars (symbols connected by solid lines).  Horizontal error bars represent the grid size (2\,cMpc) of the measurements.  The horizontal dotted and dashed lines indicate the uniformly weighted mean transmission, $\bar{T}_\alpha$, and the field-to-field weighted value, $\bar{T}_\mathrm{\alpha,wht}$, respectively.
Right panel: similar to the left panel, but with transmission curves normalized by the mean value, $\bar{T}_\mathrm{wht}$.  To assess the significance of deviations from a flat trend, the scatter (16--84th percentiles) from mock measurements under the null hypothesis is shown as shaded regions.  For comparison, hatched regions show the same scatters but assuming a simplified case where galaxies are uniformly distributed at random.
\label{fig:Tcurve_Lya}}
\end{figure*}

\subsection{Galaxy--transmission cross-correlation analysis}
\label{sec:T_curve}

We measure the cross-correlation between the transmitted flux and the positions of galaxies along the sightlines.  This represents the average IGM transmission as a function of distance from the galaxies, and was called the mean transmission curve in \citet{2023ApJ...950...66K}.  This is inherently a galacto-centric analysis, in which galaxy positions serve as the reference points for measuring IGM transmission.  As a result, environments with higher galaxy densities contribute more strongly to the average, in contrast to the volume-centric approach discussed earlier.

\subsubsection{Transmission curve measurement}

We denote the transmission curve as $\langle T(r)\rangle$ where $r$ is the three-dimensional comoving distance from galaxies, and angle brackets indicate averaging over the galaxies of interest.  The formal definition is:
\begin{equation}
    \langle T(r) \rangle = \frac{\sum_{i,j} T_{i,j}(r) l_{i,j}(r)}{\sum_{i,j} l_{i,j}(r)}
    \label{eq:Tcurve}
\end{equation}
where $i$ indexes the quasar field and $j$ indexes individual galaxies in each field.  $T_{i,j} (r)$ is the mean transmission in a radial bin at distance $r$ from galaxy $j$ in field $i$, and $l_{i,j}(r)$ is the corresponding path length of the \Lya\ forest segment contributing to that radial bin.  Note that $l_{i,j}(r)$ is not constant but varies depending on the relative geometry between the galaxy and the \Lya\ forest segment.

\begin{figure}[t]
\centering
\includegraphics[width=3.4in]{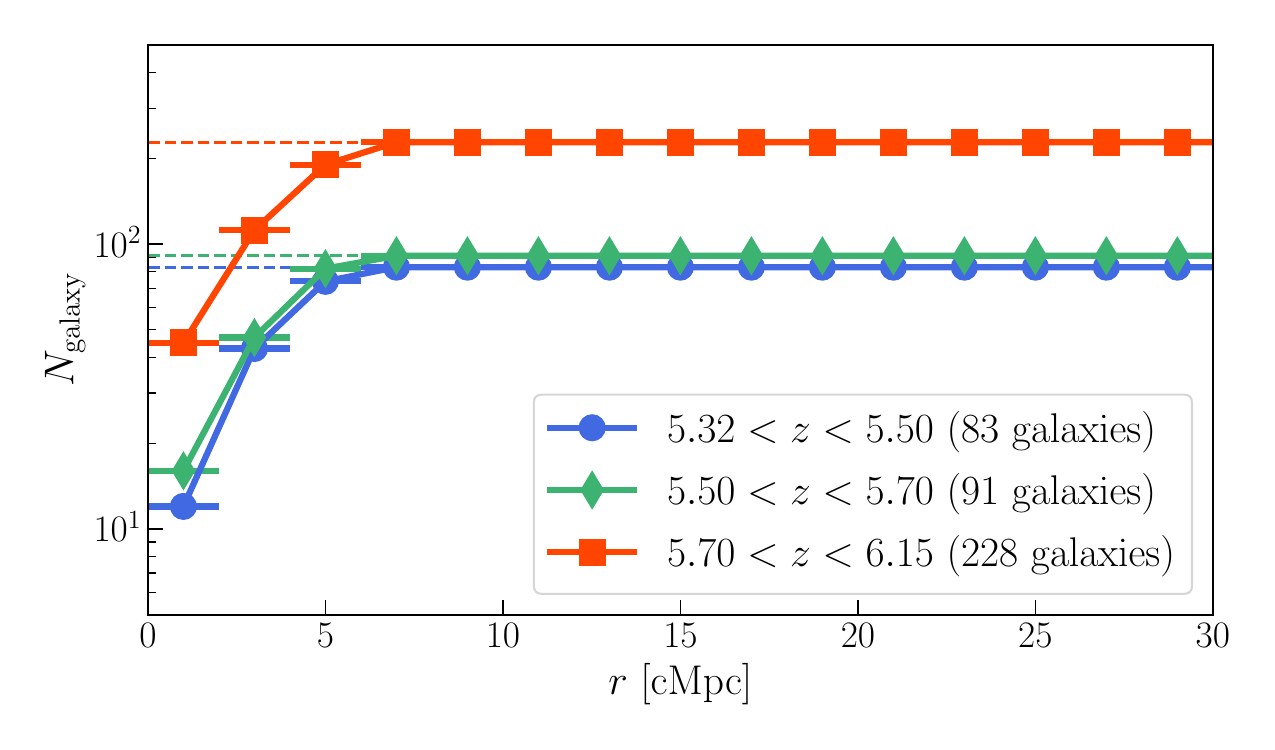}
\includegraphics[width=3.4in]{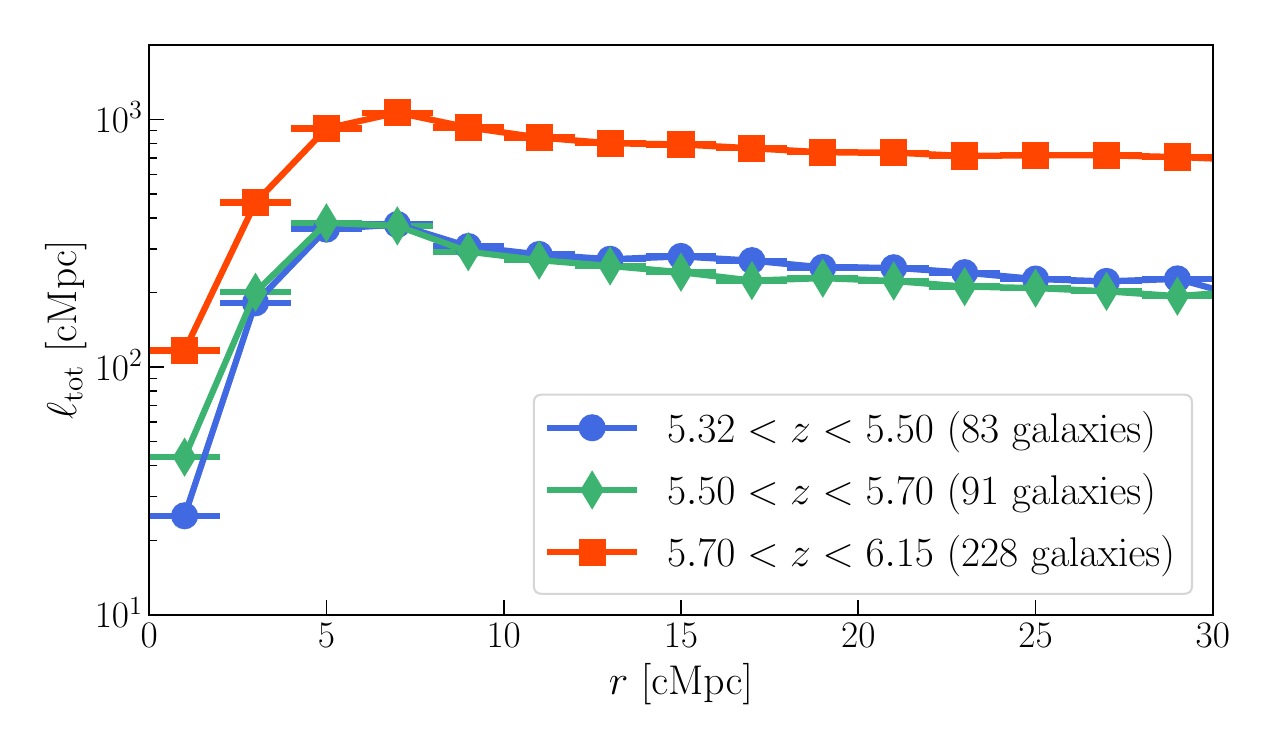}
\caption{
Upper panel: Number of galaxies contributing to the transmission curve measurement within each radial bin. Horizontal lines indicate the total number of galaxies in each redshift bin (as shown in the legend).  
Lower panel: Cumulative total path length of the \Lya\ forest, $l_\mathrm{tot}$, that contributes to each radial bin in the transmission curve measurement.
\label{fig:Tcurve_Lya_Neff}}
\end{figure}

Figure~\ref{fig:Tcurve_Lya} presents the \Lya\ transmission curves in the three redshift bins used in Section \ref{sec:gal_vs_T}.  We restrict our analysis to $r < 30\,\mathrm{cMpc}$, which corresponds to about four times the maximum transverse distance covered by our data ($\sim 7$\,cMpc at $z \sim 6$).  Note that at small radii ($< 7\,\mathrm{cMpc}$), only galaxies located near the sightlines contribute to the measurement.  Figure~\ref{fig:Tcurve_Lya_Neff} shows the number of contributing galaxies (upper panel) and the cumulative total \Lya\ forest path length (lower panel) within each radial bin.  This indicates that more than 10 galaxies contribute to the smallest radial bin, and all galaxies can contribute above $ r \approx 7\,\mathrm{cMpc}$, which corresponds to the maximum transverse distance from the center of the survey fields.

Note that uncertainties in the quasar continuum models (Section \ref{sec:QSO_spectra}) are not explicitly included this analysis.  However, their impact on the absolute transmission curves is estimated to be 2--3\%, assuming constant (random) systematic offsets in the continuum within each redshift bin.  For the normalized transmission curves (see below), the impact is even smaller--only 1--2\%--since both the observed curves and the mean transmission shift in a similar manner under a given realization of continuum model errors.

\subsubsection{Scatter estimation under the null hypothesis}

To assess the significance of features in the observed transmission curves, we compare them to expectations under the null hypothesis that the IGM transmission is completely independent of the galaxy positions.  To do this, we construct mock galaxy catalogs and cross-correlate them with the actual transmission data, following the detailed procedure described below.  Under this null hypothesis, the mock transmission curves should be flat at the mean transmission level within each redshift bin.

By using many different mock galaxy catalogues, we can also evaluate the scatter in the null hypothesis output, which procides a reasonable estimate of the statistical uncertainty of the actual measurement, and thus of the significance of any deviation from the null hypothesis.  Prior work has often estimated uncertainties assuming that galaxies are randomly distributed and spatially uncorrelated \citep[e.g.,][]{2003ApJ...584...45A,2005ApJ...629..636A,2020MNRAS.494.1560M,2021ApJ...909..117M}.  For example, errors have been approximated by dividing the standard deviation of the transmission values by $\sqrt{N}$, or by applying resampling techniques such as galaxy-by-galaxy bootstrap or jackknife.  However, such approaches do not account for the spatial clustering of galaxies, which reduces the effective number of independent measurements and can therefore lead to an overestimation of the significance.

To properly account for galaxy clustering, we construct realistic mock galaxy samples based on lightcone catalogs from the UniverseMachine Data Release 1 \citep{2019MNRAS.488.3143B}.  We convert the star formation rates (SFRs) in the catalog to [\OIII] luminosities using abundance matching to reproduce the observed [\OIII] luminosity function \citep{2023ApJ...950...67M}, assuming that the rank-ordering of SFR and [\OIII] luminosity is the same.  While this conversion is certainly an oversimplification, it is sufficiently reasonable for our purposes, as it produces a sample with the consistent number density, by construction, and at least approximately the consistent spatial clustering.

We generate 48 independent narrower lightcones, each matching the survey area of our NIRCam observations.  From these, six are randomly selected and assigned to the six quasar sightlines.  For each lightcone, we simulate grism observations using the corresponding completeness map (see Section \ref{sec:completeness}), which determines the detection probability for each mock galaxy.  This yields six mock catalogs of detected mock galaxies, which are then cross-correlated with the actual \Lya\ forest spectra (which must, by design, be completely uncorrelated with the mock catalogs).  
This random association is repeated $10^4$ times (out of almost $10^{10}$ possible combinations), and the distribution of the resulting $10^4$ transmission curves therefore allows an assessment of whether our observed curve could have been produced by chance under the null hypothesis.

As in \citet{2023ApJ...950...66K}, by normalizing the observed transmission curve by an estimate of the mean transmission,  we can obtain the normalized transmission curve, which in the null hypothesis would be unity at all radii.  The significance of deviations from unity can then be easily assessed using the set of null-hypothesis mock catalogs.  However, some care needs to be taken in constructing the appropriate baseline $\bar{T}$, as follows.

We can first easily determine a baseline, $\bar{T}$, using the ensemble of mock catalogs and Equation \ref{eq:Tcurve}.  For practical simplicity, however, we instead use a sufficiently large set of uniformly distributed random source catalogs.  These random catalogs are constructed similarly to the mock catalogs, modeling the luminosity probability distribution and completeness for each field.  Using a single large radial bin ($r=0$--15\,cMpc),\footnote{If adopting smaller fiducial binsize $\Delta r=2$\,cMpc, the transmission curve is quite flat.} this approach yields a baseline $\bar{T}$ that closely matches the simple mean over all sightlines, while properly accounting for field-to-field small variation in completeness and effective area.

The baselines $\bar{T}$ obtained in this way for the three redshift bins are shown as the horizontal dotted lines in the left hand panel of Figure~\ref{fig:Tcurve_Lya}.  A difficulty that arises is that, because $\langle T(r)\rangle$ is a galacto-centric measurement, it inherently gives greater weight to fields with more detected galaxies.  However, as has been clearly shown above, it is clear that the galaxy density correlates with the IGM transmission on large scales (Figures~\ref{fig:ngal_vs_T_LoS} and \ref{fig:gal_vs_T_Lya}).  The sign of this correlation reverses between the lowest and highest redshift bins, and essentially vanishes in the intermediate bin.   This observed correlation means that the average transmission of the set of observed EIGER galaxies will be slightly different from the estimate $\bar{T}$ obtained from the mock catalog (or randomly distributed galaxies), because these galaxies are, by construction, uncorrelated with IGM transmission at any scale.  

This mismatch will introduce a bias in the overall normalization of the observed and mock curves, potentially introducing an apparent correlation signal on small scales that could conceivably be only a manifestation of these identified correlations evident on very much large scales.  In examining the shape of the transmission curves, our interest is primarily in identifying possible small scale features, and the effects of the previously identified large scale correlations must therefore be removed.  To address this, we therefore modify the estimator of the mean transmission obtained from the mock galaxy catalogues by applying a field-to-field weighting:
\begin{equation}
    \langle T(r) \rangle_\mathrm{wht} = \frac{\sum_{i,j} T_{i,j}(r) l_{i,j} (r) N^\mathrm{obs}_i}{\sum_{i,j} l_{i,j} (r) N^\mathrm{obs}_i}
    \label{eq:Tcurve_weighted}
\end{equation}
where the summation is over all combinations of mock sources and sightlines, and $N^\mathrm{obs}_i$ is the number of real galaxies (not artificial sources) in quasar field $i$ within the redshift range.  This weighting ensures that the mock transmission curves have exactly the same average transmission as the observed ones.  The corresponding field-weighted baselines, $\bar{T}_\mathrm{wht}$, are shown in \ref{fig:Tcurve_Lya} as the dashed lines.  As expected, this is lower, relative to $\bar{T}$, in the lowest redshift bin, where there was an inverse correlation between galaxy number and mean transmission, but higher in the highest redshift bin, where the correlation reversed. It is indistinguishable in the middle redshift bin where there was little overall correlation.

\subsubsection{Observed deviations from null predictions}

We are now in a position to fairly compare the observed transmission curves with those derived from the mock catalogs.  The right panel of Figure~\ref{fig:Tcurve_Lya} shows the observed curves normalized by these field-to-field weighted $\bar{T}_\mathrm{wht}$ in each redshift range.  The shaded regions represent the 16-84th percentile range of the $10^4$ mock realizations, providing the expected scatter under the null hypothesis.   As expected, these are centered on unity.   

For comparison, we also performed the same analysis using uniformly distributed random catalogs, while still accounting for inhomogeneous completeness.  The hatched regions in Figure~\ref{fig:Tcurve_Lya} indicate the scatter expected from these random catalogs, using the same total number of sources as in the observed data.  As shown, this approach significantly underestimates the true scatter, as it neglects the spatial clustering of galaxies.

Even with this improved scatter estimate from the lightcone-based mock catalogs, the observed curves exhibit deviations that appear statistically significant.  Specifically, we find (i) a clear excess absorption at small distances ($r<8$\,cMpc) in the lowest redshift bin ($5.32<z<5.50$) and less significantly at $5.50<z<5.70$; and (ii) a marked excess in transmission at intermediate distances ($\sim 6$--20\,cMpc) in the highest redshift bin ($5.70<z<6.15$).  These features suggest the presence of physical processes that shape the IGM ionization conditions in different ways at different epochs.

\begin{figure*}[t]
\centering
\includegraphics[height=4.0in]{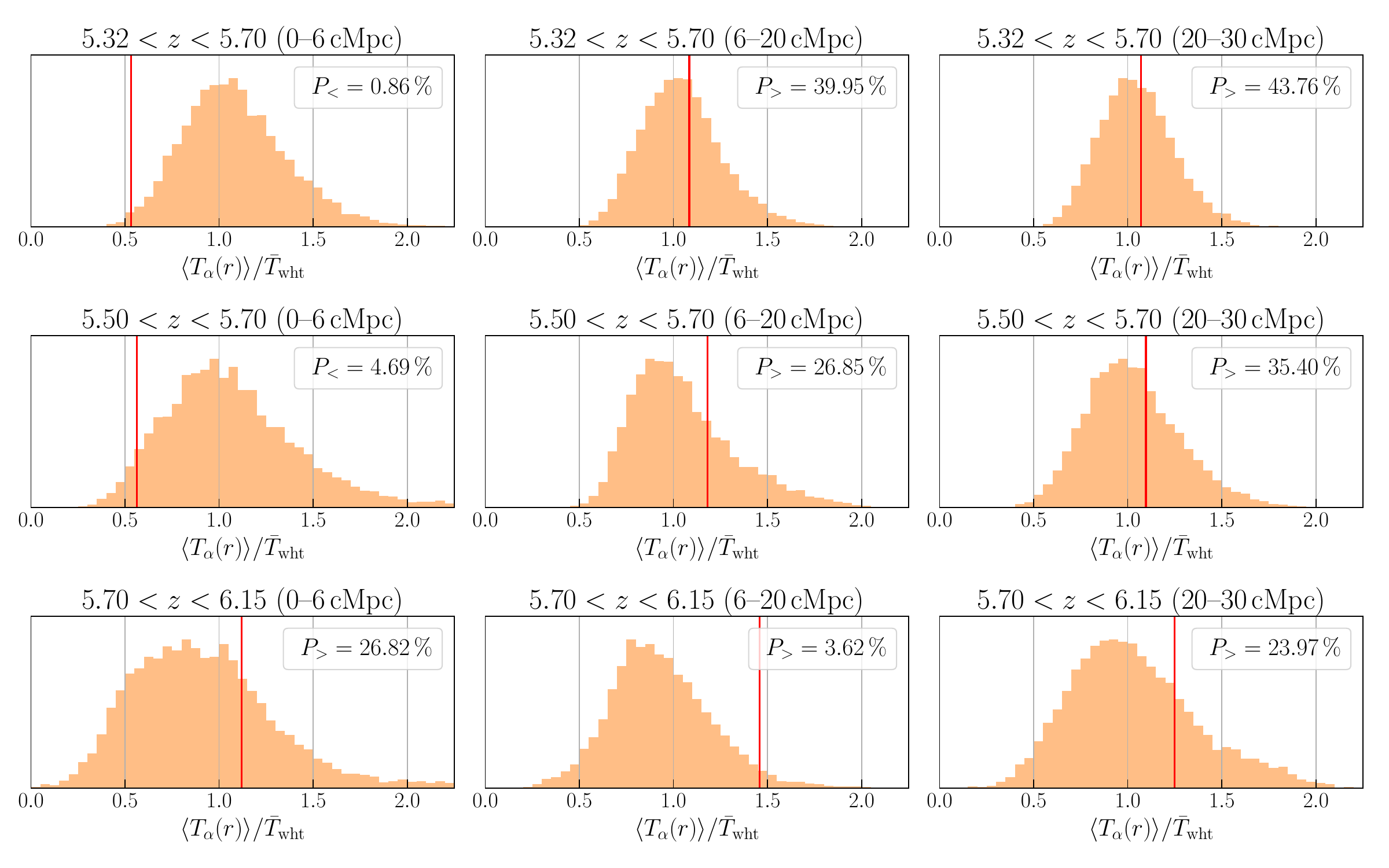}
\caption{Distributions of mock transmission values measured using Equation \ref{eq:Tcurve_weighted}, shown for different combined radial distance bins and redshift ranges.  The transmission values are normalized by the field-to-field weighted mean, $\bar{T}_\mathrm{wht}$.  The vertical red line marks the observed value.  The percentage of the mock realizations that exceed the observed value, either above ($P_>$) or below ($P_<$), are noted in each panel.
\label{fig:hist_T_norm}}
\end{figure*}

These features certainly make astrophysical sense, as will be discussed below.  However, a quantitative assessment of their statistical significance is nontrivial because adjacent radial bins are not independent, as the same transmission elements contribute at different radii for different galaxies.  This effect is present in both the real data and the mock analyses.  To address this, we combine the transmission values across broader radial ranges--specifically, $r=0$--$6\,\mathrm{cMpc}$, 6--20\,cMpc, and $r>20$\,cMpc--and compute the average transmission in each radial bin.  Figure~\ref{fig:hist_T_norm} shows the distribution of these averaged transmission values of the $10^4$ mock realizations.  

Notably, in several specific cases, the observed values fall in the tails of the distributions: (i) the excess absorption at the small radii is found in only $<1\%$ of the mock realizations for the lowest redshift bin, and $<5\%$ for the intermediate bins; and (ii) the excess transmission at the intermediate radii in the highest redshift bin occurs in just $<4\%$ of the mocks.  These values provide empirical $p$-values for testing the null hypothesis, and support our claim that the observed transmission curves reflect genuine physical correlations between galaxies and the surrounding IGM transmission properties.

We also note that, if the field-to-field weighting had not been applied when computing the mock transmission curves, the deviations become even more pronounced, incorporating large-scale galaxy--IGM correlations.  For example, the $p$-value improves to 0.04\% for the small-scale absorption in the lowest redshift bin, and to 1.05\% for the intermediate-scale transmission excess in the highest redshift bin.

\subsubsection{\Lyb\ transmission curve}

\begin{figure}[t]
\centering
\includegraphics[width=3.4in]{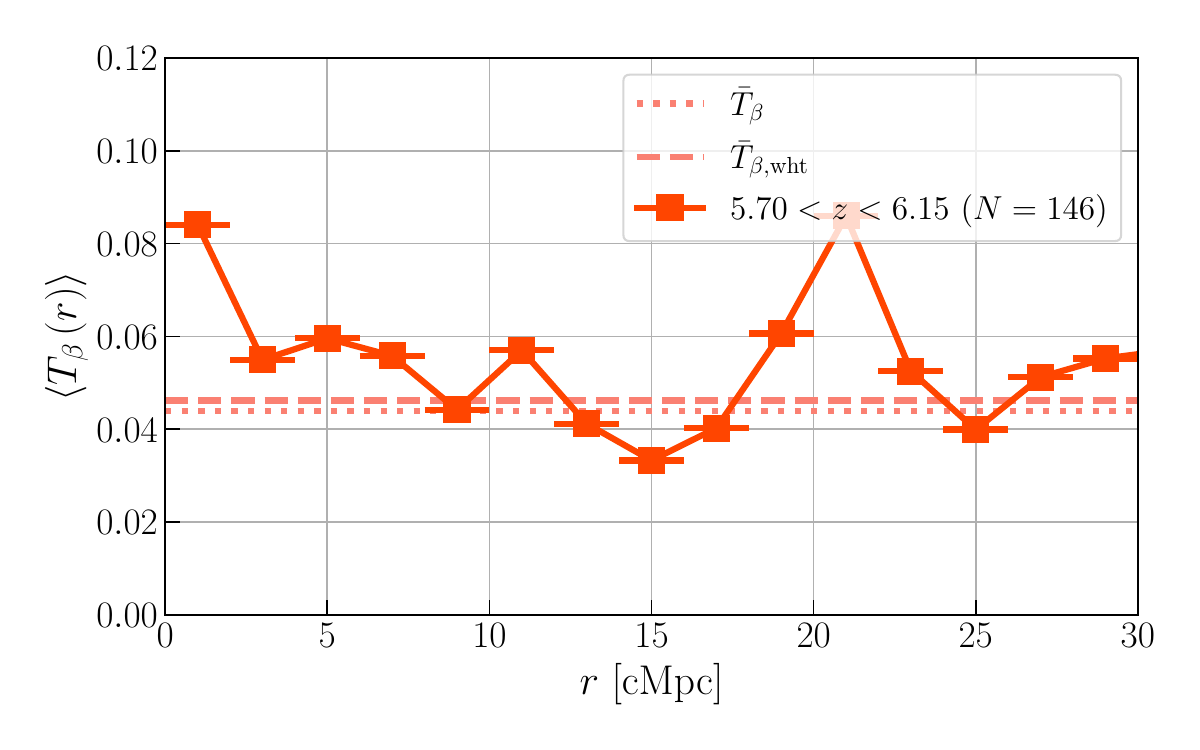}
\includegraphics[width=3.4in]{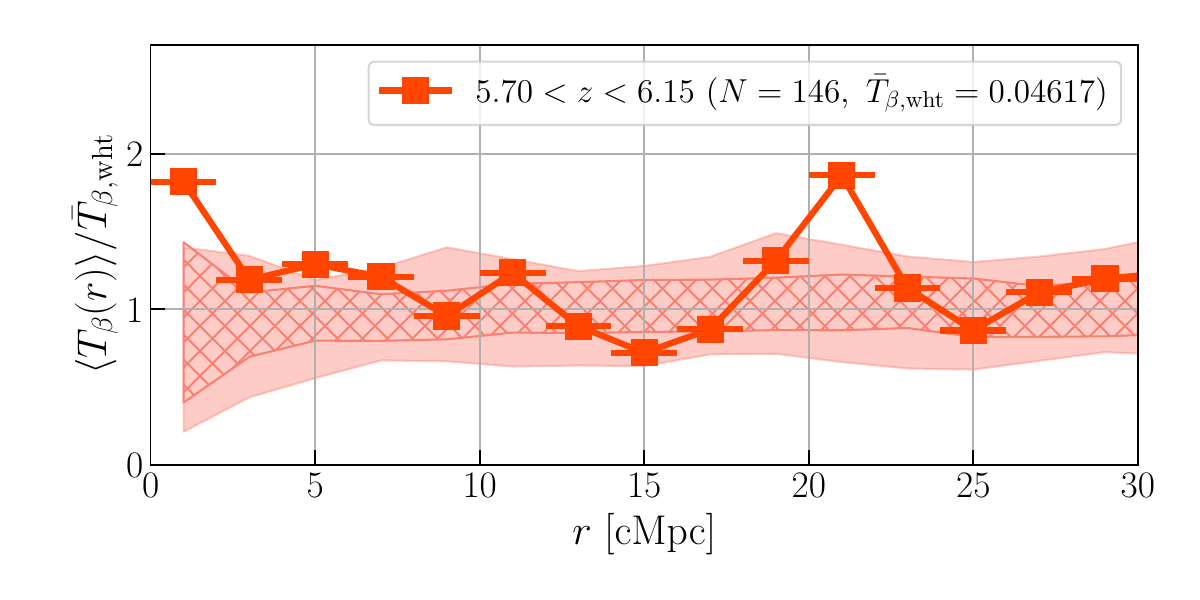}
\caption{Similar to Figure~\ref{fig:Tcurve_Lya}, but showing the \Lyb\ transmission curve, measured within $5.70<z<6.15$.
\label{fig:TransProf_Lyb}}
\end{figure}

We perform the same analysis using the \Lyb\ forest region of the spectra.  Figure~\ref{fig:TransProf_Lyb} shows the \Lyb\ transmission curve, $\langle T_\beta(r) \rangle$, for the redshift range $5.70 < z < 6.15$ (same as the highest redshift bin for \Lya), based on a sample of 146 galaxies.  Unlike the \Lya\ measurements, the \Lyb\ transmission curve does not exhibit a prominent excess transmission.  This is likely due to its reduced statistical power, stemming from a shorter total path length and contamination from foreground \Lya\ absorption at lower redshifts, which is more difficult to average out given the current sample size.  Nevertheless, the \Lyb\ curve shows a mild excess above the mean at $r \lesssim 25\,\mathrm{cMpc}$, qualitatively consistent with the trend seen in the \Lya\ curve.

We perform the same analysis using the \Lyb\ forest region of the spectra. Figure~\ref{fig:TransProf_Lyb} shows the \Lyb\ transmission curve, $\langle T_\beta(r) \rangle$, for the redshift range $5.70 < z < 6.15$ (the same as the highest redshift bin for \Lya), based on a sample of 146 galaxies. Unlike the \Lya\ measurements, the \Lyb\ transmission curve does not exhibit a prominent transmission excess. This is likely due to its reduced statistical power, stemming from a shorter total path length and contamination from foreground \Lya\ absorption at lower redshifts, which is more difficult to average out given the current sample size.

\section{Discussion}
\label{sec:discussion}

\subsection{Interpreting the evolving galaxy-transmission correlation}
\label{sec:interpretation}

\begin{figure*}[t]
\centering
\includegraphics[width=7.0in]{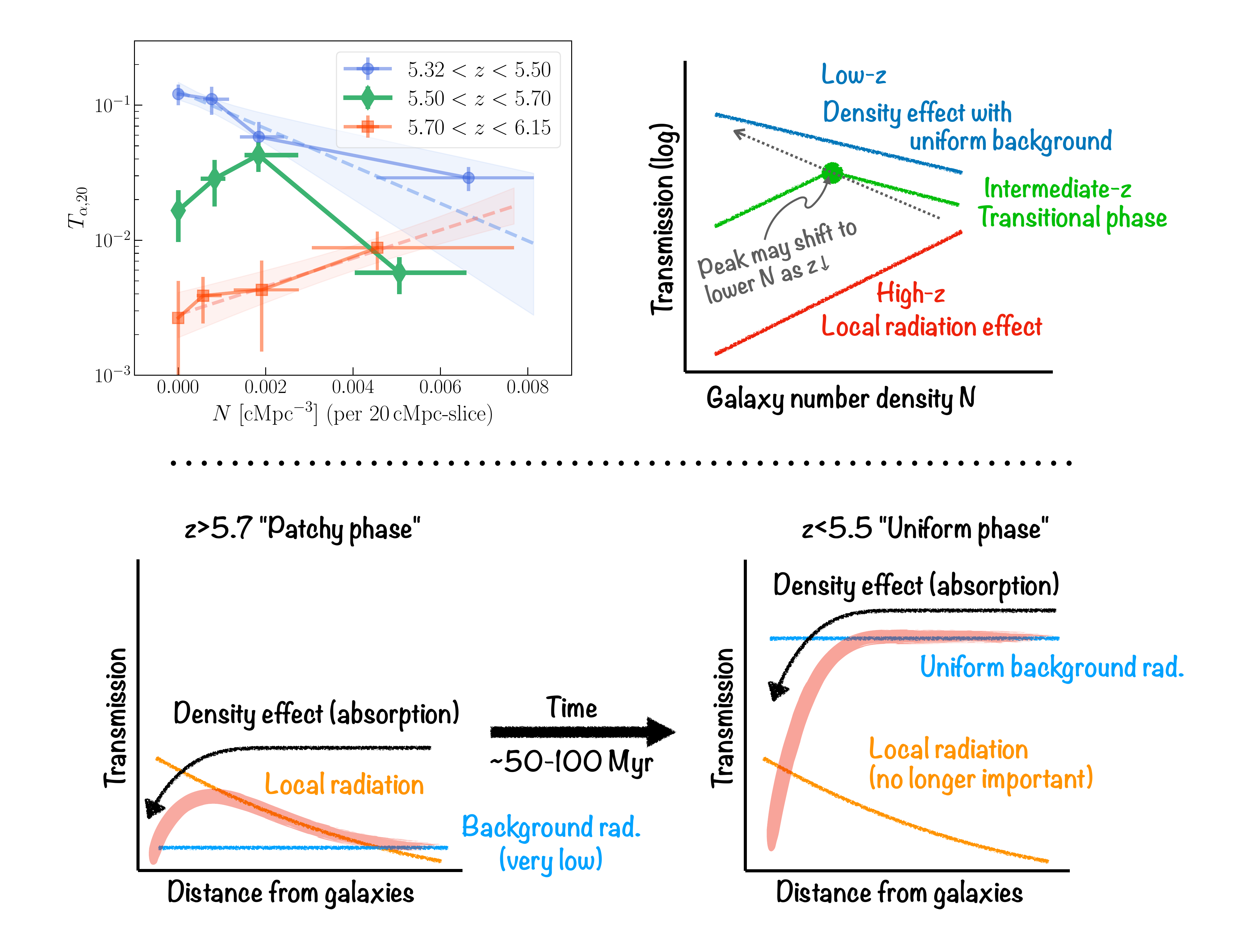}
\vspace{-5mm}
\caption{
Schematic illustration of the evolution of the galaxy-transmission correlation signals.  The upper left panel summarizes the three bottom panels in Figure~\ref{fig:gal_vs_T_Lya}, showing the correlation between mean transmission and galaxy number density averaged over 20-cMpc slices for three redshift ranges (with density binning).  In the lowest and highest redshift ranges, the data show clear monotonic negative (blue) and positive (red) correlations, respectively (dashed lines with shaded regions are the same fits as in Figure~\ref{fig:gal_vs_T_Lya}).  In contrast, the intermediate redshift bin shows a (non-monotonic) ``peak-like'' trend (green).  The upper right panel presents a qualitative interpretation of these trends, while the lower panels provide a schematic explanation for the evolution of the transmission curve (see text).
\label{fig:interpretation}}
\end{figure*}

Our analysis reveals a clear redshift evolution in the correlation between galaxies and \Lya\ transmission, based on both the ``volume-centric'' and ``galacto-centric'' metrics (Sections \ref{sec:gal_vs_T} and \ref{sec:T_curve}).  This evolution reflects a shift in the dominant physical processes that shape the IGM during the late stages of reionization.  Here, we interpret how density and radiation effects combine and compete across redshift, in shaping the observed signals, and construct a unified picture of the galaxy-IGM connection.  Figure~\ref{fig:interpretation} presents a schematic summary.

At earlier times, i.e., in the highest redshifts ($5.7 < z < 6.15$), regions of higher galaxy densities exhibit stronger \Lya\ transmission (Figure~\ref{fig:gal_vs_T_Lya}).  This positive correlation is most clearly seen with 20-cMpc binning and remain evident even in the across-sightline comparison (Figure~\ref{fig:ngal_vs_T_LoS}).  The transmission curve consistently shows a significant excess over $r \sim 5\textrm{--}25\,\mathrm{cMpc}$.  Taken together, these observations suggest that enhanced local radiation fields around galaxies dominate and effectively boost the IGM transmission, especially at intermediate scales.  The lack of a significant excess at smaller scales ($r\lesssim 5\,\mathrm{cMpc}$) implies that the effects of local radiation and increased absorption due to higher densities (coupled with elevated recombination rates) cancel each other out in the immediate vicinity of galaxies.  

At intermediate redshifts ($5.5 < z < 5.7$), the IGM enters a transitional phase.  Both density and radiation effects are at play: the transmission curve shows strong absorption at small scales and moderate excess at $r \sim 10$--20\,cMpc.  While the volume-centric correlation becomes weak across the entire range of density, a peak-like trend emerges at large scales ($\gtrsim 20$\,cMpc), with a positive correlation in low-density regions and a negative correlation in high-density regions.  This can be interpreted as a competition between local radiation dominating in low-density regions and density-driven absorption dominating in overdense ones.  During this transitional period, the mean free path of ionizing photons increases dramatically, establishing a nearly uniform radiation field across the universe.  As this transition progresses, the density effect becomes more dominant, and the ``knee'' in the density-transmission relation is expected to shift toward lower densities (see Figure~\ref{fig:interpretation}).

Then, at $z<5.5$, a strong negative correlation between galaxy number density and transmission (Figure~\ref{fig:gal_vs_T_Lya}) indicates enhanced absorption in overdense regions.  The transmission curve shows a consistent signal, with excess absorption concentrated within $\sim 8\,\mathrm{cMpc}$ of galaxies (Figure~\ref{fig:Tcurve_Lya}).  The monotonic trend of the transmission curve implies that, at this stage, ionizing photons have largely filled the universe, and the variation in transmission is governed primarily by density fluctuations under a nearly uniform background radiation field.

Our evolutionary scenario illustrated in Figure~\ref{fig:interpretation} implies a rapid increase in transparency within low-density regions, aligning with the expectation in a late-reionization scenario \citep{2020MNRAS.491.1736K}.  The transitional phase, which might have lasted for $\sim$50--100\,Myr ($\Delta z \sim 0.2$–-0.4), marks the end of the ``patchy'' phase of reionization and the onset of a more uniform ionized state of the IGM.  This is consistent with the widely discussed ``inside-out'' scenario of cosmic reionization, where high-density regions reionize first, and ionization gradually fills the voids.

Our interpretation also aligns with previous studies based on Subaru HSC observations, which show that regions of high $\taueff$ (on $\sim70$\,cMpc scales) tend to coincide with galaxy underdensities \citep{2018ApJ...863...92B,2020ApJ...888....6K,2021ApJ...923...87C}, and that $\taueff$ and galaxy density are negatively correlated \citep{2022MNRAS.515.5914I}.  A more recent study identified two low-$\taueff$ regions but with low galaxy densities at $z \sim 5.7$, suggesting substantial diversity among highly transmissive regions rather than a tight correlation \citep{2023ApJ...955..138C}. This is consistent with the predictions from cosmological hydrodynamic simulations \citep{2024arXiv241002853G,2025JCAP...03..069G}.  This behavior may reflect two aspects seen in the transmission curves: transmission tends to rise at some distances from galaxies (as the immediate surroundings are denser and more absorbing), and the transparency of low-density regions likely evolves rapidly, as discussed above.

These findings--particularly the clear positive correlation between galaxy density and transmission, and the excess in the transmission curve at high redshifts--provide strong evidence that star-forming galaxies played a central role in reionization, supporting a ``galaxy-driven'' scenario.  However, quantifying the fractional contributions of different galaxy populations--such as numerous faint galaxies versus rare, luminous ones, which likely share similar spatial distributions--cannot be determined with the current analysis.  While the contribution from AGN is considered minor, with an estimated AGN fraction of $\sim 1\%$ \citep{2024ApJ...963..129M} at $z \sim 5$, accurately assessing their role will require uncovering hidden populations.

\subsection{Qualitative discussion of some specific galaxy-transmission structures}

Following the statistical analyses presented above, we now provide several illustrative examples that highlight the connection between galaxy distribution and IGM transmission structures.  

\begin{figure}[t]
\centering
\includegraphics[width=3.4in]{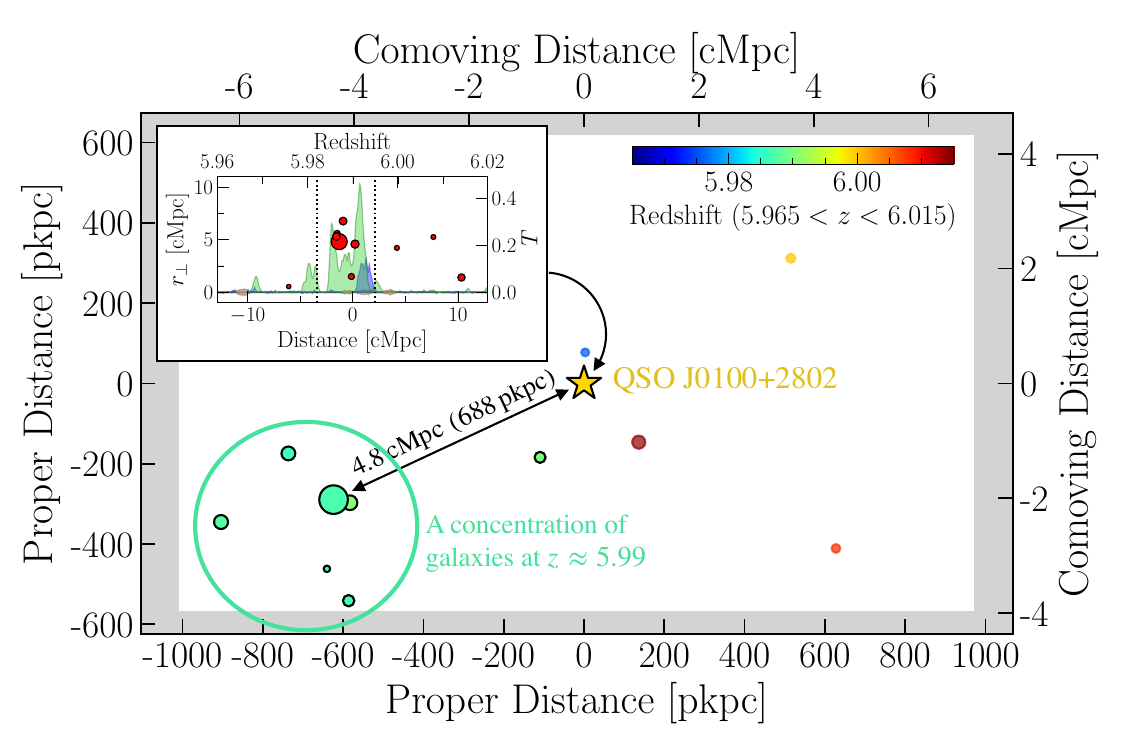}
\caption{Locations of the detected [\OIII]-emitters coincide with the strong \Lya and \Lyb\ transmission spikes at $z\approx5.99$ along the sightline toward QSO J0100 (gold star) within the survey field.  The symbol sizes represent [\OIII] luminosities, and colors indicate redshift. The inset panel shows a zoom-in of the \Lya and \Lyb\ transmission spectra around this redshift (excerpted from Figure~\ref{fig:LoS_zoom_J0100}).
\label{fig:J0100_z5.99_spike}}
\end{figure}

Along the sightline toward QSO J0100$+$2802, as previously reported by \citet{2023ApJ...950...66K}, prominent transmission spikes in both \Lya\ and \Lyb\ at $z \approx 5.99$ coincide with a group of galaxies (see Figure~\ref{fig:LoS_zoom_J0100}). Figure~\ref{fig:J0100_z5.99_spike} presents the sky positions of these galaxies relative to the quasar sightline. The clustering of \Lyb\ transmission spikes implies an ionized region extending over $\sim$5\,cMpc ($\sim$700\,pkpc) or more, assuming pure Hubble flow.  This scale is comparable to the transverse distance to the galaxy group--particularly the brightest member, suggesting that these galaxies likely created localized ionized regions giving rise to these transmission spikes.

Looking at the J1030$+$0524 sightline (Figure~\ref{fig:LoS_zoom_J1030}), we find no significant transmission spikes between $z \approx 5.72$ and $5.76$, even though a relatively high number of galaxies are detected, including some very close to the sightline ($r_\perp \lesssim 1\,\mathrm{cMpc}$).  In contrast, spikes emerge at $z \approx 5.77\textrm{--}5.81$, where we also identified a number of galaxies.  The difference between these two situations is that in the higher redshift interval the detected galaxies are located somewhat farther from the sightline ($r_\perp \gtrsim 3\,\mathrm{cMpc}$).  This suggests that the transmission is enhanced by a stronger ionizing radiation field around galaxies or galaxy groups, whereas it is suppressed in the immediate vicinity of galaxies because of the higher density.  This is consistent with the behavior of the transmission curve.

QSO J0148$+$0600 is known for its deep \Lya\ trough, spanning approximately 160\,cMpc from $z \approx 5.52$--5.81 \citep{2015MNRAS.447.3402B} (see Figure~\ref{fig:LoS_zoom_J0148}).  While no \Lya\ transmission spikes are present within this long trough, several \Lyb\ spikes have been identified, indicating that, at least, in those regions the gas is ionized enough to allow transmission of \Lyb \citep{2020MNRAS.491.1736K}.  Consistent with previous reports \citep{2018ApJ...863...92B,2020ApJ...888....6K,2021ApJ...923...87C}, we detect relatively few [\OIII]-emitters across the trough, serving as important observational support for our interpretation.  Moreover, and even more intriguingly, these few detected galaxies tend to be located near these \Lyb\ spikes.  This provides another piece of evidence for local reionization driven by galaxies within predominantly neutral regions of the IGM.

Another noteworthy feature is a galaxy overdensity identified at $z\approx 5.92$ within the proximity zone of J0148 (this redshift portion is not included in our analysis presented in Section \ref{sec:results}).  This coincides with a strong \Lya\ absorption.  Upon close inspection, a transmission spike is present at the center of this strong absorption feature, accompanied by \Lyb\ transmission.  This suggests a scenario in which the strong \Lya\ absorption is primarily driven by the high density in this region, while the nearest galaxy, and/or undetected companions, carve out a small pocket of highly ionized gas that permits the transmission of both \Lya\ and \Lyb.  This implies that the ionization state of the IGM can vary significantly over small distances around galaxies, and that average transmission curve measurements alone may be insufficient to fully capture the complexity of the galaxy--IGM connection on such scales.

These examples are individual, noteworthy relationships between galaxies and the structures of the IGM transmission that truly exist in the universe.  They provide us with consistent insights into the role of galaxies in cosmic reionization, reinforcing the picture we developed based on our statistical analyses using the full dataset.
However, not all situations observed across the quasar sightlines can be interpreted as clearly as these particular examples.  For example, while not every galaxy can be unambiguously associated with a specific transmission spike, some discrete spikes lack identified galaxies in their vicinity, for instance, strong \Lya\ spikes at $z\approx 5.85$--5.86 in the J1030 field, and at $z\approx 5.86$ in J1120.  The latter cases may suggest that the ionizing sources are galaxies located outside the field coverage or nearby faint galaxies below the detection limit.

\subsection{Comparison with the THESAN simulation}
\label{sec:thesan}

To gain further insights, we compare our measurements to the THESAN simulation \citep{2022MNRAS.511.4005K,2022MNRAS.512.3243S,2022MNRAS.512.4909G}\footnote{\url{www.thesan-project.com}}, a large-volume ($L_\mathrm{box} = 95.5\,\mathrm{cMpc}$) radiation-magneto-hydrodynamic simulation that precisely captures the interactions between ionizing photons and intergalactic gas.  It adopts the well-tested IllustrisTNG galaxy formation model \citep{2018MNRAS.473.4077P}.  THESAN models the escape of ionizing photons from individual galaxies, incorporating a subgrid, birth-cloud escape fraction, which was tuned to a constant value to reproduce a realistic reionization history \citep{2022AAS...24012604Y}, consistent with the observed evolution of neutral fraction \citep{2017MNRAS.465.4838G}.
The transmission curve for the THESAN simulation is computed for each snapshot at different redshifts.  Therefore, lightcone effects are not accounted for in this comparison.  

Here we focus on the local effects of galaxies on their surrounding IGM, as captured by the behavior of transmission curves normalized by the field-to-field weighted average (Equation \ref{eq:Tcurve_weighted}).  It is important to note that, in order to facilitate a fair comparison, we select simulation snapshots whose volume-averaged transmission most closely matches the observed field-to-field weighted mean in each redshift bin, rather than simply using the snapshots closest in redshift.  
The simulation cannot reproduce the large-scale galaxy–-transmission correlations seen in the observational data (Figure~\ref{fig:ngal_vs_T_LoS}) due to its finite box size, but this approach appropriately restricts the comparison to spatial scales that are accessible within the simulation volume.  It also helps avoid uncertainties related to differences in the reionization history between the simulation and the real universe, as well as potential biases caused by our limited set of sightlines.

\begin{figure}[t]
\centering
\includegraphics[width=3.4in]{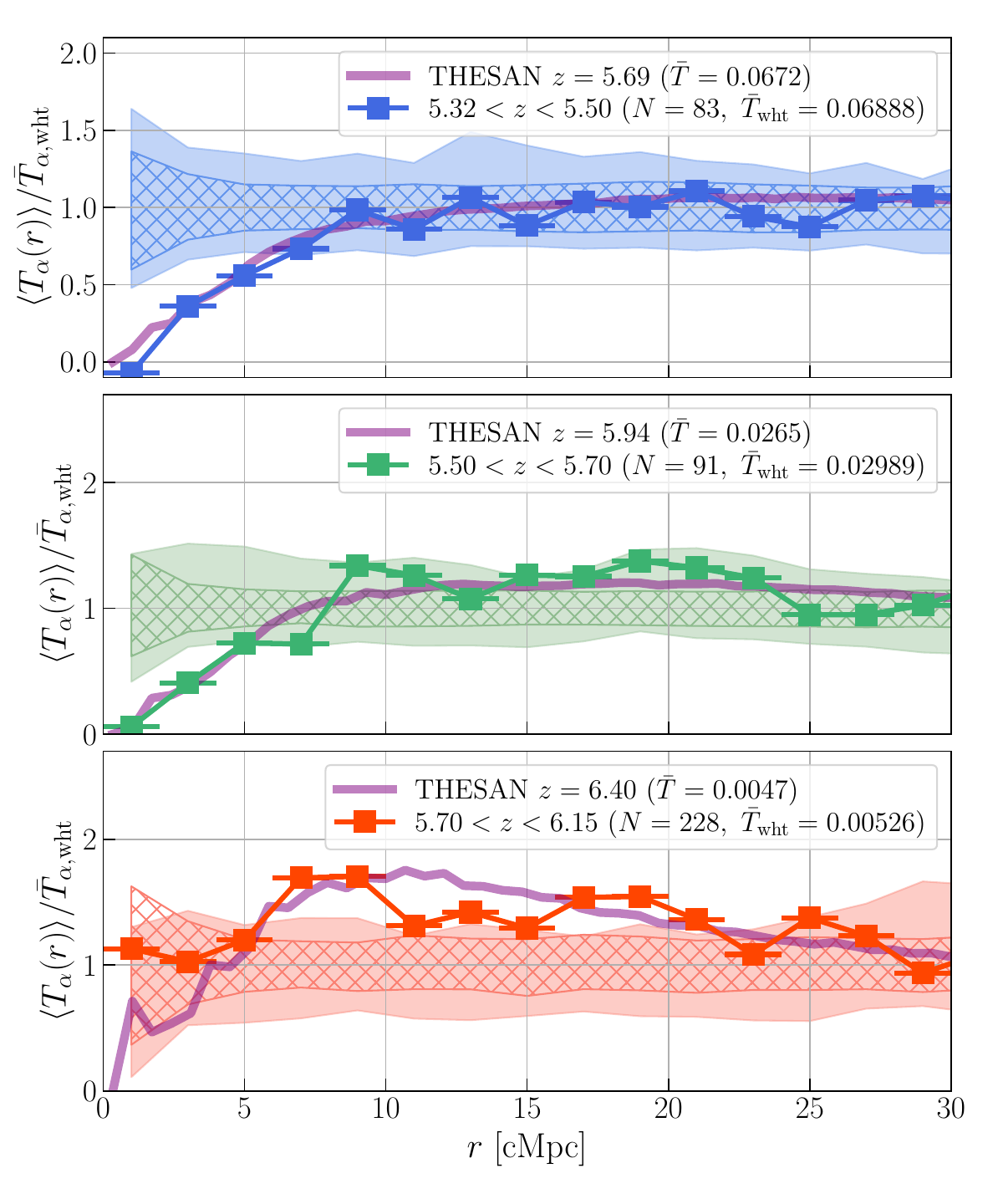}
\caption{Same as the right panel of Figure~\ref{fig:Tcurve_Lya}, but comparing with the predictions from the THESAN simulation for the sample of galaxies at $M_\star > 10^8\,M_\odot$ (purple solid lines).  From top to bottom, the redshift range is $z=[5.32, 5.50], [5.50, 5.70]$ and $[5.70, 6.15]$, respectively.
\label{fig:Tcurve_Lya_THESAN}}
\end{figure}

Figure~\ref{fig:Tcurve_Lya_THESAN} presents the normalized \Lya\ transmission curves obtained from our observations (same as in the right panel of Figure~\ref{fig:gal_vs_T_Lya}), compared with those predicted in the fiducial THESAN-1 run \citep[][private communication]{2022MNRAS.512.4909G}.  The THESAN predictions are based on a sample of galaxies with stellar masses above $> 10^8\,M_\odot$.  This stellar mass range broadly agrees with our sample \citep{2023ApJ...950...67M}.  Note that comparisons with samples with $>10^9\,M_\odot$ and $>10^7\,M_\odot$ show no significant differences in the conclusions.

The chosen snapshots' redshifts are $z=5.69$, 5.94, and 6.40 for the three redshift bins from lowest to highest. The corresponding volume-weighted neutral hydrogen fractions ($\bar{x}_\mathrm{HI}$) in these snapshots are 0.016, 0.060, and 0.20, respectively.  These estimates are in broad agreement with up-to-date constraints based on modeling inhomogeneous radiation field using a post-processing method (\citealt{2023MNRAS.525.4093G}; see also \citealt{2025PASA...42...49Q}).  While necessarily model-dependent, these estimates offer useful insight into the ionization state of the regions probed by our six sightlines.  In particular, the high $\bar{x}_\mathrm{HI}$ value in the highest redshift bin implies that a substantial fraction of the volume remains neutral (i.e., locally $x_\mathrm{HI} \sim 1$), consistent with the patchy view of reionization.

At all redshifts, the transmission curve predicted in the simulation aligns remarkably well with the observation.  In our lowest redshift bin, the decline toward $r = 0$ is closely matched in THESAN, including its scale. This consistency is also evident in the intermediate redshift bin, where the simulation also reproduces the modest excess transmission at larger $r$ with reasonable accuracy.  In the highest redshift bin, the magnitude of the transmission excess predicted by the simulation is in excellent agreement with the observations.  

Our analysis, based on six sightlines, enables a level of comparison unprecedented in precision.  This demonstrates that observations of the galaxy--IGM interplay are now reaching a stage where they can be quantitatively compared with state-of-the-art reionization simulations.  As future observations expand the statistical sample, more precise comparisons will become feasible, allowing tighter constraints on galaxy evolution models and the physical processes governing the escape of ionizing photons into the IGM.

\section{Summary}
\label{sec:summary}

We presented a comprehensive analysis of the interplay between galaxies and the intergalactic medium (IGM) during the tail end of the reionization, utilizing the complete dataset from the EIGER survey.  Using JWST NIRCam grism spectroscopy, we constructed a sample of 948 spectroscopically-confirmed [\OIII]-emitting galaxies spanning $5.33 < z < 6.97$ within six quasar sightlines.  This study particularly focused on the correlation between galaxy density and transmission, as well as the so-called galaxy-transmission cross-correlation, across three redshift ranges ($5.32<z<5.50$, $5.50<z<5.70$, and $5.70<z<6.15$).  

We observed significant redshift evolution in the galaxy density--transmission correlation (Section \ref{sec:results}).  At the lowest redshift range, we found a negative correlation between galaxy density and \Lya\ transmission.  Consistently, the transmission curve exhibits strong \Lya\ absorption within $r<8\,\mathrm{cMpc}$ of galaxies.  These findings can be attributed to the elevated neutral hydrogen density in overdense environments.  

At the highest redshift range, the correlation reverses, with \Lya\ transmission increasing in high-density regions.  The transmission curve revealed coherent transmission excess over $r\sim5\textrm{--}25\,\mathrm{cMpc}$, indicating an enhanced local radiation field generated by galaxies.  

The intermediate redshift range marks a transitional period, where the effects of overdensities and radiation field cancel each other out, leading to the absence of significant correlation between galaxy density and transmission.   This represents the conclusion of the patchy phase of reionization and the achievement of a uniform ionization state of the IGM.

Overall, these results clearly demonstrate that local ionizing radiation from galaxies is responsible for transmission of the IGM at $z\gtrsim 5.7$, and thus provide convincing evidence for galaxy-driven reionization.  This study underscores the importance of spectroscopic galaxy surveys with quasar sightline observations for probing the reionization process.  Future investigations combining even deeper and multiwavelength galaxy surveys along a large number of sightlines will be pivotal for mitigating cosmic variance and refining our understanding of the fractional contributions of different galaxy populations to reionization.

\begin{acknowledgments}

We thank Enrico Garaldi, Rahul Kannan, Aaron Smith and the \textsc{THESAN} team for providing us with their simulation data, and Norbert Pirzkal for helpful advice on our NIRCam observing program.

This work is based on observations made with the NASA/ESA/CSA James Webb Space Telescope.  The data were obtained from the Mikulski Archive for Space Telescopes at the Space Telescope Science Institute, which is operated by the Association of Universities for Research in Astronomy, Inc., under NASA contract NAS 5-03127 for \textit{JWST}.  These observations were taken under GTO program \# 1243.   

The participation of S.J.L. as an Interdisciplinary Scientist in the JWST Flight Science Working Group 2002-2022 has been supported by the European Space Agency.  His earlier involvement in the development of the JWST 1996-2001 was supported by the Canadian Space Agency.

This work is partially based on observations collected at the European Organisation for Astronomical Research in the Southern Hemisphere under ESO programmes 084.A-0360(A), 084.A-0390(A), 086.A-0162(A), 086.A-0574(A), 087.A-0607(A), 089.A-0814(A), 093.A-0707(A), 098.B-0537(A), and 286.A-5025(A).
Some of the data presented herein were obtained at Keck Observatory, which is a private 501(c)3 non-profit organization operated as a scientific partnership among the California Institute of Technology, the University of California, and the National Aeronautics and Space Administration. The Observatory was made possible by the generous financial support of the W. M. Keck Foundation.  This paper includes data gathered with the 6.5 meter Magellan Telescopes located at Las Campanas Observatory, Chile. 

This work has been supported by JSPS KAKENHI Grant Number JP21K13956 (D.K.).  J.M. is supported by the European Union (ERC, AGENTS, 101076224).

\end{acknowledgments}

\facilities{JWST (NIRCam), VLT:Kueyen (X-Shooter), Magellan:Baade (FIRE), Keck:I (MOSFIRE), Keck:II (ESI)}

\software{
Python,
numpy \citep{harris2020array},
scipy \citep{2020SciPy-NMeth},
Astropy \citep{2013A&A...558A..33A},
Matplotlib \citep{4160265},
PypeIt \citep{2019ascl.soft11004P}
}

\bibliography{ads}
\bibliographystyle{aasjournal}

\appendix

\section{Comparisons to our previous results}

\begin{figure}[t]
\centering
\includegraphics[width=3.4in]{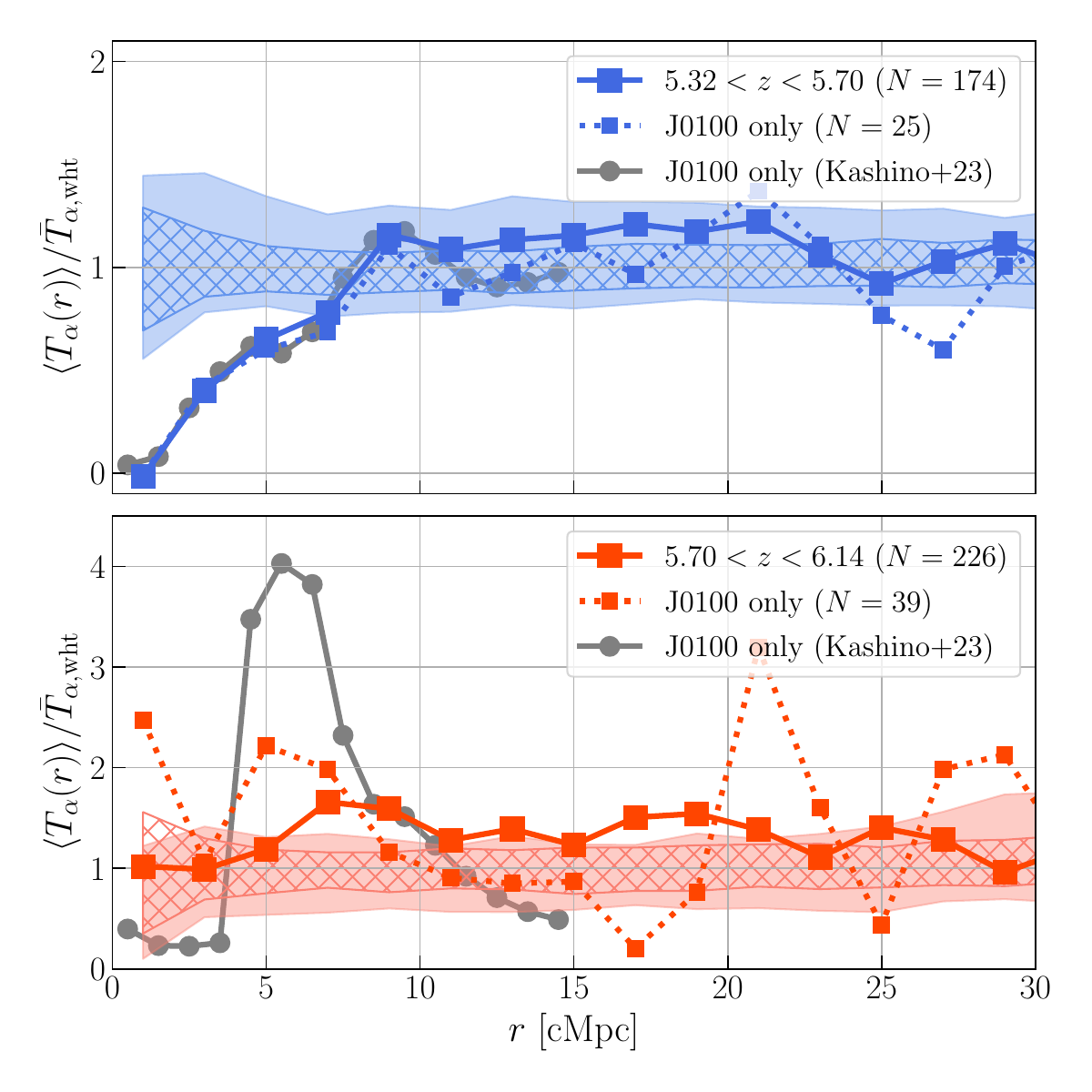}
\caption{Comparison between the early results from \citet{2023ApJ...950...66K}, based solely on the J0100 field, and the latest measurements.  Normalized transmission curves are shown for two redshift intervals: the upper panel for $5.32<z<5.70$ and the lower panel for $5.70<z<6.14$.  Larger squares connected by solid lines show current measurements using all six quasar fields (five for the lower redshift bin, excluding J1120).  Smaller squares connected by dotted lines represent updated measurements using only the J0100 field.  Gray squares indicate the previous version of the J0100-only result, where the methodology for computing the curves differs slightly.
\label{fig:Tcurve_Lya_J0100comp}}
\end{figure}

We compare the current findings with those presented in \citet{2023ApJ...950...66K}, which were based solely on the single quasar field J0100.  That earlier study performed a cross-correlation analysis in three redshift intervals: $5.32 < 5.70$, $5.70 < z < 6.14$, and $6.15<z<6.26$.  The highest bin corresponds to the near zone of the quasar and is excluded from the present comparison, as the current study does not include near-zone regions.  For consistency, we remeasure the transmission curves in the same two lower redshift bins, using both the full set of six quasar fields and only the J0100 field, now based on the updated galaxy catalog.  The results are shown in Figure~\ref{fig:Tcurve_Lya_J0100comp}.

We note that the methodology for deriving transmission curves differs between the two studies.  In particular, the previous analysis applied Gaussian smoothing with $\sigma = 2\,$cMpc to the transmission spectra and normalized the transmission by the cosmic mean as a function of redshift \citep{2018ApJ...864...53E}, prior to computing the correlation.  Given these differences, we focus on qualitative comparisons of the overall trends, rather than attempting a detailed quantitative comparison.

In the lower redshift bin, \citet{2023ApJ...950...66K} reported strong absorption at $r \lesssim 8\,\mathrm{cMpc}$, which is consistent with our current results, whether using all six fields or only the J0100 field, in both amplitude and the radial extent of the signal.

In the higher redshift bin, the earlier study identified a prominent transmission peak at $r\sim6\,$cMpc, reaching up to four times the mean transmission level.  This was interpreted as strong evidence for local ionization of the IGM by nearby galaxies.  The reanalysis using the J0100 field and updated catalog still shows a peak at similar scales, though with reduced amplitude.  In addition, this updated J0100-only result shows excess transmission at the smallest radial bin, a feature not seen in the original study.  These differences are likely due to changes in normalization method and improved detection of fainter galaxies in the revised analysis.

In contrast, our new measurements based on multiple sightlines do not reveal strong peaks at specific scales.  Instead, they exhibit a more gradual and extended transmission excess over a broad range of distances, which may reflect a more representative average over cosmic variance.

It is important to recognize that transmission curves derived from a single sightline, such as J0100, are sensitive to cosmic variance \citep[see][]{2024arXiv241002850G}.  Indeed, the strong peak seen in the earlier result is largely driven by a specific group of galaxies and a concentration of transmission spikes at $z=5.99$ along this sightline (see Figure~\ref{fig:LoS_zoom_J0100}).  While this does not represent the cosmic average, it provides direct evidence for the spatial coincidence of galaxies and locally ionized regions, offering strong support for the scenario in which galaxies drive reionization.  Rather than contradicting the earlier results, the addition of new sightlines and refined measurements in this study allows us to better characterize the ensemble-average signal of the galaxy-IGM connection during reionization.

\end{document}